\documentclass[aps,prb,reprint,twocolumn,oneside,floatfix,superscriptaddress,showpacs]{revtex4-1}
\usepackage{graphicx,xcolor}
\usepackage[utf8]{inputenc}
\usepackage{amssymb,bbm}
\usepackage[reqno]{amsmath}
\usepackage[english]{babel}
\usepackage{ulem}

\newcommand{\mbf}[1]{\mathbf{#1}}
\newcommand{\dw}{\downarrow}
\newcommand{\up}{\uparrow}

\usepackage[colorlinks,bookmarks=false,citecolor=blue,linkcolor=red,urlcolor=blue,hyperfootnotes=true]{hyperref}

\begin{document}

\title{Variational cluster approach to superconductivity and magnetism in the Kondo lattice model}

\author{Benjamin Lenz}
\affiliation{Institut f\"ur Theoretische Physik, 
  Universit\"at G\"ottingen, 37077 G\"ottingen, Germany}
\affiliation{Centre de Physique Th\'{e}orique, Ecole Polytechnique, CNRS, Universit\'{e} Paris Saclay, 91228 Palaiseau, France}
\author{Riccardo Gezzi}
\affiliation{Institut f\"ur Theoretische Physik, Universit\"at G\"ottingen, 37077 G\"ottingen, Germany}

\author{Salvatore R. Manmana}
\affiliation{Institut f\"ur Theoretische Physik, Universit\"at G\"ottingen, 37077 G\"ottingen, Germany}

\date{\today}                 

\pacs{}

\begin{abstract}
We investigate in detail antiferromagnetic (AF) and superconducting (SC) phases as well as their coexistence in the two-dimensional Kondo lattice model on a square lattice,  which is a paradigmatic model for heavy fermion materials.
The results presented are mainly obtained using the variational cluster approximation (VCA) and are complemented by analytical findings for the equations of motion of pairing susceptibilities. 
A particularly interesting aspect is the possibility to have s-wave SC near half-filling as reported by Bodensiek \textit{et al.} [Phys. Rev. Lett. \textbf{110}, 146406 (2013)]. 
When doping the system, we identify three regions which correspond to an AF metallic phase with small Fermi surface at weak coupling, an AF metal with a different Fermi surface topology at intermediate coupling, and a paramagnetic metal with a large Fermi surface at strong coupling.
The transition between these two AF phases is found to be discontinuous at lower fillings, but turns to a continuous one when approaching half-filling.
In the quest for s-wave superconductivity, only solutions are found which possess mean-field character. 
No true superconducting solutions caused by correlation effects are found in the s-wave channel.
In contrast, we clearly identify robust d-wave pairing away from half-filling.
However, we show that only by treating antiferromagnetism and superconductivity on equal footing artificial superconducting solutions at half-filling can be avoided.
Our VCA findings support scenarios previously identified by variational Monte Carlo approaches
and are a starting point 
for future investigations with VCA and further approaches such as cluster-embedding methods.
\end{abstract}

\maketitle


\section{Introduction.}
Strong correlations lead to unconventional behavior. 
This is in particular true for heavy fermion materials, where elements with partially filled f-shells contribute strongly localized $f$-electrons that are perceived as local moments by the conduction electrons of the s-, p- or d-shells.  
It is this interaction between conduction electrons and $f$-electrons, which leads to the emergence of a variety of unconventional phases.
Two questions which are still debated concern the nature of the quantum critical point, which is found between a magnetically ordered and a disordered phase in some of these materials \cite{SAC+00,SRIS01,GS03,Si06,GSS08,HOK+10,Si10,SS10,GSGB15} and the nature of superconductivity in others \cite{ABB+89,TKI+98,SAM+01,BFH+02,YGD+03,HGdN+07,NSW+10,SAF+11,SSW+13,Ste14,ESM+16,SW16}.
From a theoretical point of view these questions are often addressed within a paradigmatic model of heavy fermion systems, namely the Kondo lattice model (KLM) \cite{Don77}, whose phase diagram on a two-dimensional (2D) square lattice at $T=0$ is the topic of this paper.

In this model, the coupling between localized f-spins and conduction electrons results at weak coupling in an effective RKKY interaction between the f-spins, which leads to an antiferromagnetic (AF) ordering.
In the limit of strong coupling, the local spins are screened by the conduction electrons and local Kondo singlets between conduction electrons and f-spins are formed.
At zero temperature, the AF order is destroyed at a critical coupling strength $J_c$, which amounts to a quantum critical point (QCP).
For this QCP, two scenarios are discussed.
In case of a local QCP \cite{SRIS01,GS03,ZGS03} the breakdown of antiferromagnetic order coincides with the absence of Kondo screening (so-called Kondo breakdown).
If only the AF long-range order vanishes at the QCP, it is a QCP of Hertz-Millis-Moriya type \cite{Her76,Mil93,MT95,GSS08}.

The second important phenomenon treated in our paper is the emergence of unconventional (non-phonon-mediated) superconductivity (SC) encountered in certain heavy fermion systems. 
Such phases are often found in the vicinity to an antiferromagnetic QCP \cite{SS10}, but superconductivity due to magnetic spin fluctuations associated to other types of order have also been reported \cite{MGM+96,BFH+02,HGdN+07,HW15}.
In some compounds, superconductivity and antiferromagnetism are even reported to coexist \cite{YKM+07,NSW+10}.
Within the KLM the aforementioned RKKY interaction leads to an antiferromagnetic QCP, and numerical approaches have indeed reported d-wave SC close to this QCP recently \cite{ABF13,AFB14,WT15,Ots15}.

The scope of this paper is to investigate these aspects by obtaining and characterizing the phase diagram of the KLM via the variational cluster approximation (VCA). 
Since it is a cluster method, it is able to take into account the $k-$dependence in Green functions, in contrast to previous dynamical mean-field theory (DMFT) studies. 
In particular, we want to gain further insights into the realization of d-wave SC and the possibility of stabilizing s-wave SC as reported by Bodensiek \textit{et al.} \cite{BZV+13}.
Where possible, the results will be compared to the scenarios obtained by other methods, such as dual Fermions \cite{Ots15}, dynamical cluster approximation (DCA) \cite{MA08,MBA10}, real-space DMFT (rDMFT) \cite{PK15} and (variational) Monte Carlo ((V)MC) approaches \cite{Ass99,ABF13,AFB14}.

The paper is organized as follows: In the next section the VCA method, which we adopt to study the KLM and the physical quantities of interest, is introduced.
Next, in Sec.~\ref{sec:phasedia} we discuss the phase diagram for fillings $0.8 \lesssim n \lesssim 1$, which is the main result of this paper.

In Sec.~\ref{KLM_PM} we illustrate the results of the paramagnetic phase at and away from half-filling.
Section~\ref{KLM_AF} is then devoted to the investigation of magnetism in this region of the phase diagram.
In particular, we discuss the ground state and the characteristic Fermi surfaces for three distinct regimes at weak, intermediate and strong coupling.
	
In Sec.~\ref{KLM_SC} we study superconductivity for local s-wave and nodal d-wave order. 
Concerning s-wave SC, we find only mean-field-like solutions and no local superconductivity is induced by correlation effects. 
In contrast, robust d-wave SC is found and analyzed in detail.
In Sec.~\ref{EOM}, we complement this numerical study by considering the equations of motion (EOM) for the pairing susceptibility. 
Finally, we discuss the interplay of d-wave SC and AF in Sec.~\ref{SCd+AF}.
Summary and Outlook in Sec.~\ref{Summary} conclude the paper.

\section{Model and Technique}
In this paper we study the Kondo-lattice model (KLM) 
\begin{equation} \label{eq:KLM}
\begin{split}
\mathcal{H} =& -\sum_{\langle i,j\rangle, \alpha} {t}_{ij}(\hat{c}^{\dagger}_{i\alpha} \hat{c}^{}_{j\alpha} + \mathrm{H.c.} ) - \mu \sum_i \hat{n}^{}_i \\&+ 
\frac{J}{2}  \sum_{i,\alpha,\beta} \mathbf{\hat{S}}_i^{} \cdot \hat{c}_{i\alpha}^{\dagger} \mathbf{\sigma}_{\alpha\beta}^{} \hat{c}_{i\beta}
\end{split}
\end{equation}
using the variational cluster approximation (VCA) at zero temperature.
Here, the creation and annihilation operators of a conduction electron on site $i$ with spin $\sigma$ are denoted by $c_{i\sigma}^{(\dagger)}$ and the spin operators of the localized spin on site $i$ by $\mathbf{\hat{S}}_i$.
The first two terms describe a tight binding band with nearest neighbor hopping ${t}_{ij}$ on a square lattice and a chemical potential $\mu$, which controls the filling of the system. 
Throughout the  paper, we choose an isotropic hopping on the lattice, i.e. $t_{ij} = t$ for neighboring sites $i$ and $j$, and ${t}_{ij}=0$ else.
The last term of Eq.~\eqref{eq:KLM} is the antiferromagnetic spin-spin Heisenberg interaction between localized spins ($f$-electrons)
and conduction band electrons ($c$-electrons), where $\mathbf{\sigma}$ represents the vector of Pauli matrices.

The VCA is a well-established cluster method for strongly correlated electron systems \cite{PAD03,DAH+04,SLMT05,AAPH06a,NAAvdL12,MY15}.
It is based on the framework of self-energy functional theory \cite{Pot03a}, where the self-energy functional (SEF) 
\begin{equation}
\Omega(\Sigma) = F(\Sigma) + \text{Tr} \ln\left( G^{-1}_0-\Sigma \right)
\label{Eq:SEF1}
\end{equation}
is used to calculate the grand potential $\Omega$.
Here, $G_0$ denotes the non-interacting Green function and $F(\Sigma)=\Phi^{\mathrm{LW}}(G(\Sigma)) - \text{Tr}(\Sigma G(\Sigma))$ is the Legendre-transformed Luttinger-Ward functional $\Phi^{\mathrm{LW}}$ \cite{LW60,Pot06a}.
$\Omega$ is obtained at the stationary point of $\Omega(\Sigma)$ with respect to all possible self-energies, which means that $\delta\Omega(\Sigma)/\delta\Sigma=0$.
Since the Luttinger-Ward functional is universal in the sense that it only depends on the interaction, it is the same for a reference system with identical interaction terms. 
Using such a reference system, one can rewrite the self-energy functional and obtains
\begin{equation}
\Omega(\Sigma^{\prime}) = \Omega^{\prime}(\Sigma^{\prime}) + \text{Tr}\ln{(G_0^{-1}-\Sigma^{\prime})}-\text{Tr}\log{G^{\prime}},
\label{Eq:SEF2}
\end{equation}
where all quantities of the reference system have been denoted by a prime and $G^{\prime}$ is the interacting Green function of the reference system.
The approximation of VCA consists in choosing a tiling of the original system into identical clusters as a reference system.
For the cluster system the grand potential $\Omega^\prime$, the self energy $\Sigma^\prime$ and the interacting Green function $G^\prime$ can be calculated at zero temperature using exact diagonalization.
This approximation amounts to restricting the variational space of self-energies in Eq.(\ref{Eq:SEF1}) to those, which can be realized on a cluster of finite size. 

In practice, the one-body terms of the cluster are varied and lead to a change in the cluster self-energies.
One determines them such that the SEF is stationary with respect to their corresponding cluster self-energy.
Out of the large set of one-body terms that could possibly be added to the cluster one chooses a subset as variational parameters.
A more detailed derivation and discussion of the technique can be found in References \onlinecite{Pot03,Pot03a,NST08,Pot12}.

Although VCA has been used on a variety of purely electronic models so far, it turned out to be difficult to treat spin interactions directly \cite{FP14}${}^,$\footnote{Instead, \textit{Laubach et al.} showed in Ref. \onlinecite{LJR+16} that it is possible to treat the Hubbard model in the limit $U\rightarrow\infty$ in order to study the Heisenberg model within VCA. However, the enlarged local Hilbert space of the Hubbard model as compared to the Heisenberg model prevented the authors from studying larger clusters.}.
So far, in the context of heavy fermion systems, it  has only been used to study the periodic Anderson model \cite{MY15}, which is related to the KLM in the limit of infinitely large Coulomb repulsion \cite{SN02}. 
In this work, however, we apply it to the KLM, Eq.~\eqref{eq:KLM}, which is an electronic system with pure spin interactions, though local ones.
Here, VCA is applied to the KLM in its standard form, i.e., quantities that enter the calculation of the SEF like the non-interacting Green function of the lattice $G_0$ and the interacting Green function of the cluster $G^{\prime}$ are Green functions only of the electronic part of the KLM.
In this way, the on-site spin interactions are included in the cluster self-energies.
However, propagation of the $f$-spins is not included and would require an extension of the technique, e.g. based on an adapted Luttinger-Ward functional \cite{CPR05}.

In order to fix the density $n$ to a preset value, the grand potential (approximated by the self-energy functional) is Legendre transformed to the free energy \cite{BP10} and $\mu$ is used as a variational parameter.
As further variational parameters we then choose the hopping on the cluster $t^{\prime}_{ij}$
, the chemical potential of the reference 
system $\mu^{\prime}$ to ensure thermodynamic consistency \cite{AAPH06}, and the strengths of potential Weiss fields, further discussed below.
In the main part of the paper, we use an isotropic hopping on the cluster, which is why we skip the site indices and refer to the variational parameter by $t^{\prime}$.
The restriction to one isotropic variational parameter is justified as discussed in Appendix \ref{App:Hopping}.

\subsection{Observables of Interest within VCA}
The focus of VCA is the calculation of one-body expectation values.
The electron density $n$ is obtained by computing 
\begin{equation}
\langle{n}\rangle=-\frac{\partial\Omega}{\partial\mu}\stackrel{\mu^{\prime}=\mu^{\prime}_{\text{opt}}}=\frac{1}{N}\int_{\mathcal{C}<}\frac{dz}{2\pi i}\text{Tr}\left[ \delta_{RR^{\prime}}\delta_{\sigma\sigma^{\prime}}\mbf{G}(\mbf{k},z)\right],
\end{equation} where $\mathbf{G}$ denotes the one-particle Green function and the contour of the integration surrounds the negative real frequency axis counterclockwise. 
The vector $\mathbf{R}$ runs over the cluster sites, $\sigma$ denotes the value of the spin, and $\mu^\prime_{\rm opt}$ is the value of the cluster chemical potential at the saddle point of the SEF.

As usual in VCA, to allow for the possibility of long-range order on the cluster one has to add a fictitious Weiss field to the cluster Hamiltonian only.
In the case of magnetism it takes the form
\begin{equation}
\mathcal{H}_{\text{AF}}=M\sum_{\mathbf{R}} e^{i\mathbf{Q}\cdot\mathbf{R}}(n_{R\uparrow}-n_{R\downarrow}),
\end{equation}
where the wave vector $Q=(\pi,\pi)$ corresponds to N\'eel antiferromagnetism and $\mathbf{R}$ runs over the cluster sites.\footnote{Although it would be interesting to investigate spin density waves with incommensurate ordering wavevectors, within VCA one usually uses commensurate $Q$ vectors. The reason is that the reference system is made up of identical clusters, which treat short-ranged spatial correlations inside the clusters exactly. A way to treat longer-ranged, incommensurate ordering vectors might consist in using supercluster constructions.
} 
The strength $M$ of this field is then used as a variational parameter. 
Finally, the staggered magnetization of c-electrons can be obtained at the stationary point as 
\begin{equation}
m^{c}=\frac{1}{N}\int_{\mathcal{C}<} \frac{dz}{2\pi i}\text{Tr} [e^{i\mathbf{Q}\cdot\mathbf{R}}(-1)^{\sigma}\mathbf{G}_{\mathbf{R}\sigma,\mathbf{R}\sigma}(z)].
\end{equation}
Note that the staggered magnetization of f-spins is only available on the cluster as the Green function of the system does only contain excitations with respect to the conduction electrons, 
\begin{equation}
m^{f}_{cl}=\frac{1}{N}\sum_{\mbf{R}}e^{i\mbf{Q}\cdot\mbf{R}}\langle n^f_{\mbf{R},\uparrow}-n^f_{\mbf{R},\downarrow}\rangle.
\label{eq:struc}
\end{equation}

Superconductivity (SC) is captured in a similar way by adding the Weiss field
\begin{equation}
\mathcal{H}_{\text{SC}} = D\sum_{i,j}\left(\Delta_{ij} c_{i\uparrow}^{}c_{j\downarrow}^{} + \mathrm{H..c.} \right),
\end{equation}
where $D$ is the strength of the Weiss field and $\Delta_{ij}$ the geometric factor which is adapted to the specific superconducting channel studied.
In particular, in this paper we focus on local s-wave SC with
\begin{equation}
\Delta_{ij}=\delta_{ij},
\end{equation}
and on d-wave SC with
\begin{equation}
\Delta_{ij}=\begin{cases} +1: r_i-r_j=\pm e_x \\
  -1: r_i-r_j=\pm e_y, \end{cases}
\end{equation}
in the case of $d_{x^2-y^2}$ SC, where $e_{x/y}$ denote unit vectors along the lattice directions.


\begin{figure}[b!]
	\begin{center}
		\includegraphics[width=0.495\textwidth]{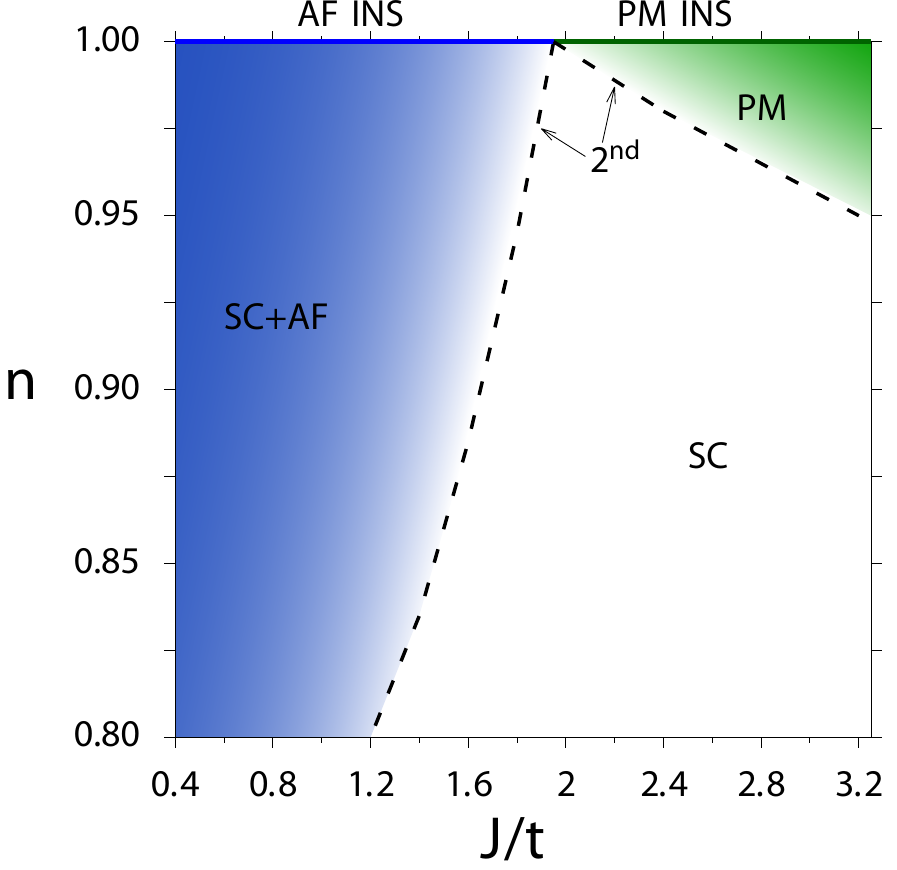}
	\caption{Schematic phase diagram as obtained with VCA using a $3\times2$ cluster. At half-filling the system is an antiferromagnetic insulator for $J<J_c=1.94t$ and a paramagnetic insulator for $J>J_c$. Away from half-filling there are three phases: A paramagnetic metal close to half-filling at strong coupling, a d-wave superconductor and a coexistence region of antiferromagnet and d-wave superconductor. 
	Furthermore, when suppressing SC, within the AF phase at $J/t\sim 1.24$ an additional transition line between AF phases with different Fermi surface topology appears (not shown), as discussed in Sec.~\ref{KLM_AF_OffHF}.}
		\label{fig:PD_sketch}
\end{center}
\end{figure}

\section{Phase Diagram}
\label{sec:phasedia}

The main findings of this work are summarized in the phase diagram sketched in Fig. \ref{fig:PD_sketch}.
At half-filling the system is insulating for all coupling strengths $J/t$.
Two different insulating phases are found, namely an antiferromagnetic (AF) phase at small couplings $J<J_c$ which is induced by the effective RKKY interaction and a paramagnetic phase for larger couplings $J>J_c$ caused by Kondo screening.
The transition between both phases is continuous, which renders the transition point $J_c$ to be a quantum critical point.

As soon as the system is doped away from half-filling, three different phases occur.
Close to half-filling and for strong couplings $J/t$, the system is in a paramagnetic metallic phase with a large Fermi surface.
It extends down to the critical coupling strength $J_c$, but when doping further away from half-filling, d-wave superconducting order builds up.
At weak couplings $J<J_c$, doping the system results in a coexistence of antiferromagnetism and d-wave superconductivity.
Inside this phase, both phenomena show within the VCA cooperative behavior for $J/t\gtrsim1.2$ and competitive behavior for smaller couplings as discussed in Sec.~\ref{SCd+AF}. 
When keeping the ratio $J/t$ fixed, the more the system is doped away from half-filling the more the antiferromagnetic order is reduced and it finally vanishes continuously at a critical electron filling $n_c(J)$.
There, the coexistence phase goes over to a pure d-wave superconductor.

When only considering AF, at the aforementioned value $J/t \sim 1.2$, the Fermi surface changes its topology when keeping the filling $n$ constant and varying $J/t$, as discussed in Sec.~\ref{KLM_AF_OffHF}.
At lower fillings within the AF phase, this is accompanied by a jump in the value of the AF order parameter, indicating for a discontinuous transition, which becomes continuous at fillings $n \gtrsim 0.97$.

In the forthcoming sections we will describe in detail how this phase diagram has been obtained.
Furthermore, putative s-wave SC, pure antiferromagnetic and d-wave SC solutions, as well as their interplay will be discussed.

\section{Paramagnetic phase of the KLM}
\label{KLM_PM}
In this section, we start our investigation of the phase diagram by treating the simplest possible approach in the VCA.
It consists in only considering paramagnetic (PM) solutions, i.e., we neglect possible long-range order at this first stage. 
As VCA does not allow for phases with broken symmetries unless proper Weiss fields are added, it is possible to investigate such PM solutions at all coupling strengths irrespective of the 'true' ground state of the system.
As we will see later in Secs.~\ref{KLM_SC_d} and \ref{SCd+AF}, this phase is the correct physical solution at large coupling close to half-filling.
However, even for other parameter regimes, the paramagnetic solution serves as a starting point and is always considered a reference solution: 
Even if additional solutions with broken symmetries such as AF or SC occur, one needs to check, whether their energy is lower than this normal state solution without broken symmetry.

\subsection{Kondo insulator at half-filling}

\begin{figure}[b!]
	\begin{center}
		\includegraphics[width=0.495\textwidth]{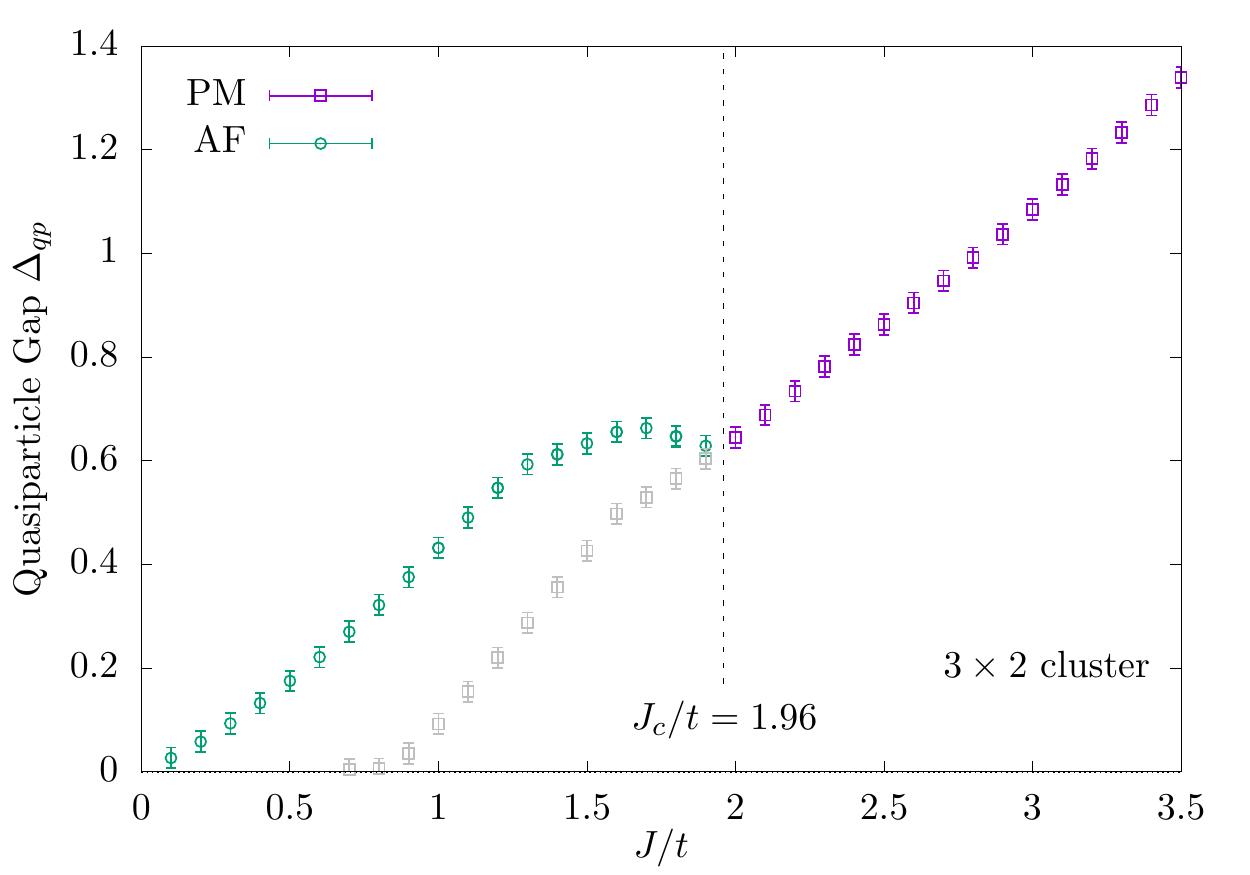}\newline
		\caption{Quasiparticle gap $\Delta_{\mathrm{qp}}$ at half-filling in the paramagnetic ($J>J_c$) and antiferromagnetic $(J<J_c$) insulator obtained by using a $3\times2$ cluster. The solutions for the pure paramagnet (PM) without addition of an AF Weiss-field as well as the solution when including AF are shown.  The quantity plotted amounts to $\mu({n=1.001})$ (for details see text).
		\label{fig_HF_AFPM_GAPS}}
	\end{center}
\end{figure}

The simplest type of insulator which can be encountered here is the (atomic) Kondo insulator at half-filling that consists of local singlets between f-spins and conduction electrons. 
Here, we investigate the paramagnetic solution by doping the system away from half-filling and determine the spectral gap $\Delta\mu=\Delta_{qp}/2$. 
One characteristic of an insulator is a vanishing electronic compressibility $\chi_e = \frac{1}{n^2} \frac{\partial n}{\partial \mu}$.
This corresponds to a plateau at half-filling in a $n$-versus-$\mu$ plot, which we investigate as a function of $J/t$ for the paramagnetic solution. 
Instead of calculating the electron filling as a function of chemical potential it is possible to obtain the quasiparticle gap by doping the system slightly away from half-filling.
Using $\Delta_{\text{qp}}=\lim_{\epsilon\rightarrow0^+} \mu(n=1+\epsilon)$ it is possible to extract the quasiparticle gap more efficiently.
In order to determine the quasiparticle gap one then calculates the stationary points at $n=1+\epsilon$ as a function of $J/t$.
The choice of a finite $\epsilon$ leads to an error in $\Delta_{\text{qp}}$, but for fillings close to half-filling (e.g. for $n=1.001$) the system is metallic and the chemical potential nearly coincides with the quasiparticle gap $\Delta_{\text{qp}}$ at half-filling.
The so-obtained quasiparticle gap is shown in Fig. \ref{fig_HF_AFPM_GAPS}.
As expected from the strong-coupling picture of a Kondo insulator the quasiparticle gap grows linearly in $J/t$ for large coupling.
However, at intermediate couplings $J/t\lesssim 1.8$ deviations from this behavior are found and the gap reduces much faster.
Finally, at $J/t\lesssim 0.8$ the gap vanishes and the system becomes metallic.

Based on the exact, finite-size extrapolated QMC results of Ref.~\onlinecite{Ass99}, one expects to find a finite quasiparticle gap for all positive (antiferromagnetic) values of $J/t$ if considering the correct ground state for every value of $J/t$. 
This is an important aspect, as only for $J>J_c$ we expect the PM solution to be the ground state. 
Below $J_c$ the system orders antiferromagnetically (see next section) and the gapless paramagnet is not the ground state solution at $T=0$. 

 \subsection{Paramagnetic phase away from half-filling}
 \label{PMDoped}
\begin{figure}[t!]
	\begin{center}
		\includegraphics[width=0.495\textwidth]{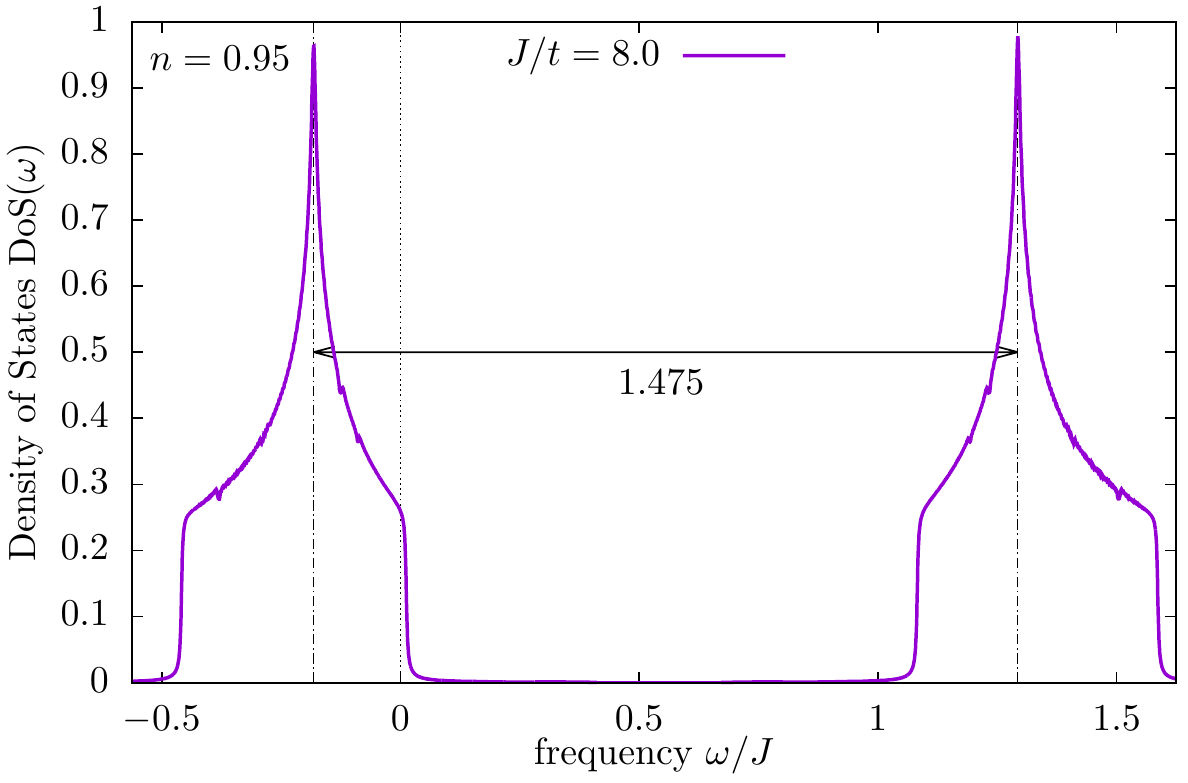}
		\caption{Local density of states [DOS, see Eq.~\eqref{eq:dos}], at large coupling $J/t=8.0$ close to half-filling ($n=0.95$). The two resonances are separated by roughly $3/2J$ and correspond to the break-up of Kondo singlets.}
		\label{fig_FS_PM_offHF}
	\end{center}
\end{figure}	

In the previous section we saw that for strong coupling $J/t$ the ground state of the Kondo lattice model at half-filling is paramagnetic. 
More precisely, it consists of singlets between the local spins and the conduction band electrons.
At large $J/t$ all electrons are bound into such singlets and the ground state is hence insulating.

This picture changes once the system is doped away from half-filling, as Kondo singlets are broken up and electrons can again move through the system.
This induces a metal which naturally displays a Fermi surface (FS).
We will discuss the behavior of the FS in the various regions of the phase diagram in detail in Sec.~\ref{KLM_KB}.
An example for the PM solution discussed here is the right panel of Fig.~\ref{fig_3x2_AF_OffHF_FS}, which shows the Fermi surface at $J/t=3.0$ and an electron filling of $n=0.95$.
Despite the low electron filling the Fermi surface is large and measuring its area gives a value of $n_{\mathrm{FS}}\approx 1.95$.
Such a large value is expected from the Friedel sum rule \cite{LA61,Lan66,Mar82}, as the local f-spins take part in the charge transport in form of the Kondo singlets and therefore contribute with $n_{\mathrm{FS}}^f=1$ to the Fermi surface volume.

The existence of the Kondo singlets away from half-filling is also clearly seen in the local density of states (DOS)
\begin{equation}
\label{eq:dos}
\mathrm{DOS}(\omega) = -\frac{1}{\pi} \lim_{\eta\rightarrow 0^+}\mathrm{Im}~G(\omega+i\eta) ,
\end{equation}
where $G(\omega)=\sum_k G(k, \omega)$, and which is 
shown in Fig.~\ref{fig_FS_PM_offHF} for large coupling $J/t=8.0$ close to half-filling.
It shows two peaks, which are separated by roughly $3J/2$, which is roughly twice the energy of a singlet, $|E_{\rm singlet}| = 3J/4$.  
The peaks hence can be related to the break-up of Kondo singlets, which we expect from the Kondo insulator at strong couplings and half-filling \cite{Col15a}. 
Instead of two isolated sharp peaks, due to the finite contribution by the hopping term, the structures are broadened.
Since the system is doped away from the symmetry point at half-filling, the two signals are not perfectly symmetric with respect to the center of the gap between them.

\section{Magnetic Phase diagram}
\label{KLM_AF}

The competition between Kondo singlet formation and RKKY interaction sets the stage for Doniach's phase diagram \cite{Don77}. 
One possibility to study this competition is the investigation of antiferromagnetic correlations, as RKKY interaction mediates an antiferromagnetic ordering of the f-spins, which induces such an ordering for the conduction electrons, too. 
At half-filling, these antiferromagnetic correlations are expected to be suppressed from a certain $J_c/t$ on as at this point Kondo singlet formation is energetically favorable and absorbs free electrons which could mediate the antiferromagnetic ordering. 

\subsection{The antiferromagnet at half-filling}
\label{KLM_AF_HF}

We start our investigation of the AF solution of the KLM at half-filling.
In this case, it is possible to use clusters with up to eight physical sites ($4\times2$) when considering only the strength of the AF Weiss field and of the isotropic hopping on the cluster as minimal set of variational parameters.
This allows to study a finite-size extrapolation of the critical coupling strength $J_c$ and to compare the extrapolated results at infinite cluster size to the exact QMC results of Ref.~\onlinecite{Ass99}.

In the previous section the paramagnet has been analyzed at half-filling and showed discrepancies from exact results at small and intermediate coupling strengths.
This leads to deviating quasiparticle gaps at intermediate couplings and even to metallic behavior at weak coupling in case of the $3\times2$ cluster.
However, such a scenario is an artifact caused by suppressing AF order in the VCA:  
When including and optimizing the SEF for an additional AF Weiss field, at small couplings clearly an AF phase with lower energy than the PM is stabilized.
The resulting AF ordering and the perfect nesting of the corresponding Fermi surface leads to an AF gap, even at very small $J$, see Fig.~\ref{fig_HF_AFPM_GAPS}.
Hence, an insulating phase is stabilized, which, however, is conceptually different from the paramagnetic insulator, which is caused by spin singlet formation.

This happens at the critical coupling strength $J_c/t$, at which the magnetization vanishes and the SEF shows only one stationary point, namely the 'trivial' one corresponding to the previously discussed paramagnetic insulator.
Therefore, VCA at half-filling correctly shows an insulator for all finite values of $J$.
However, in contrast to the QMC results of Ref.~\onlinecite{Ass99}, the quasiparticle gap shows a peak when decreasing the coupling strength below $J_c/t$ instead of a monotonous decrease for all coupling strengths $J<J_c$.
This feature has also been reported for other approximative techniques such as dynamical cluster approximation\cite{MBA10}. 

\begin{figure}[t!]
	\begin{center}
		\includegraphics[width=0.45\textwidth]{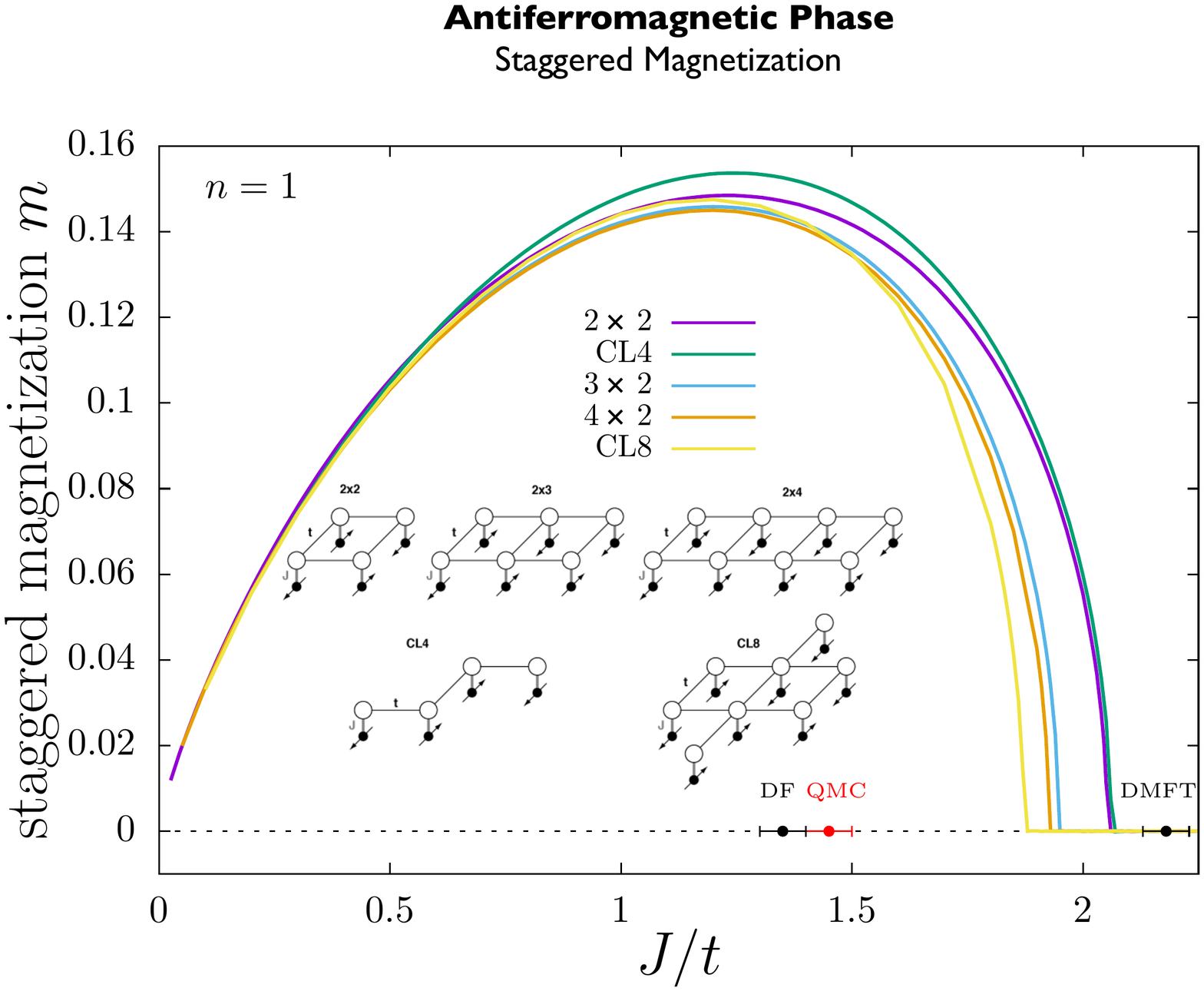}
		\includegraphics[width=0.45\textwidth]{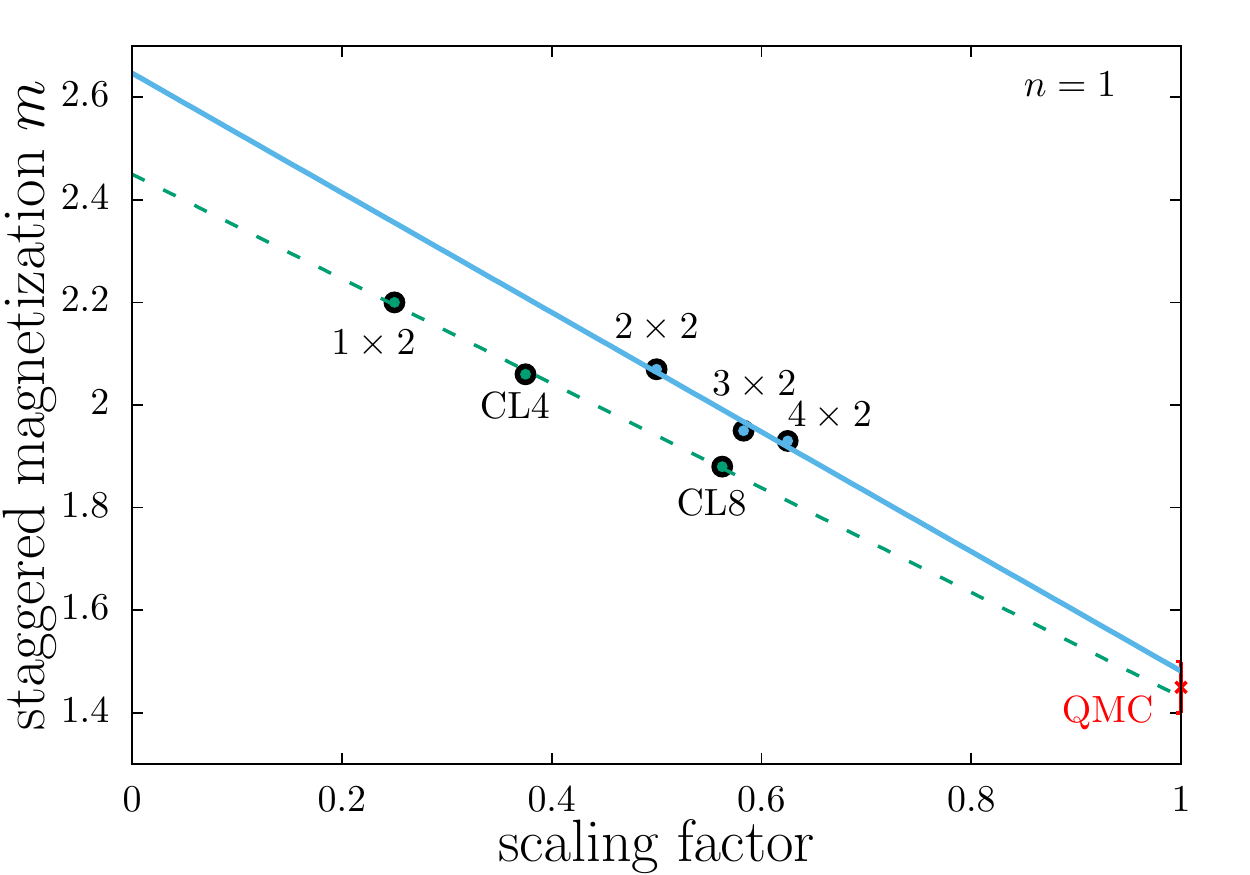}
		\caption{Top: staggered magnetization $m$ for the different cluster sizes and geometries shown. 
		As a reference, values of $J_c/t$ obtained with dual fermions \cite{Ots15}, quantum Monte Carlo \cite{Ass99}, and DMFT+NRG \cite{Bod13} are shown. Bottom: finite-size extrapolations of the value of the critical coupling $J_c/t$ to the infinite-size cluster limit. The blue (solid) line takes into account only the regularly shaped clusters, as indicated, while the green (dashed) line considers the remaining cluster types. The scaling factor equals the number of intra-cluster links divided by twice the number of cluster sites.} 
		\label{fig_AF_m_clusters}
	\end{center}
\end{figure}

We now turn to the AF order parameter.
Figure \ref{fig_AF_m_clusters} shows in the top panel the staggered magnetization $m$ of the conduction band electrons.
Note, however, that the staggered magnetization of the localized spins $m_f$ can be calculated on the cluster only as the Green function contains only information on the conduction electrons.
Hence, in contrast to $m$, $m_f$ is not calculated as the expectation value using the system's Green function. 
For this reason, we turn to the staggered magnetization of the conduction band electrons $m$ instead.
As the choice of sign is somewhat arbitrary (the self-energy functional is symmetric with respect to the Weiss field strength $M_c$), in subsequent plots we will only show positive values of $m$.
The strength of $m$ increases for small couplings with the coupling strength and reaches a maximum at a value $J_{\mathrm{max}/t}$ which depends on the cluster size.
For larger values of $J/t$ it decreases rapidly and vanishes smoothly at the critical coupling $J_c$, where the system stays insulating, but without magnetic order.

Compared to the DMFT results of Ref.~\onlinecite{Bod13}, where a NRG solver was used to precisely capture the low energies inherent to Kondo (lattice) systems, the magnetization curve has the same characteristics, but the absolute values differ.
In particular, the value of the critical coupling strength $J_c/t$ is lowered. 
This can be understood by recalling that the electronic fluctuations inside the $4\times2$ cluster are a natural antagonist to AF order.
By increasing the size of the cluster, the spatial extent of the fluctuations which enter the reference self-energy grows and thereby changes the variational space for the determination of stationary points.
Hence, it is useful to systematically increase the cluster size and in this way to determine the value of $J_c/t$ after a finite-size extrapolation.

Such a finite-size scaling would be best controlled when using clusters of different sizes but the same geometry, e.g., quadratic clusters.
Unfortunately, the accessible clusters are limited to small total sizes and one needs to also resort to other cluster shapes, e.g., ladder geometries.
In Ref.~\onlinecite{S08}, S\'en\'echal \textit{et al.} compared different scaling factors and tested the outcome of a finite-size extrapolation in case of the Hubbard model on a two-dimensional square lattice, where larger clusters could be used due to the smaller local basis.
The most promising scaling factor of this comparison, which is also applied in the lower panel of Fig.~\ref{fig_AF_m_clusters}, consists in taking the number of links on the cluster and dividing it by twice the number of cluster sites.
It scales to one in the limit of infinite cluster size and takes the ratio of boundary and bulk into account.
\newline
When taking into account all available cluster sizes for the KLM, also quite pathological ones such as $1\times2$ and $CL4$ (see Fig.~\ref{fig_AF_m_clusters}), which include dangling sites, it is not a surprise that the critical value $J_c/t$ spreads a lot depending on the cluster geometry.
However, an extrapolation by considering only the "ladder" clusters $2\times2$, $3\times2$ and $4\times2$ (blue fit in the lower panel of Fig.~\ref{fig_AF_m_clusters}) results in an infinite-cluster-size value of $J_c/t=1.48\pm0.28$, which is very close to the QMC results of Ref.~\onlinecite{Ass99} which give $J_c/t = 1.45 \pm 0.05$ in the thermodynamic limit.
The VCA results are therefore in agreement with this numerically exact approach within our estimated error bars. 
However, larger clusters are needed to confirm the result of the ladder cluster extrapolation and to improve on this rough fit. 

Nevertheless, it is remarkable that the critical value obtained by this rough fit is much better than the one obtained using DMFT+NRG in Ref.~\onlinecite{BZV+13}.
In a recent study, Ref.~\onlinecite{Ots15}, Otsuki obtained similar values for $J_c/t$ by applying the dual fermion approach to the KLM, further indicating that it is important for the physics of the KLM to allow for spatial fluctuations beyond the single impurity ansatz used in DMFT.

\begin{figure}[b!]
		\includegraphics[width=0.45\textwidth]{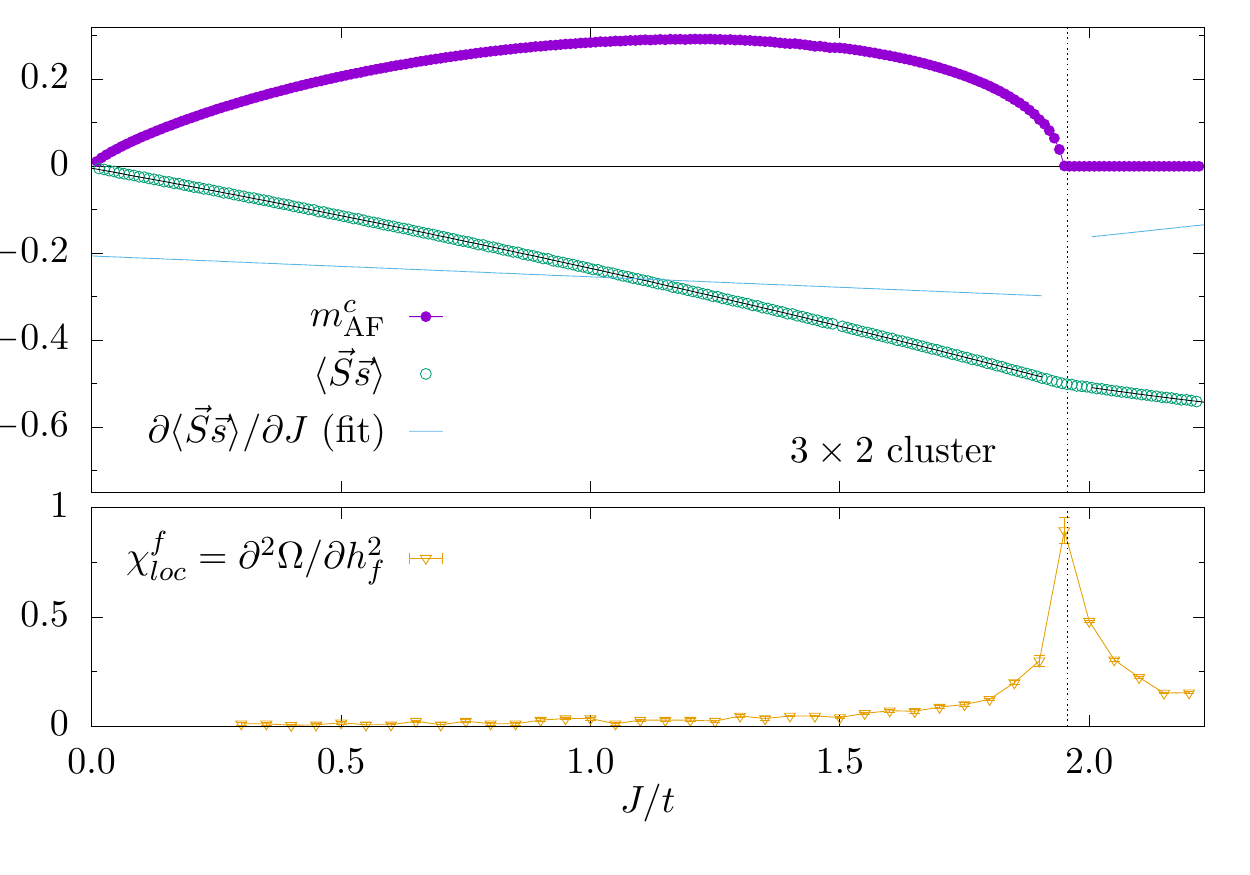}
		\caption{Staggered magnetization, expectation value of the Spin-spin correlator $\langle\vec{S}\cdot\vec{s}\rangle=\partial \Omega/\partial J$ and local $f$-spin susceptibility $\chi^f_{\text{loc}}=\partial^2\Omega/\partial h_f^2\left\vert_{h_f\rightarrow0}\right.$ as a function of $J/t$ at half-filling using a $3\times2$ cluster. Black lines are fits of the spin-spin correlator, blue lines denote their derivatives. They have been fitted with second order polynomials in the AF and PM regions up to $J_c/t\pm0.05$. The set of variational parameters includes $\{M_c,t^{\prime}\}$.}
		\label{fig_AF_GS_Sisi}
\end{figure}

\subsubsection{Further properties of the ground state at half-filling}

After having identified the AF properties of the KLM at half-filling, we briefly present additional aspects of the ground state which will become relevant in the following sections.
A detailed discussion can be found in Appendix \ref{App:Details_PM_HF}.

Two quantities that are sensible to Kondo singlet formation between $f$-spins and $c$-electrons are the local spin-spin correlator $\langle \vec{S_i}\cdot\vec{s_i}\rangle = \frac{\partial \Omega}{\partial J}$ and the local magnetic susceptibility of the $f$-spins $\chi^f_{\text{loc}}$, see Eq.~\eqref{Eq:Chiloc}.

In Fig.~\ref{fig_AF_GS_Sisi} we show our results for both quantities obtained for a $3\times2$ cluster.
Qualitatively, the results stay the same for the $4\times2$ cluster so that we will base our discussion on the results for the smaller clusters.
Also note that, due to the higher computational cost, for the $4\times2$ clusters only $M_c$ could be used as variational parameter.
In Appendix \ref{App:Hopping}, the influence of also including the cluster hopping $t^{\prime}$ into the set of variational parameters for the $3\times2$ clusters is discussed.

In a recent rDMFT study \cite{PK15} the spin-spin correlator $\langle S_+\cdot s_-\rangle$ was used to investigate the magnetic phase transitions in the doped KLM.
It showed small discontinuities at first order transitions and had no features in case of second order transitions.
As far as the transition from a paramagnetic to an antiferromagnetic insulator is considered, the correlator $\langle \vec{S_i}\cdot\vec{s_i}\rangle$ seems to be a good indicator for the transition point $J_c$, too:
At the value of $J_c/t$ at which the staggered magnetization becomes finite, the gradient of $\langle \vec{S_i}\cdot\vec{s_i}\rangle$ jumps and its curvature changes sign.

In the strong coupling limit, the spin-spin correlator converges to a value of $-3/4$ per site which is expected for pure Kondo singlets.
When approaching zero coupling, the correlator goes to zero as well, which means that electrons and f-spins are uncorrelated.

To investigate Kondo singlet formation between $f$-spins and conduction electrons one often studies the local magnetic susceptibility $\chi_{\text{loc}}$ \cite{HK13}. 
Compared to the spin-spin correlator, which is finite for all non-zero $J/t$, the local susceptibility allows for a clearer identification of a putative Kondo breakdown via a divergence. 
In Fig.~\ref{fig_AF_GS_Sisi} we show results for the local magnetic susceptibility of the $f$-spins, which is calculated via the self-energy functional at the stationary point.\footnote{
Since we do not have access to the $f$-spins via the Green function, this is the only way to access this susceptibility.
To calculate the total local magnetic susceptibility (including the contribution of the $c$ electrons) one needs to use spin-dependent variational parameters \cite{BP10}, which strongly increases the computational cost.
As both the $f$-spins and the $c$-electrons are part of the Kondo singlets at strong coupling, it is sufficient to investigate the local susceptibility of one of its constituents.
}
To calculate the local $f$-spin susceptibility, we added a small magnetic field term to the Hamiltonian acting locally on one of the $f$-spins: $h_fS_i^z$.
For small field strengths $\vert h_f\vert\leq0.03$, we extracted\footnote{In practice, we determined $\chi$ from a polynomial fit of $\Omega(h_f)$ in the AF and from $\Omega(h_f)\sim \ln(\cosh(h_f))$ in the PM. The latter corresponds to the expected behavior of the magnetization in case of a paramagnet, $m\sim \cosh(h_f)$.} the local susceptibility 
\begin{equation}
\chi^f_{\text{loc}}=\left.\frac{\partial^2\Omega(h_f)}{\partial h_f^2}\right\vert_{h_f\rightarrow0}.
\label{Eq:Chiloc}
\end{equation}
Just as for $\langle\vec{S}\cdot\vec{s}\rangle$, the local susceptibility allows for a clear identification for the onset of AF at $J_c$, but does not show any clear indications for changes within the AF phase.
This indicates that there is no Kondo breakdown within the AF phase.
The question remains, as to whether the divergence seen in Fig.~\ref{fig_AF_GS_Sisi} could be indicative for Kondo breakdown. 
However, one generically expects such a divergence at a continuous transition from a PM to AF phase, so that it remains difficult to address this point using the local susceptibility.

\begin{figure}[b!]
	\begin{center}
		\includegraphics[width=0.45\textwidth]{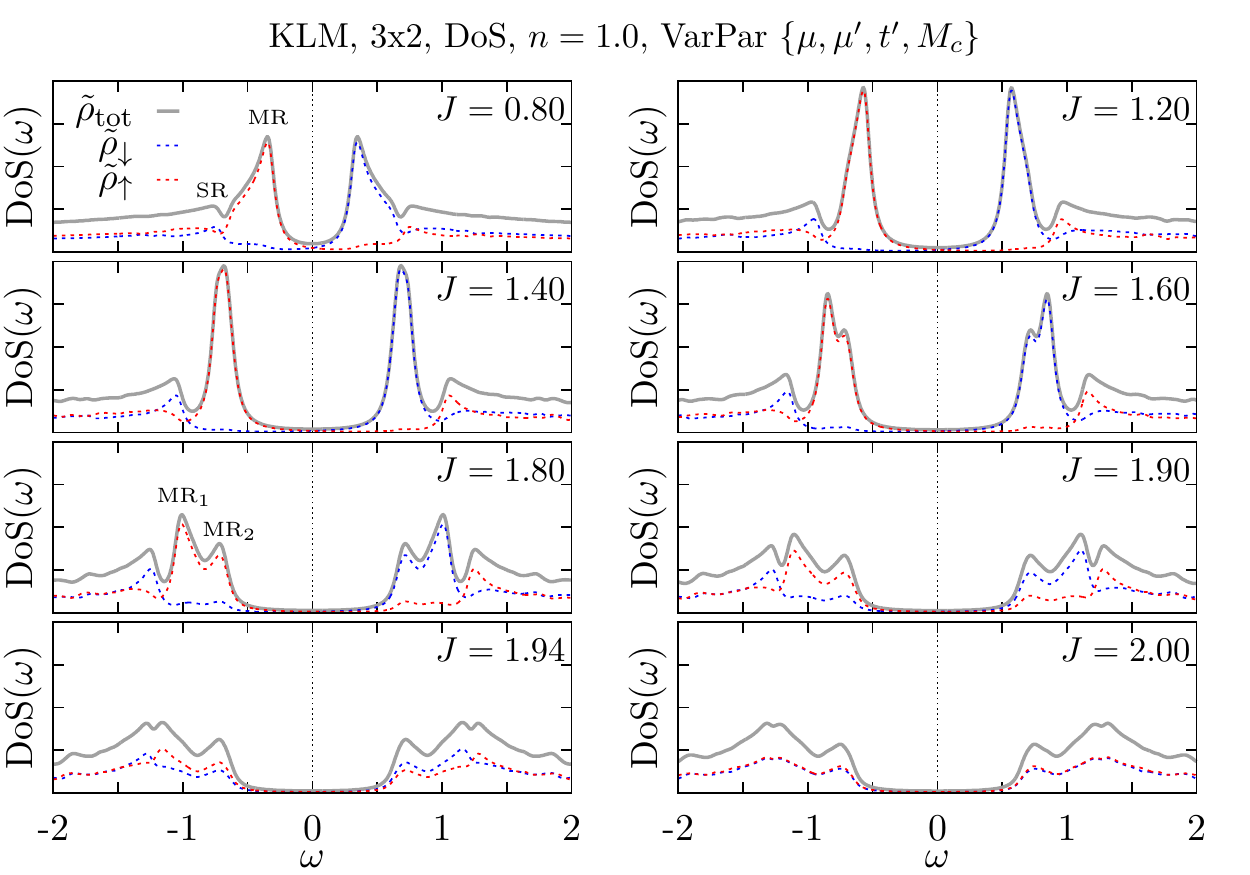}				
		\caption{Density of states (DOS) at half-filling for several $J/t$ in the antiferromagnetic phase. The rest of the parameters are the same as in Fig.~\ref{fig_3x2_AF_HF_DoS_J1} and the scale of $\mathrm{DOS}(\omega)$ is kept the same for all panels.}
		\label{fig_3x2_AF_HF_DoS}
	\end{center}
\end{figure}

Another quantity which shows the difference between the AF ground state and the PM alternative solution at half-filling for $J<J_c$ is the spin-dependent local DOS as shown in Fig.~\ref{fig_3x2_AF_HF_DoS}.
It is obtained by a staggered average over the sites inside the cluster,
\begin{equation}
\tilde{\rho}_{\sigma}(\omega)=\frac{1}{N} \sum_{\mathbf{R}} e^{i\mathbf{Q}\cdot\mathbf{R}}\rho_{\mathbf{R}\sigma}(\omega),
\end{equation}
where $\rho_{\mathbf{R}\sigma}(\omega)$ denotes the local density of states on cluster site $R$.

At small interaction strengths two pronounced main resonances (MR) at the band edges are visible, which are separated by the quasiparticle gap, and right next to them small side resonances (SR) can be found (see, e.g., as shown for the value $J/t=0.8$ in Fig.~\ref{fig_3x2_AF_HF_DoS}).
If one limits the discussion to the filled part of the DOS for one of the sublattices, the main resonance is mainly made up of the majority-spin electrons and the side resonance predominantly of minority-spin electrons, although with a considerable admixture of majority-spin electrons. 
Increasing $J$ affects the main and side resonance differently: the weight of the main resonance first increases but then decreases again with a maximum at $J/t\approx 1.2$, which coincides with the point of maximal staggered magnetization.
For $J/t>1.2$, the main peak starts to split into two peaks $\mathrm{MR}_{1/2}$, which have initially similar weight.
When approaching the critical interaction strength $J_c$, the weight of the peaks diminishes and redistributes between $\mathrm{MR}_1$ and $\mathrm{MR}_2$ such that the peak $\mathrm{MR}_1$ closer to the side resonances retains more weight.
The side resonance moves to larger frequencies when $J/t$ is increased and gains a bit of weight.

At the transition, the side resonance of the minority electrons and $\mathrm{MR}_1$ of the majority electrons merge to one new side resonance and the resonance $\mathrm{MR}_2$ is made up equally from both spin-up and -down electrons.

This reshuffling of weight when approaching the phase transition reflects the competition between the two different mechanisms that are responsible for the formation of a quasiparticle gap.
In the antiferromagnetic region close to the transition there are already strong precursors of the paramagnetic insulator visible in form of a notable contribution of minority spin electrons to the peak at the gap edges.
In contrast, for $J/t \lesssim 1.2$ this contribution is not clearly visible, and the gap is stabilized by AF fluctuations.

The energy of the side resonance roughly agrees with the value of $3/4J$.
When identifying this resonance with a Kondo singlet peak, it is thereby possible to trace the existence of Kondo singlets even back to the weak interaction regime deep in the antiferromagnetic phase.
In other words, the antiferromagnetic order is far from perfect and the ground state in the antiferromagnetic region can be interpreted as still having a finite amount of Kondo singlets.
This is in agreement with the behavior of the local susceptibility, which in Fig.~\ref{fig_AF_GS_Sisi} does not indicate any change within the AF phase, and will be discussed in more detail in Sec.~\ref{KLM_KB}.

\subsection{Doping the antiferromagnet}
\label{KLM_AF_OffHF}

\begin{figure}[t!]
	\begin{center}
		\includegraphics[width=0.45\textwidth]{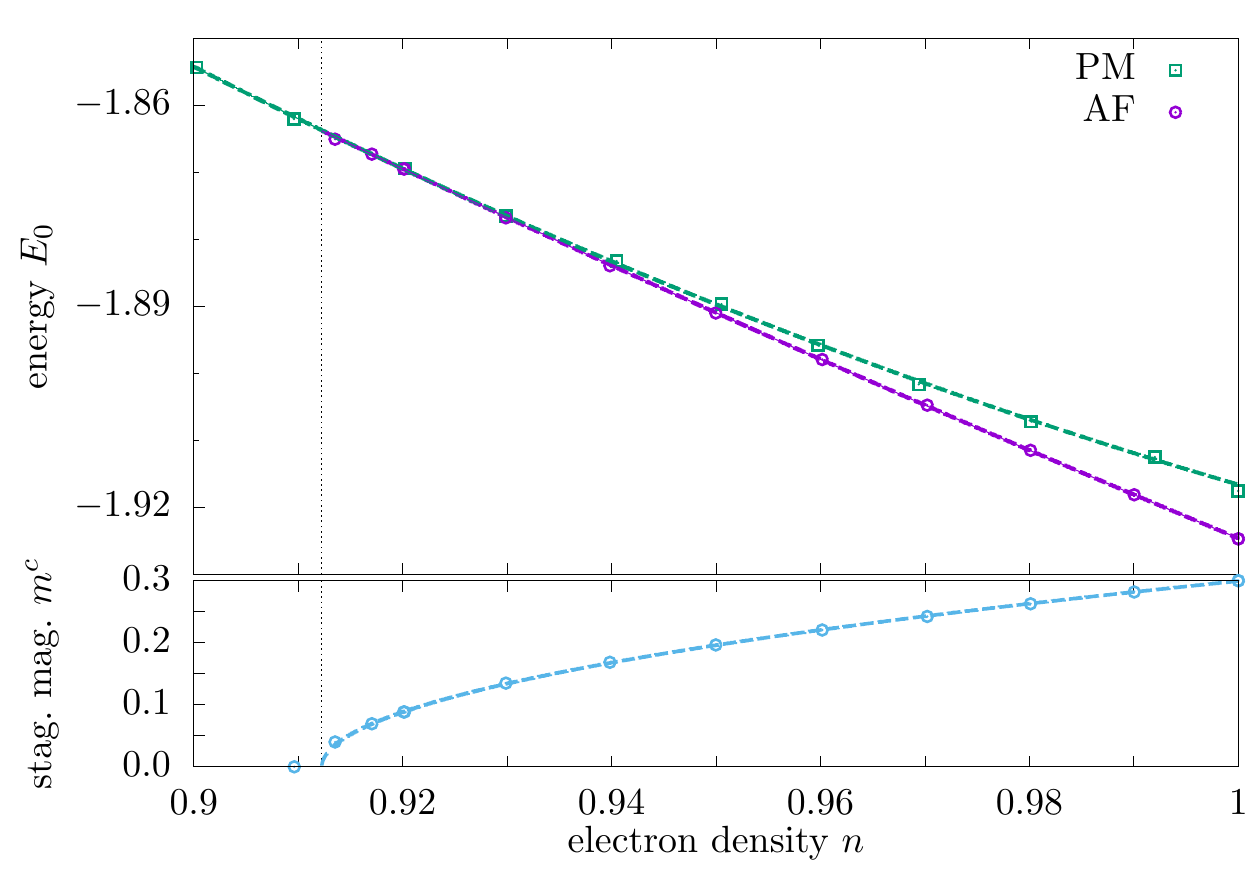}
		\caption{Comparison of the PM and the AF solution at $J=1.6t < J_c^{\mathrm{AF}}$ as a function of electron filling $n$. 
		For $n> n_c\approx0.91$ (vertical dashed line in the plot) the AF solution is energetically preferred (see upper panel).
		The lower panel shows the value of the staggered magnetization of the antiferromagnetic solution. 
		For $n<n_c$ the staggered magnetization $m^c$ is zero and the solution is therefore paramagnetic. 
		The dashed lines are fits to the data points and only a guide to the eye.}
		\label{fig_AFvsPM_Off_HF}
	\end{center}
\end{figure}

The situation in the half-filled Kondo lattice model is quite special: in the ground state, every local moment can be screened in the large $J$ limit with exactly one electron to form a singlet at each site. 
Removing conduction electrons from the half-filled system creates unpaired local moments which can be interpreted as spinfull c-holes. In the paramagnetic case, Kondo singlets can still be formed for sufficiently large Kondo coupling, although the electrons gain mobility due to vacancies in the conduction band (see Sec.~\ref{PMDoped}).
For smaller values of $J$ doping the system is also interesting as the number of electrons which mediate the antiferromagnetic order of the local spins via RKKY interaction is then reduced. 
Naively, one would hence expect, that the antiferromagnetic correlations diminish when the system gets doped. 

\begin{figure}[t!]
	\begin{center}
		\includegraphics[width=0.495\textwidth]{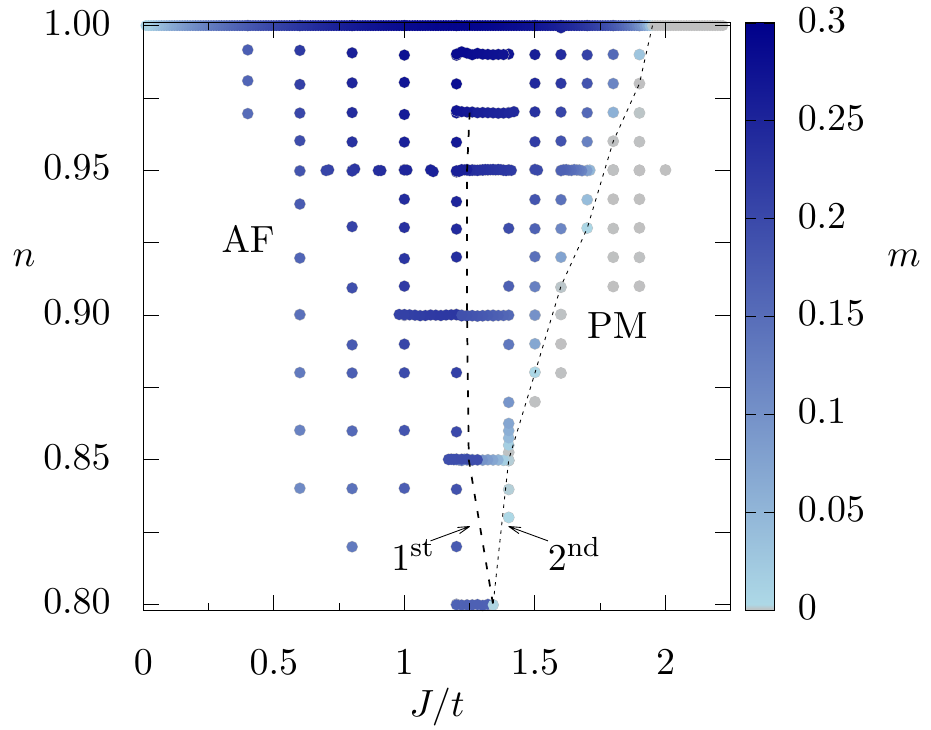}
		\caption{Staggered magnetization of the AF solution as a function of electron filling $n$ and $J/t$ using a $3\times2$ cluster with variational parameters $\mu,\mu^{\prime},t^{\prime}$ and $M_c$. Circles denote data points, their filling is color-coded and shows the value of the staggered magnetization $m$. The dashed lines denote transitions. The one on the left label by ``1st'' indicates a discontinuity of the value of $m$ when keeping $n$ fixed and varying $J/t$. At fillings $n\gtrsim0.97$ this becomes continuous (not shown). The dashed line to the right labeled by ``2nd'' indicates the continuous phase transition between AF and PM metal.}
		\label{fig_3x2_AF_OffHF_PD}
	\end{center}
\end{figure}

In Fig.~\ref{fig_AFvsPM_Off_HF}  the energy of the AF and PM solutions as a function of electron filling are compared.
It can be seen that the AF solution exists down to a critical electron filling of $n_c\approx0.915$.
When approaching this filling from above, the staggered magnetization of the AF solution vanishes continuously and the corresponding energy approaches the one of the PM solution.
As its energy is always lower than the one of the PM, the AF metallic phase is realized close to half-filling and goes over to a PM phase continuously when approaching $n_c$.
\newline
In the DMFT study of Ref.~\onlinecite{OKK09}, Otsuki \textit{et al.} identified an AF ground state at $J=0.2W$ (with $W$ the bandwidth, which in the 2D case treated here is $W=8t$) down to fillings $n_c(J=0.2W) \approx 0.9$. 
In our VCA approach, we obtain a similar result at $J/t \approx 1.6$, as shown in Fig.~\ref{fig_3x2_AF_OffHF_PD}, which shows the magnetic phase diagram close to half-filling, which we obtained using a $3\times2$ cluster.

In this figure, the gray regions denote PM solutions, while the AF region is shown in blue, where the staggered magnetization is color coded.

Aside from the magnetic properties of the system it is also interesting to investigate the metal-insulator transition as a function of electron density $n$.
In the half-filled case it was shown in the previous subsection that the AF insulator goes over to a Kondo insulator at $J_c$.
When doping the PM (Kondo) insulator it was also shown that the system turns metallic once the system is doped away from half-filling.
\newline
Here, the ``new'' transitions are the one from AF insulator to AF metal for $J<J_c$ at $n=1\rightarrow 1-\epsilon$ and from AF to PM metal for $J<J_c$ at $n_{c,\mathrm{AF}}$.
Note that so far we have been neglecting superconductivity.
Therefore, one needs to test for stability against SC order, which is further discussed in Sec.~\ref{SCd+AF}.

\begin{figure}[b!]
	\begin{center}
		\includegraphics[width=0.45\textwidth]{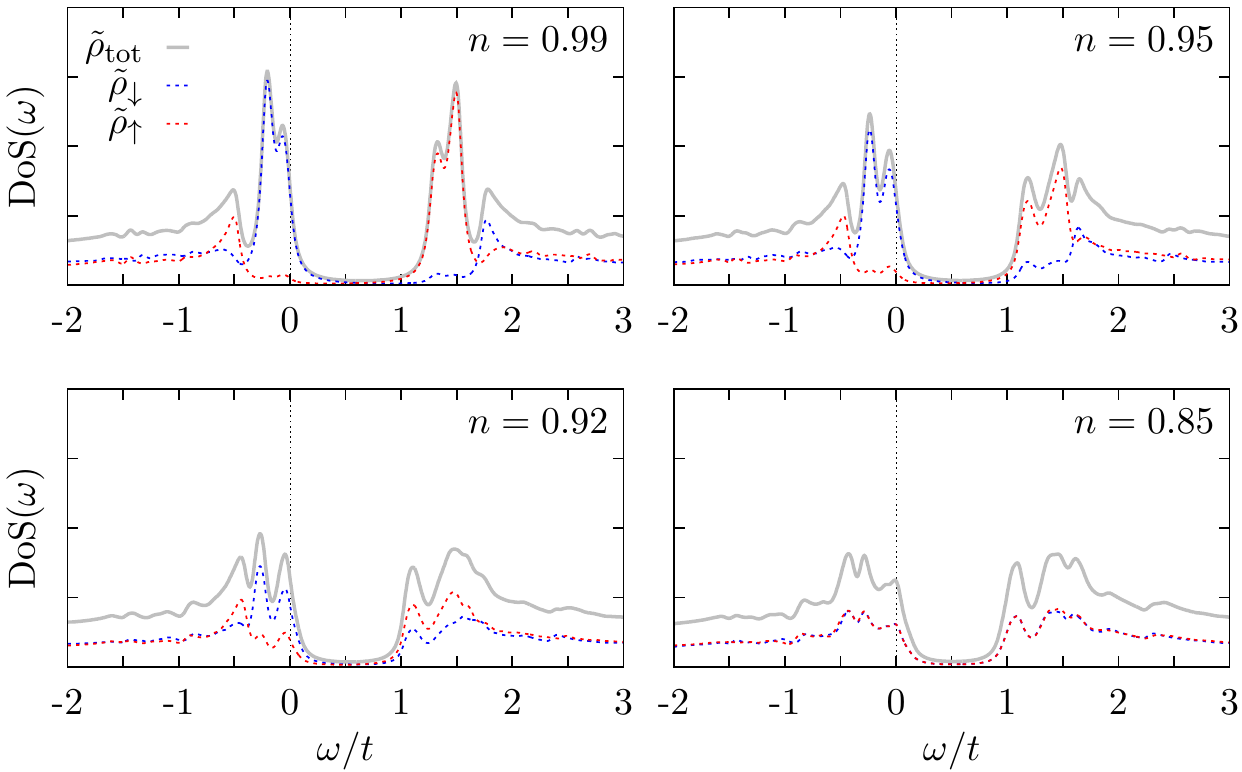}
		\caption{Density of states (DOS) for $J/t=1.6$ using the $3\times2$ cluster with $\{\mu,\mu^{\prime},t^{\prime},M_c\}$ as variational parameters. The electron fillings $0.91< n <1$ correspond to the AF metal phase. The panels show fillings close to the AF insulator ($n=0.99$), close to the PM metal ($n=0.92$) and right in the middle of the phase. The results obtained at $n=0.85$ are deep in the PM metal phase. The DOS has been obtained on a $k$-grid of $100\times100$ points and an artificial broadening of $\eta=0.05$ has been added.}
		\label{fig_3x2_AF_OffHF_DoS}
	\end{center}
\end{figure}

In order to make the difference between AF and PM metals more visible, Fig.~\ref{fig_3x2_AF_OffHF_DoS} shows the DOS at $J/t=1.6$ for different electron densities $n$. 
In contrast to Fig.~\ref{fig_3x2_AF_HF_DoS}, where the transition from AF to PM was studied at half-filling as a function of coupling strengths, the spectrum is not particle-hole symmetric.
Still, the redistribution of electron density between up- and down-electrons when approaching the magnetic order-disorder transition is similar.
The relative position of the three peaks does not change much when reducing the electron filling, but their size changes from the pronounced two main resonance structure known from the half-filled case to a roughly equally sized three-peak structure close to the transition.
Already at $n=0.99$ the main resonances are not entirely made up from the majority spins, but also carry little weight from the minority spins.
The difference between the position of the third peak and the middle of the gap at positive frequencies is at $n=0.99$ roughly $1.15$ which is close to the value of $3/4J=1.2$.
When doping further away from half-filling the relative peak position changes and differs more and more from $3/4J$.
This is somewhat expected as the value of $3/4J$ was obtained by assuming a perfect Kondo insulator.
Once electrons are removed even at strong coupling the singlets are mobile and the resulting dispersion changes.
Finally, in the paramagnetic region electrons of both spins contribute at each energy equally to the density of states.
\newline
This redistribution has consequences for the composition of the Fermi surface, which is discussed in the following subsection.

\subsubsection{Changing Fermi surface topology}
\label{KLM_KB}
Once the system is doped away from half-filling it is metallic and hence possesses a Fermi surface.
Deep in the paramagnetic phase (strong coupling $J$)
mobile Kondo singlets form, which leads to a large Fermi surface (see Sec.~\ref{PMDoped}).
It is interesting to consider the changes of the FS at the onset of AF order and at the transition within the AF phase. 
Since the FS is closely related to the spectral function $A(\mathbf{k},\omega)=-\frac{1}{\pi}\lim_{\eta\rightarrow0^+}\mathrm{Im}~G(\mathbf{k},\omega+i\eta)$, we will mainly discuss this quantity here, as it is directly accessible by the VCA.
In Appendix~\ref{App:LSR} we will discuss the results for the FS obtained from further analyzing $A(\mathbf{k},\omega)$ and see that it reflects the same behavior.
\newline

\begin{figure}[t!]
	\begin{center}
		\includegraphics[width=0.45\textwidth]{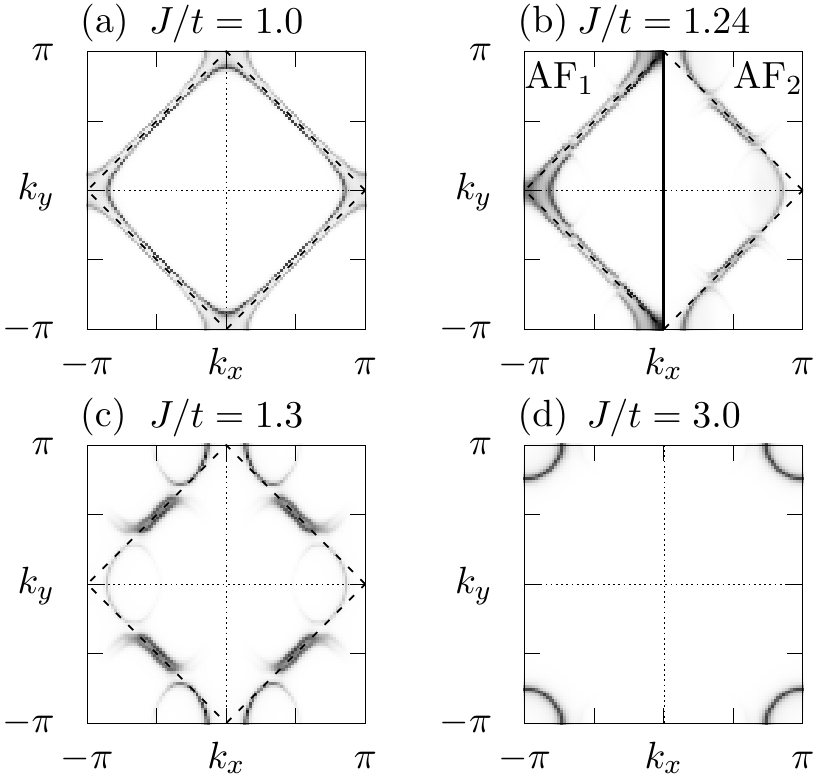}
		\caption{Characteristic spectral function $A(\mathbf{k},i\eta)$ for the three regimes that can be identified in the AF solution off half-filling, here shown at filling $n=0.90$ with broadening $\eta=0.01$: PM metal with a large Fermi surface (d) and AF metal with small Fermi surface (a). At intermediate coupling the Fermi surface volume jumps to a much higher value (b), but the system stays AF (c).}
		\label{fig_3x2_AF_OffHF_FS}
	\end{center}
\end{figure}

Figure \ref{fig_3x2_AF_OffHF_FS} shows characteristic results for $A(\mathbf{k}, \omega = 0+i \eta)$ (with broadening $\eta = 0.01$) close to half-filling ($n=0.90$) at the Fermi energy for three different regimes.
Aside from the already described PM case, one can identify two distinct phases by looking at the spectral function of the resulting metallic solutions.
In the AF phase the Brillouin zone halves, which is indicated in the figure by a dashed line.
At small coupling strengths, 
$A(\mathbf{k},i \eta)$ shows a closed structure,
 and when focusing on the inner sheet 
 , it resembles the small closed Fermi surface that is found in the weak coupling region by 
  other numerical techniques such as dual fermions \cite{Ots15} or rDMFT \cite{PK15}.
The structures in $A(\mathbf{k},i \eta)$ 
 at strong and weak coupling are essentially those found before in the emission spectrum by VMC \cite{ABF13} and in the FS by DCA \cite{MBA10} studies.
For larger coupling strengths close to the transition to the PM, the surface topology is different [Figs.~\ref{fig_3x2_AF_OffHF_FS}(b) and \ref{fig_3x2_AF_OffHF_FS}(c)].

Small closed structures appear, but the doubling of the surface that corresponds to AF long-range order still persists.
Figure~\ref{fig_3x2_AF_OffHF_FS}(b) shows the drastic change in topology at $J_{c,\mathrm{FS}}/t=1.24$ from one open (AF${}_2$) to two closed sheets in AF${}_1$.
We interpret the fact that the structures in AF${}_2$ are not symmetric in the $x$- and $y$-direction as a finite-size effect of the asymmetric $3\times2$ cluster used in Fig.~\ref{fig_3x2_AF_OffHF_FS}. 
Since the correlation length diverges at the critical point it is not surprising that it exceeds the cluster size in its vicinity, which then causes finite-size effects. 
However, the AF${}_2$ phase is stable within all three cluster sizes, which strongly suggests that it persists for larger clusters.
An intermediate AF phase was also observed within VMC and DCA \cite{ABF13,MBA10}, although with different topology.
Still, in those studies the FS also contained closed structures in form of hole pockets.
Since the topology in the AF${}_2$ phase still changes for different cluster sizes, a more systematic study of this phase using larger cluster sizes would be useful.

\begin{figure}[t!]
		\includegraphics[width=0.495\textwidth]{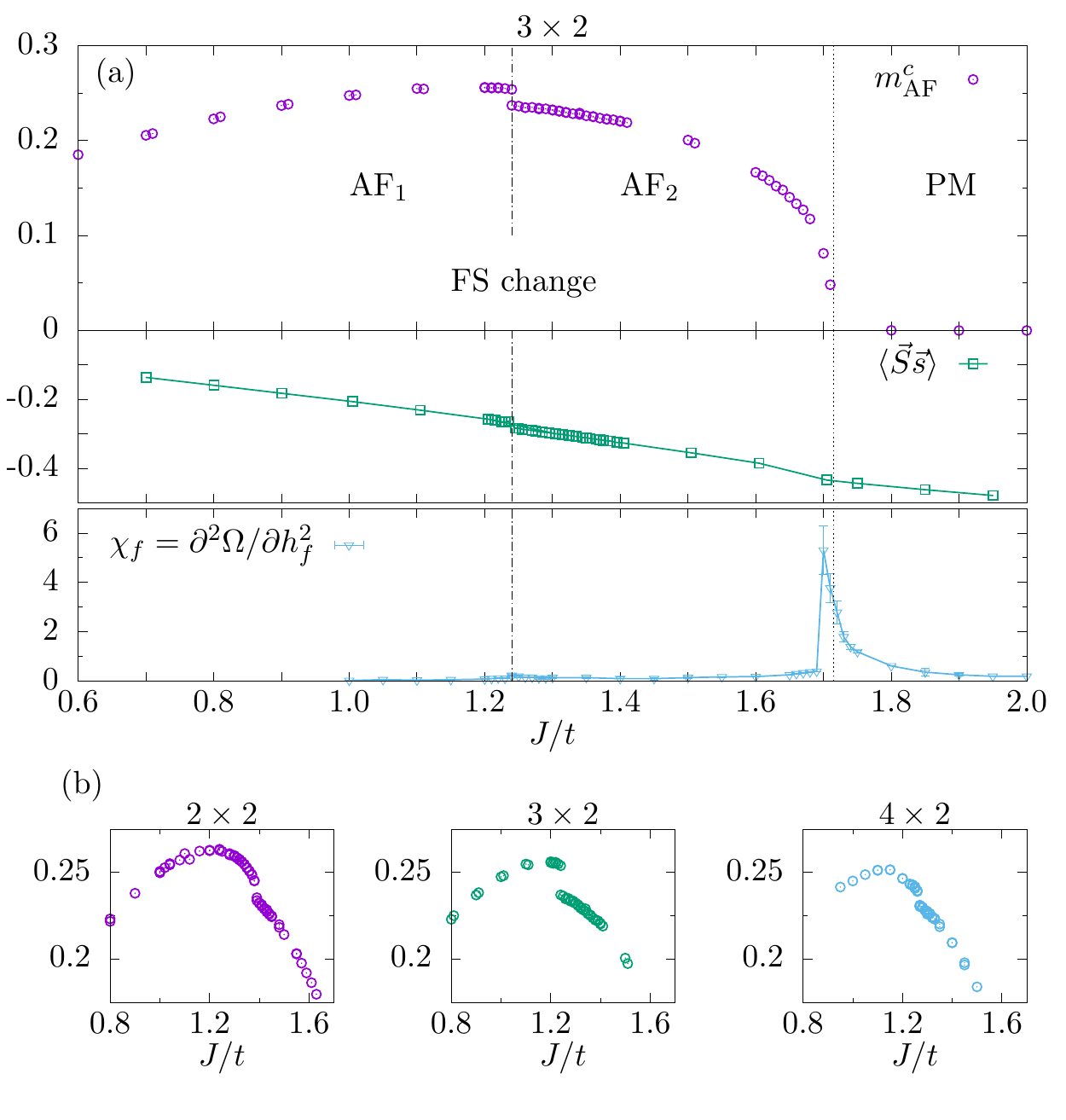}
		\caption{(a) Staggered magnetization $m^c_\text{AF}$, expectation value of the Spin-spin correlator $\langle\vec{S}\cdot\vec{s}\rangle=\partial \Omega/\partial J$ and local susceptibility $\chi^f_{\text{loc}} = \partial^2 \Omega/\partial h^2_f|_{h_f \to 0}$ 
		as a function of $J/t$ at $n=0.95$ using a $3\times2$ cluster. (b) Staggered magnetization around the AF${}_1$-AF${}_2$ transition for the three indicated clusters.} 
		\label{fig_AF_GS_Sisi_n095}
\end{figure}

It is this very region close to the transition to the paramagnet where the spin-spin correlator $\langle \vec{S}_i \cdot \vec{s}_i \rangle$ at half-filling suggests a considerable admixture of Kondo singlets in the ground state.
Figure~\ref{fig_AF_GS_Sisi_n095}(a) shows the staggered magnetization, the spin-spin correlator and the local magnetic susceptibility of the $f$-spins for a filling of $n=0.95$.
Compared to the half-filled case, the antiferromagnetic phase is divided into two regions with the aforementioned different (Fermi surface) topologies.
The transition between both AF phases at $J_{c,\mathrm{FS}}/t=1.24$ is discontinuous as both $m$ and $\langle\vec{S} \cdot \vec{s}\rangle$ jump at the transition.
For smaller electron densities the transition between the AF phases remains to be discontinuous down to an electron density of $n=0.8$.
There, the second AF solution $\mathrm{AF}_2$ vanishes and when increasing the coupling strength in the AF metal $\mathrm{AF}_1$ the system directly jumps to the PM metal solution.
Therefore, the transition between AF metal and PM metal is of first order for $n\lesssim 0.8$.
When approaching half-filling, the jump in the staggered magnetization reduces until the transition between both AF phases becomes continuous at $n=0.97$.
As at half-filling, the local magnetic susceptibility $\chi^f_{\text{loc}}$ shows a divergence at the onset of AF.
However, inside the AF the local susceptibility has an additional increase at $J_{c,\mathrm{FS}}$.

In Fig.~\ref{fig_AF_GS_Sisi_n095}(b) the jump in the staggered magnetization is shown for the three cluster sizes studied here.
As can be seen, the jump size at the transition is finite for all cluster sizes, but 
 not systematic, so that a finite-size extrapolation of both, the jump size and the jump position, are not possible with the available cluster sizes. 
A discontinuous transition between the AF phases was also found using VMC approaches \cite{WO07,ABF13}, but within the framework of DCA \cite{MBA10} and rDMFT \cite{PK15} the transition between the two AF phases is found to be continuous instead.
{Note that} only the VMC works at zero temperature like the VCA, so that the continuous nature of the transitions seen by DCA and rDMFT could be due to finite temperatures.

\begin{figure}[b!]
	\begin{center}
		\includegraphics[width=0.495\textwidth]{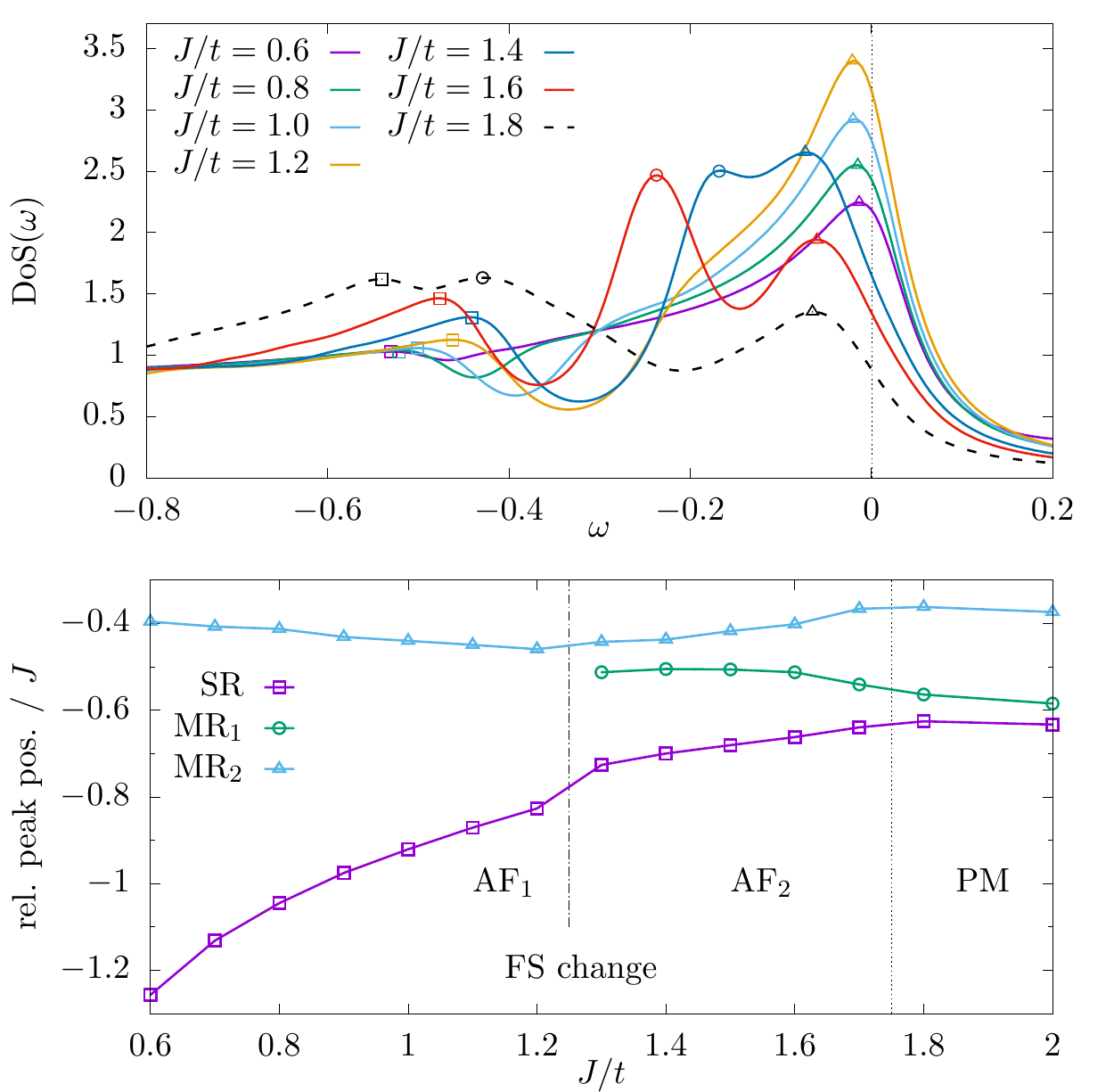}				
		\caption{DOS (top panel) and relative position of the resonances in the DOS (bottom panel) at $n=0.95$ for different coupling strengths in the antiferromagnetic region obtained from the AF solution of the VCA. A change of the peak structure close to the Fermi energy at $\omega=0$ happens at $J/t\gtrsim 1.2$: the main peak splits and an additional side resonance develops. Close to $J_c$ the structure already resembles the three peak structure known from the Kondo insulating region for $J>J_c$.}
		\label{fig_3x2_AF_OffHF_DoS_J}
	\end{center}
\end{figure}

The top panel of Fig.~\ref{fig_3x2_AF_OffHF_DoS_J} shows part of the DOS close to the Fermi surface for different values inside the AF phase at $n=0.95$.
At the largest value of the coupling strength shown, which is close to the critical value $J_c/t \approx 1.75$, three distinct peaks are visible which are similar to the characteristic three-peak structure of the paramagnet at $J>J_c$.
When reducing $J$, the two peaks closest to the Fermi energy approach each other and finally seem to merge at $J/t\approx1.2$.
However, the bipartite character of this main resonance can be still seen in a shoulder at $\omega\approx-0.18$.
Further decreasing the coupling strength results in a more and more broadened shoulder which at small coupling strengths such as $J/t=0.6$ can only be guessed to still exist.

More importantly, the leftmost peak which might be identified at half-filling as being an indicator for Kondo singlet formation survives even down to small coupling strengths.
The relative peak position with respect to the center of the gap to the right of MR$_2$, plotted in the bottom panel, is reduced for decreasing coupling strength and jumps at the Fermi surface change between the two AF phases.
However, it is difficult to quantify the peak height of this signal from the DOS as it would be necessary to remove the background which, is \textit{a priori} unknown.

At this point, we return to the question of whether the Kondo breakdown happens at finite or at zero coupling strengths.
In particular, the local susceptibility and the spin-spin correlator allow to investigate whether the Kondo breakdown coincides with the discontinuous transition between the two AF phases.
In the $\mathrm{AF}_1$ phase close to the critical coupling strength $J_{c,\mathrm{FS}}/t$, the spin-spin correlator is still smaller than $-1/4$, which indicates that the contribution of Kondo singlets is still finite.
However, for weak coupling $\langle\vec{S}\cdot\vec{s}\rangle>-0.25$ so that no definite statement of the presence of Kondo singlets can be made.
In case of a transition from itinerant to localized heavy fermions (ILT), the local magnetic susceptibility was found to diverge in a study by \textit{Hoshino} and \textit{Kuramoto} \cite{HK13}.
There, the Kondo-Heisenberg model was treated within DMFT and the ILT was only observed for finite non-local spin-interactions $J_H$ whereas the heavy fermions were found to be itinerant for the plain Kondo lattice model ($J_H=0$) \cite{HK13}.
However, here we find a divergence of the local susceptibility only at the onset of AF and not at $J_{c,\mathrm{FS}}$.
This further supports the absence of a breakdown of Kondo singlets in the AF region, in particular at the transition point between the two AF phases.

To conclude this discussion, it is difficult to fully exclude the Kondo breakdown scenario for very small coupling strengths.
The density of states, the spin-spin correlator and spectra indicate that in the AF region close to $J_c$ Kondo singlets make part of the charge carriers. 
This speaks against a local quantum critical point, where Kondo breakdown and onset of antiferromagnetic order coincide at $J_c$. 
Although a jump in the staggered magnetization indicates a phase transition within the AF phase, the absence of a divergence in $\chi_f$ does not suggest a change in the composition of the heavy-fermion state at $J_{c,\mathrm{FS}}$.
Nevertheless, the topological differences between the Fermi surface in the weak-coupling and intermediate-coupling regimes are evident.

By using larger clusters it would be interesting to check, to what amount finite-size effects enter in the FS structures that have been discussed here and how the Kondo singlet peak evolves as a function of cluster size.
Another possibility to further investigate putative Kondo breakdown scenarios would be to include $f$-propagators in the VCA by basing it on an adapted Luttinger-Ward functional \cite{CPR05}.
It also has to be noted that other techniques which work directly in $k$-space, such as DCA \cite{MA08,MBA10} or rDMFT \cite{PK15}, might be able to investigate the Fermi surface evolution more precisely.

\section{Superconductivity}
\label{KLM_SC}
In this section, we investigate both local s-wave and nodal $d_{x^2-y^2}$ superconductivity in the KLM.
Section \ref{SCs} presents the main results obtained by VCA for s-wave SC, and some additional details of the calculations can be found in Appendices \ref{Details_swave} and \ref{swave_AF_off}.
It will be shown that only mean-field-like solutions and no local SC due to correlation effects are present.
However, robust d-wave SC is found and investigated in Sec.~\ref{KLM_SC_d}.
Finally, the interplay of d-wave SC and AF is analyzed in Sec.~\ref{SCd+AF}.
Section~\ref{EOM} complements these numerical results by considering the EOM for the pairing susceptibilities.

For small clusters, VCA is known to prefer superconducting solutions even at half-filling as seen in the Hubbard model \cite{LTP+15}, which is used as a model system for high-temperature superconductors such as the cuprates \cite{SLMT05}.
This occurs especially if the system only has a small gap as then allowing for pairing to another quantum sector results in an energy gain which may be sufficient to overcome this gap. 
Nevertheless, VCA allowed for a qualitative study of superconductivity in the Hubbard model \cite{SLMT05,AAPH06}, motivating us to apply this technique to the KLM.

\begin{table*}[t!]
\begin{center}
	\begin{tabular}{|l|c|c|c|l|}\hline
		No.&$D$&$\mu^{\prime}$&$t^{\prime}$&Characteristics\\ \hline\hline
		1&$\dw$&$\up$&$\up$&Two solutions with $t^{\prime}\approx t$\\\hline
		2&$\dw$&$\up$&$\dw$&Two solutions: One has $t^{\prime}=0$, the other one is thermodynamically unstable \\\hline
		3&$\dw$&$\dw$&$\up$&One solution with $t^{\prime}\approx t$, but thermodynamical stability violated off half-filling\\\hline
		4&$\dw$&$\dw$&$\dw$&One solution with $t^{\prime}=0$\\\hline
	\end{tabular}
	\caption{Different types of stationary points investigated around half-filling at $J/t=1.5$ when varying $D_s,\ \mu^{\prime}$ and $t^{\prime}$ when considering s-wave SC. The arrows in columns 2-4 indicate whether $\Omega$ has a maximum ('$\uparrow$') or a minimum ('$\downarrow$') with respect to the variational parameter.}
	\label{Tab_SCs_J15_tc}	
\end{center}
\end{table*}

\subsection{Absence of local s-wave superconductivity in the KLM}
\label{SCs}
Our starting point for the study of s-wave superconductivity 
is the observation of such a phase in Ref.~\onlinecite{BZV+13}.
By using DMFT with a NRG solver, \textit{Bodensiek et al.} identified a broad region off half-filling and for coupling strengths $J/W\gtrsim0.1$ where the anomalous expectation value $\Phi_s :=\langle c_{\uparrow}c_{\downarrow}\rangle$ had a very small, but finite value.

Local pairing was already observed in the KLM by mean-field approaches \cite{HKS12,MY13}, but the superconducting state found within DMFT is conceptually different as the pairing does not occur between $c$- and $f$-electrons.
Instead, the superconductivity is only mediated by the antiferromagnetic spin fluctuations and pairs are formed in the conduction band only.

In contrast to this DMFT study, we here examine the existence of this unconventional scenario by including spatial fluctuations, as treated by the VCA.

\subsubsection{Half-filling}
At half-filling, adding only an s-wave SC Weiss field leads to a stable stationary point of the SEF.
However, this solution has a large Weiss field strength and including the intra-cluster hopping strength $t^{\prime}$ in the set of variational parameters reveals the local nature of the solution:
the hopping $t^{\prime}$ is zero and the reference system consists of decoupled, locally superconducting sites.
This artificial mean-field-like solution is the only stable non-trivial stationary point at half-filling.
Other true many-body solutions, where s-wave superconductivity is caused by the interaction, are not found.
A detailed analysis of s-wave SC at half-filling can be found in Appendix \ref{swaveHF}.\\

\subsubsection{s-wave superconducting solutions off half-filling}
\label{swaveOffHF}
When changing the chemical potential to leave half-filling, the SEF shows multiple stationary points with respect to the variational parameters.
It is therefore important to decide, which stationary point corresponds to the physical solution.
The set of variational parameters is four dimensional and includes both the chemical potential of the lattice and the cluster, the SC Weiss field, and the cluster hopping strength.

Due to the large number of variational parameters, the $2\times2$ cluster is used to find different stationary points and to discard unphysical solutions.
Afterwards, the $3\times2$ cluster is used for calculations on the remaining solution only.

In order to take the 'correct' quantum space, one has to choose one combination (of often at least two possible combinations) of chemical potentials $\mu$ and $\mu^{\prime}$ . 
In general, one expects to be able to tune the filling by changing the chemical potential $\mu$ around the 'natural cluster fillings' $n_{\text{cl}}=N_e/L$. 
Even a quite weak Weiss field, which couples two adjacent quantum sectors (e.g. with $N_e$ and $N_e-2$), could lead to an energy gain which supersedes an energy gap between these sectors. 
Especially for weakly gapped systems this could lead to artificial superconducting solutions due to this overcompensation effect. 
When including an s-wave superconducting Weiss field and following a stationary point, one often encounters situations where the self-energy jumps or shows a kink as a function of one of the variational parameters.
At these points, which have been encountered before in VCA \cite{BP10}, convergence to the correct stationary point is not ensured anymore. 
Overall, the addition of superconducting Weiss fields poses the problem of choosing the right stationary point, which here means to also choose out of different quantum sectors.

One has to first of all search and review all the possible saddle points and then decide which ones have to be considered. 
For stable solutions the SEF should have maxima or minima with respect to the four variational parameters leading to eight possible types of stationary points.
Focusing on the stationary points where the SEF is minimal with respect to the Weiss field strength leads to the four types of stationary points listed in Table~\ref{Tab_SCs_J15_tc}.

When evaluating different stationary points, the following criteria are used to assess the solutions.
One criterion is the value of the cluster hopping parameter $t^{\prime}$.
'Atomic' solutions with $t^{\prime}=0$ amount to reference systems with decoupled cluster sites that locally form a superconducting singlet state.
They are considered to be artificial mean-field solutions and not to represent superconductivity due to many-body effects, and hence will be discarded.

Another important criterion is thermodynamical stability.
The electron filling $n$ can be obtained either by calculating the derivative of the (approximated) grand potential or by calculating the trace of the VCA Green function.
In order to have thermodynamical stability, both ways of calculating $n$ should lead to the same value. 
Despite having included the cluster chemical potential in the set of variational parameters, in some cases thermodynamic stability is violated.

These two criteria already reduce the number of realistic solutions of Table~\ref{Tab_SCs_J15_tc} with $t^{\prime}\neq0$ and thermodynamic stability to the one solution of type 1.
There, the stationary point of the SEF is a maximum with respect to $\mu^{\prime}$ and $t^{\prime}$ and a minimum with respect to $D_c$.

Since the $2\times2$ cluster shows anomalies in the hopping parameter $t^{\prime}$ on the cluster already for intermediate values of $J/t$, the $3\times2$ cluster is considered in order to investigate the most promising stationary points off half-filling. 

\begin{figure}[t!]
	\begin{center}
	\includegraphics[width=0.45\textwidth]{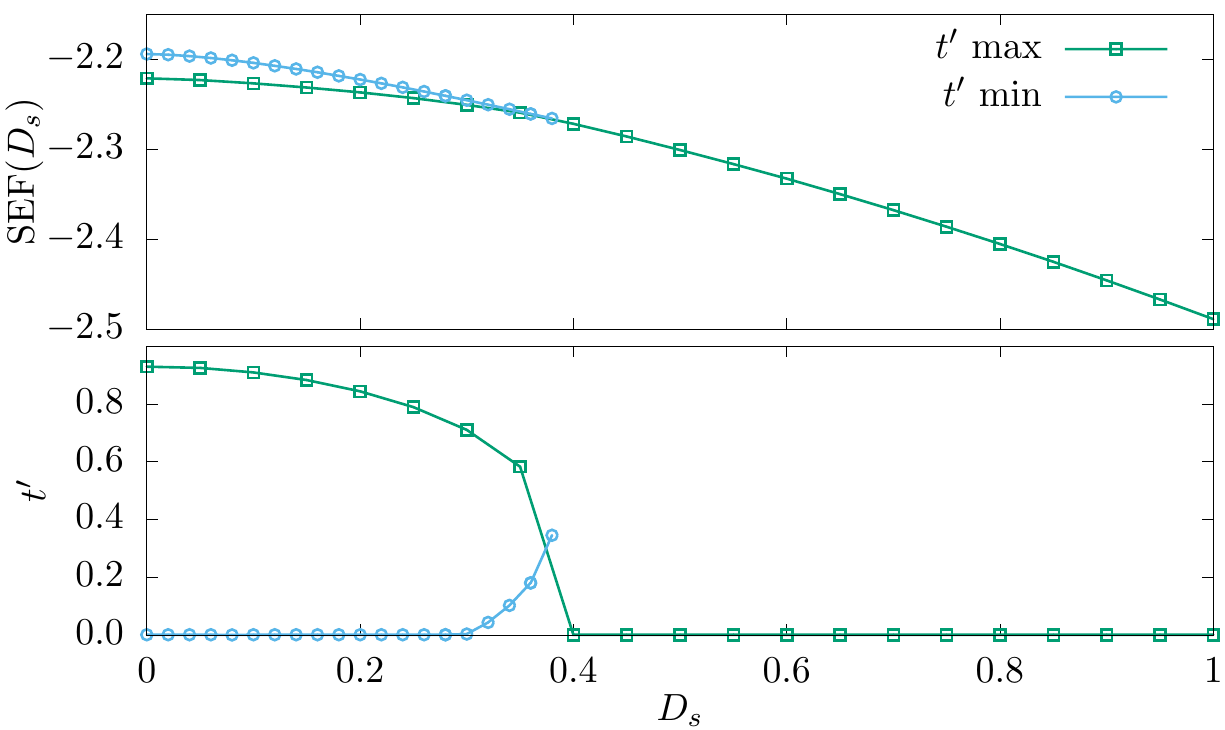}
	\caption{SEF and cluster hopping strength $t^\prime$ at the stationary point of the SEF with respect to $\mu,\mu_c$ and $t^{\prime}$ as a function of the Weiss field strength $D_s$. Plotted are two solutions for $J/t=2.4$ at $n=0.9$, one corresponding to a maximum with respect to $t^{\prime}$ (filled symbols), the other one corresponding to a minimum (empty symbols). Note the SEF having an extremum in either case at $D_s =0$, indicating no s-wave SC order being stabilized.} 
	\label{fig_J24_SCs_SEF_t}
	\end{center}
\end{figure}

In contrast to our previous analysis, where no (clearly physical) s-wave solution was found, we now take the converged parameters of the paramagnetic solution at a filling of $n\approx0.95$ as a starting point and add an s-wave superconducting Weiss field to the set of variational parameters.
Again, various stationary points are found, but most of them are identified as unphysical according to the above criteria and therefore neglected.

In Fig.~\ref{fig_J24_SCs_SEF_t} the cluster hopping parameter $t^{\prime}$ after maximization (solution 1) or minimization (solution 2) is plotted as a function of the superconducting Weiss field strength $D_s$.
If one wants to exclude 'unphysical' solutions, where the cluster consists of isolated sites and a resulting local self-energy enters the calculation of the SEF, one can restrict the search to small values of $D_s$ shown in this figure.
As can be seen from Fig.~\ref{fig_J24_SCs_SEF_t}, no stationary point is found in this region ($D_s<0.4$) except the non-superconducting solution at $D_s=0$.

The qualitative picture shown in Fig.~\ref{fig_J24_SCs_SEF_t} also holds for smaller coupling strengths and fillings around $n=0.90$ and we believe that it prevails for even smaller fillings.
For small $D_s$, a solution with reasonable intra-cluster hopping strength $t^{\prime}\sim t$ is found, but already for comparably small values of $D_c\sim 0.4-0.5$ the cluster hopping drops to zero.
At this point, the cluster hopping of the second solution, which is zero for small Weiss field strengths, diverges.
The solution breaks down and one is left with the case of decoupled sites inside the reference cluster that was discussed before.

We show in Appendix \ref{swave_AF_off} that treating both s-wave SC and AF does not stabilize a (different) s-wave SC solution.
The only stable solution showing s-wave SC is the one with vanishing hopping $t^{\prime}$ on the reference system.
An s-wave SC solution caused by correlations is not found.
In the following, we consider d-wave SC instead.

\subsection{Nodal d-wave superconductivity}
\label{KLM_SC_d}

Before treating the possible coexistence of AF and SC order, in the spirit of the investigation so far, we will only add a SC Weiss field to the paramagnetic case as additional variational parameter to check for the possible existence of d-wave SC at all.
This is in particular interesting, since d-wave superconductivity is often found experimentally in heavy fermion systems \cite{SW16} and also numerical studies of the Kondo lattice model indicate the existence of a d-wave superconducting phase \cite{AFB14,Ots15}.
Although in the recent study of Ref.~\onlinecite{Ots15} by Otsuki a p-wave superconductor was found for coupling strengths around the critical point $J_c$, 
we will not further investigate this type of SC order due to the already very large number of variational parameters and the accompanying complexity of our treatment.  
Instead, we focus in this section on superconductivity with $d_{x^2-y^2}$ symmetry and leave the investigation of this further interesting SC channel with VCA to future studies.

First, the paramagnetic solution will be taken as a starting point to investigate superconductivity by adding a Weiss field with d-wave symmetry.
In Sec.~\ref{SCd+AF} antiferromagnetism will be treated on equal footing with superconductivity and the interplay of both symmetry-broken phases will be discussed.

In the case of s-wave SC, the Cooper pairs form locally and clusters are affected in a uniform way by the Weiss field.
The geometry and size of the cluster enter the calculation through the intra-cluster hopping and in case of antiferromagnetism through the mediated effective RKKY interaction only. 
This changes for the case of extended pairing, such as the non-local $d_{x^2-y^2}$ SC - due to its geometry, the $2\times2$ cluster is for instance known to favor $d$-wave pairing, which might bias the result.
For this reason the $2\times2$ cluster will only be briefly discussed and mainly used as a reference to the $3\times2$ cluster, for which most of the results will be shown.

As long as no AF Weiss field is used in addition to the SC one, the paramagnetic phase diagram at half-filling has to be used as a starting point.
In contrast to the 'full' phase diagram which by including AF shows an insulator at arbitrary coupling strength $J/t$ at half-filling, the paramagnetic phase diagram also shows a metallic phase at half-filling.
To be more precise, the system is metallic in the weak coupling region and becomes insulating when the coupling exceeds some value $J>J_{c,INS}$, where Kondo screening is large enough to form an insulator consisting of the Kondo singlets.

\begin{figure}[t!]
	\begin{center}
		\includegraphics[width=0.45\textwidth]{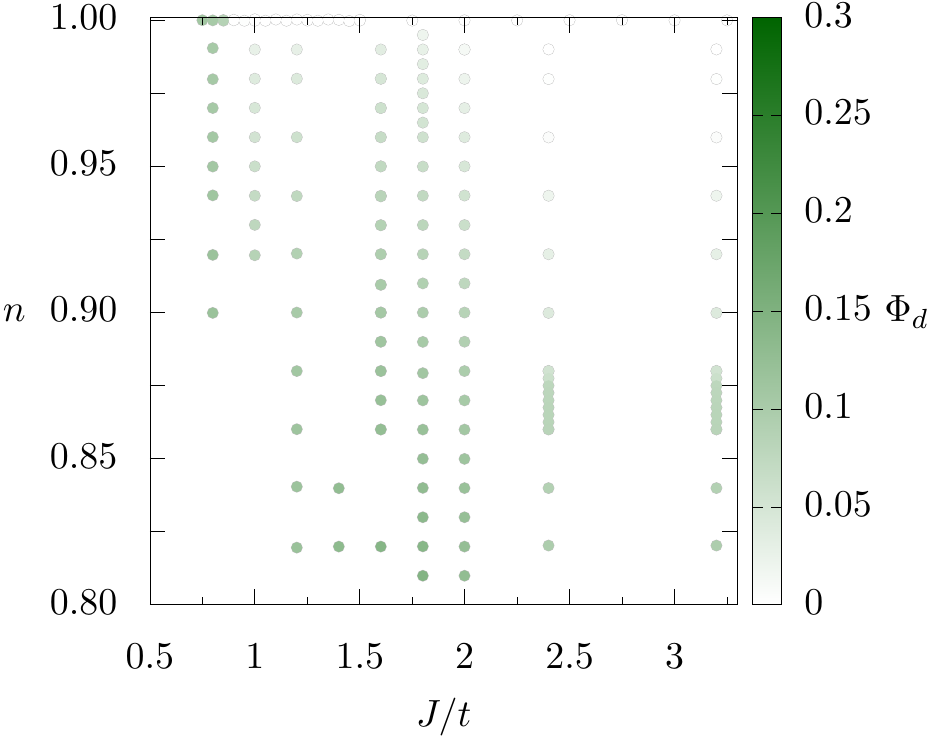}		
		\caption{Anomalous expectation value when considering d-wave SC. The figure shows $\Phi_d$ obtained using a $3\times2$ cluster and variational parameters $\mu^{\prime}, t^{\prime}$ and $D$. The circles are data points, the color code indicates the value of $\Phi_d$ at these points. For $J\leq J^{SC}_{ano}\approx0.9$ there is a superconducting solution even at half-filling, but one has to be careful as this region coincides with an anomaly in $t^{\prime}$. The interaction strength $J$ where the anomaly occurs is quite close to the one found in the paramagnetic solution ($J^{PM}_{ano}\approx 0.84$).}
		\label{fig_SC_d_PD_3x2}
	\end{center}
\end{figure}

\begin{figure}[t!]
	\begin{center}
		\includegraphics[width=0.495\textwidth]{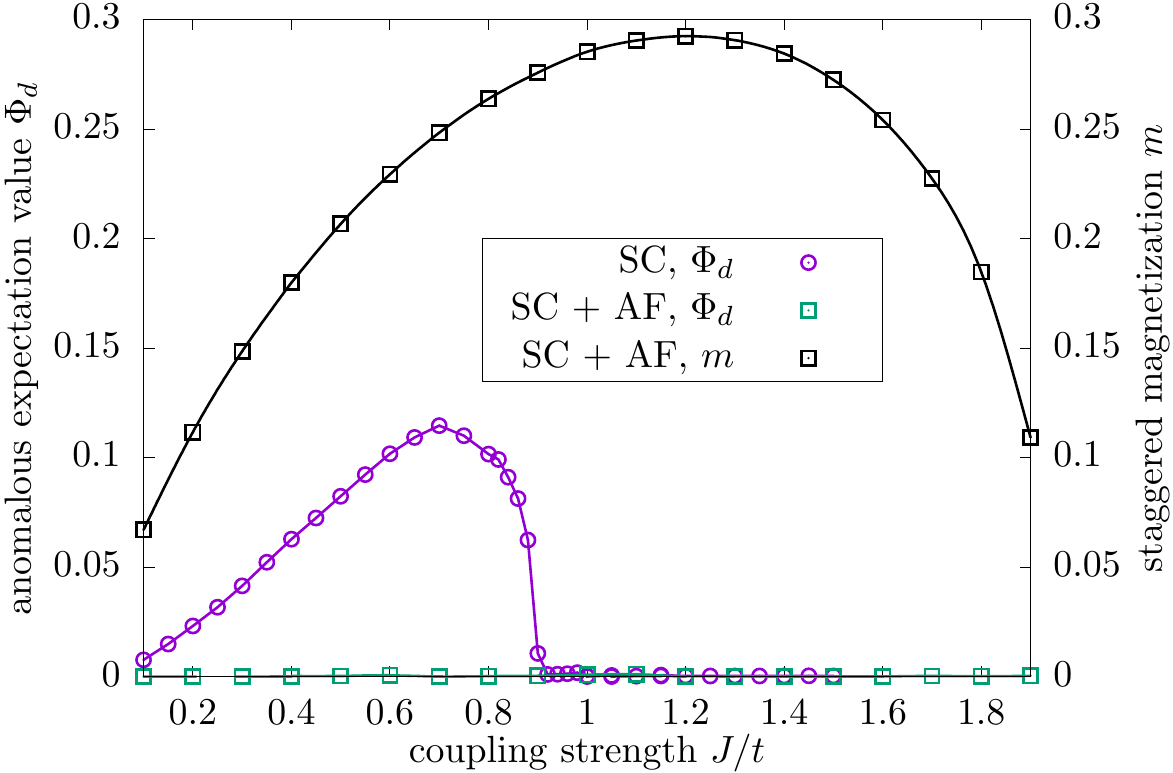}
		\caption{Anomalous expectation value $\Phi_d$ as a function of $J/t$ at half-filling for the pure SC solution (blue) and when including also AF (green) using a $3\times2$ cluster. For the latter case the staggered magnetization is shown in black.}
		\label{fig_SCd_AF}
	\end{center}
\end{figure}

When looking at the anomalous expectation value
$\Phi_{d_{x^2-y^2}}=\langle c_{i\uparrow}c_{i\pm e_x\downarrow} - c_{i\uparrow}c_{i\pm e_y\downarrow}\rangle$,
which serves as the order parameter of a superconducting phase, it is not surprising that no superconductivity is found in the insulating region (see Fig.~\ref{fig_SCd_AF}).
In contrast to the expectation that there should not be any superconductivity at all at half-filling because of the insulating phases, one finds d-wave superconductivity for weak coupling up to $J/t\approx 1.1$.
This is the region where the incomplete (paramagnetic) treatment of the system showed an anomaly in the cluster hopping.
For the $2\times2$ cluster the cluster hopping has a minimum at $J/t\sim1.4$ and starts to grow for smaller coupling; in case of the $3\times2$ and $4\times2$ clusters $t^{\prime}$ even showed a kink when plotted as a function of $J$ at $J/t\sim0.8$, which marked the phase transition to a metal for smaller coupling strengths.
As will be shown below, for the $3\times2$ cluster the onset of superconductivity is comparable to this coupling strength $J_{c,INS}$.
Although this reveals the need of including antiferromagnetism into the calculations, it still provides the correct starting point for the strong coupling region, i.e. the paramagnetic region $J>J_{c,AF}$.
There, leaving half-filling should be valid as antiferromagnetism is not realized at or off half-filling as was shown in Sec. \ref{KLM_AF}.  

Leaving the study of the region with coupling $J<J_c\approx 2.05 t$ to the next subsection where antiferromagnetism is included in the investigation, it remains to consider here the region with $J>J_c$.
Still, it is interesting that the maximum of the anomalous expectation value ($J/t\sim2.1$) amounts to the region where antiferromagnetic fluctuations lead to the onset of AF long-range order if one permits this type of ordering.
The corresponding electron density at the maximum is roughly $n\approx0.65$.
Close to half-filling the paramagnetic metal persists at coupling stengths $J>J_c$ and only doping of $\epsilon\sim0.1-0.2$ leads to a finite $\Phi_d$.
When lowering the electron density further, the size of the anomalous expectation value diminishes and finally goes to zero at small $n$.

Before including an antiferromagnetic Weiss field, d-wave superconductivity is investigated in the region around $J_c$ using a $3\times2$ cluster (see Fig.~\ref{fig_SC_d_PD_3x2}).
For this cluster, $J_c/t\approx1.95$.
The overall phase diagram compares qualitatively well to the one of the $2\times2$ cluster and it even gives quantitatively similar results.

\subsection{Equations of motion for the pairing susceptibility}
\label{EOM}

As discussed in Sec.~\ref{SCs}, there is no evidence for local s-wave SC in the VCA treatment of the KLM. 
Here, we consider a complementary approach by studying generic features of the EOM for the pairing susceptibilities. 
EOM in the context of the KLM have been used before, e.g., for small clusters or in combination with a mean-field approach \cite{HRN03,WG00}.
Here, however, we adapt the approach specifically to treat SC. 

Since we expect SC to be induced by the interaction, we here sketch the main results for the EOM of the interaction term of the KLM and leave further details to Appendix~\ref{EOM:appendix}. 
It is convenient to rewrite the interaction part of the KLM Hamiltonian, Eq.~\eqref{eq:KLM}, in terms of annihilation (creation) operators for the electrons in the conduction band $c^{(\dag)}_{i,\sigma}$ and the localized electrons in the $f$-band $f^{(\dag)}_{i,\sigma}$, respectively, giving \cite{CLGI03}
\begin{equation}\label{KLM}
H_I= \sum\limits_{i,\sigma} \biggl[\frac{J_z}{4} \big( n_{i,\sigma} N_{i,\sigma} - n_{i,\bar \sigma} N_{i,\sigma}\big) + 
\frac{J_{\perp}}{2} \big( c^{\dag}_{i,\sigma} c_{i,\bar \sigma} f^{\dag}_{i,\bar \sigma} f_{i,\sigma}\big) \biggl].
\end{equation}
The operators $n_{i,\sigma}$ ($N_{i,\sigma}$) represent the on-site occupation number of the conduction band ($f$-band) electrons with spin $\sigma$ on site $i$.
For the latter, the constraint $N_{i,\sigma} + N_{i,\bar{\sigma}}=1$ has to hold.
In case of the isotropic KLM treated here, the coupling strengths are $J_z=J_{\perp} \equiv J$.

The EOM for the pairing susceptibility is obtained by considering
\begin{equation}\label{Gap}
C(t) = \langle \Delta_{ij}^{\dagger}(t)\Delta_{i^{\prime}j^{\prime}}(0)\rangle 
\end{equation}
for the pairing operators
$$\Delta_{ij}^{\dagger} = F(i,j)c^{\dagger}_{i\uparrow}c^{\dagger}_{j\downarrow}.$$ 
The function $F(i,j)$ takes into account the geometry of the pairing order parameter.
For the various channels of interest, it reads as
\begin{equation}\label{Delta}
F(i,j) := 
\left
\{
\begin{array}{cl}
  \delta_{i,j} & \text{s-wave} \\
  &\\
    +1: r_i-r_j=\pm e_x, \\ 
    +1: r_i-r_j=\pm e_y, & \text{extended s-wave.} \\
    &\\
  +1: r_i-r_j=\pm e_x, \\
  -1: r_i-r_j=\pm e_y, & \text{d${}_{x^2-y^2}$.} \\
  &\\
  +1: r_i-r_j=\pm (e_x+e_y), \\ 
  -1: r_i-r_j=(e_x-e_y), & \text{d${}_{xy}$.}  \\ 
\end{array}
\right.
\end{equation}

The time derivative of $\Delta^{\dag}_{ij}(t)$ is given by the Heisenberg equation
\begin{equation}\label{EOM_Gap}
\frac{d}{dt} \Delta^{\dag}_{ij}(t)  =-i  \left[c^{\dag}_{i,\sigma}(t)c^{\dag}_{j,\bar \sigma}(t) , H_{I} \right]. 
\end{equation}
Details of the calculation and the resulting expressions are found in App.~\ref{EOM:appendix}.
One finds that
\[
\frac{d}{dt} \Delta^{\dag}_{ij}(t)  =0 \qquad \mbox{for s-wave pairing.}
\]
This is an interesting observation, since it shows that the dynamical properties of the  pairing susceptibility in the s-wave channel will not depend on the interaction, and hence are the same as the ones of free electrons.
Note that this result does not completely rule out the possibility for having s-wave SC.
However, it restricts the possible mechanisms which might stabilize it to ones, in which the frequency dependence of the pairing susceptibility does not play a role.
This needs further investigations, which 
we leave for future research.
However, together with the lack of evidence for s-wave pairing in our VCA treatment and also in further numerical approaches \cite{Ots15}, this further restricts the possibility of realizing such a phase in the KLM. 

However, this does not hold for the extended s-wave and the d-wave channels.
Note that the results for the d-wave channel [e.g., Eq.~\eqref{Delta_dwawe}] contain an interesting aspect: 
the time derivative of $\Delta_{ij}^\dag$ contains expressions which are a product of two creation operators (and hence a pair of two fermions) and of the staggered magnetization of the local moments.
This indicates that AF order should directly contribute to the dynamics of  the SC response function, so that a coexistence or even a cooperative interplay between AF and d-wave SC order might come into appearance.
In the next section such a possible cooperative interplay between SC and AF order is further investigated using the VCA.
\subsection{Competition of antiferromagnetism and superconductivity}
\label{SCd+AF}

In the previous subsection we have seen, that away from half-filling superconducting phases can manifest in the phase diagram of the Kondo lattice model.
References \onlinecite{MSV86,SLH86,BBE86} show that antiferromagnetic spin fluctuations can assist anisotropic even-parity pairing such as the $d$-wave superconductivity investigated here.
Also the EOM treatment in 
{Sec.}~\ref{EOM} 
{brought} up the question for possible cooperative interplay of AF and d-wave SC order.
Here, we are going to investigate this aspect in detail.

Especially for small couplings $J/t$ another symmetry breaking enters in the form of antiferromagnetic ordering of the conduction electrons. 
The interplay of these two effects are known to be important for d-wave superconductivity in the Hubbard model as possibly realized in high-temperature superconductors \cite{SLMT05}. 
In principle, there are three scenarios which are possible. 
The most improbable is that the two phases are independent and, hence, considering magnetism and superconductivity leads to the same magnetization and anomalous expectation value as treating these effects separately.
It is also possible that both phases coexist and that they either compete, which means that the onset of antiferromagnetism reduces superconductivity and vice versa, or that they cooperate, in that case the superconductivity would be enhanced due to the antiferromagnetic ordering.

However, considering the results of the previous section, approaching this question within VCA might seem to be tricky as we have encountered problems for weak coupling.
Nevertheless, as this is the very region where at least in the normal phase antiferromagnetism dominates, one has to consider both broken symmetries together to properly address this weak- to intermediate-coupling regime.
As shown in Sec.~\ref{KLM_AF}, antiferromagnetism already sets in for intermediate interaction strengths where the divergence of $t^{\prime}$ does not yet pose a problem, but one has to bear in mind that any doping of the system reduces the antiferromagnetic correlations.
Hence, doping the system sufficiently in order to observe superconductivity might already be too much doping to observe antiferromagnetism. 
Especially for intermediate-coupling strengths close to $J_c^{AF}$ this means that one has to investigate a very narrow $\mu$ window corresponding to small doping.

We have used a $3\times2$ cluster to revisit the half-filled system at $J<J_c^{AF}$, this time using both a d-wave superconducting and an antiferromagnetic Weiss field at the same time. 
For all coupling strengths the solution coincides with the AF insulator that was already found in Sec.~\ref{KLM_AF}.
While the stationary point with $M=0$ still exists, a comparison of the corresponding energies shows that the antiferromagnetically ordered phase is  always lower in energy.
Especially at weak coupling allowing both for superconductivity and for antiferromagnetism results in an antiferromagnetic insulator and no superconducting solution with lower energy is found at half-filling. 

\begin{figure}[b!]
	\begin{center}
		\includegraphics[width=0.495\textwidth]{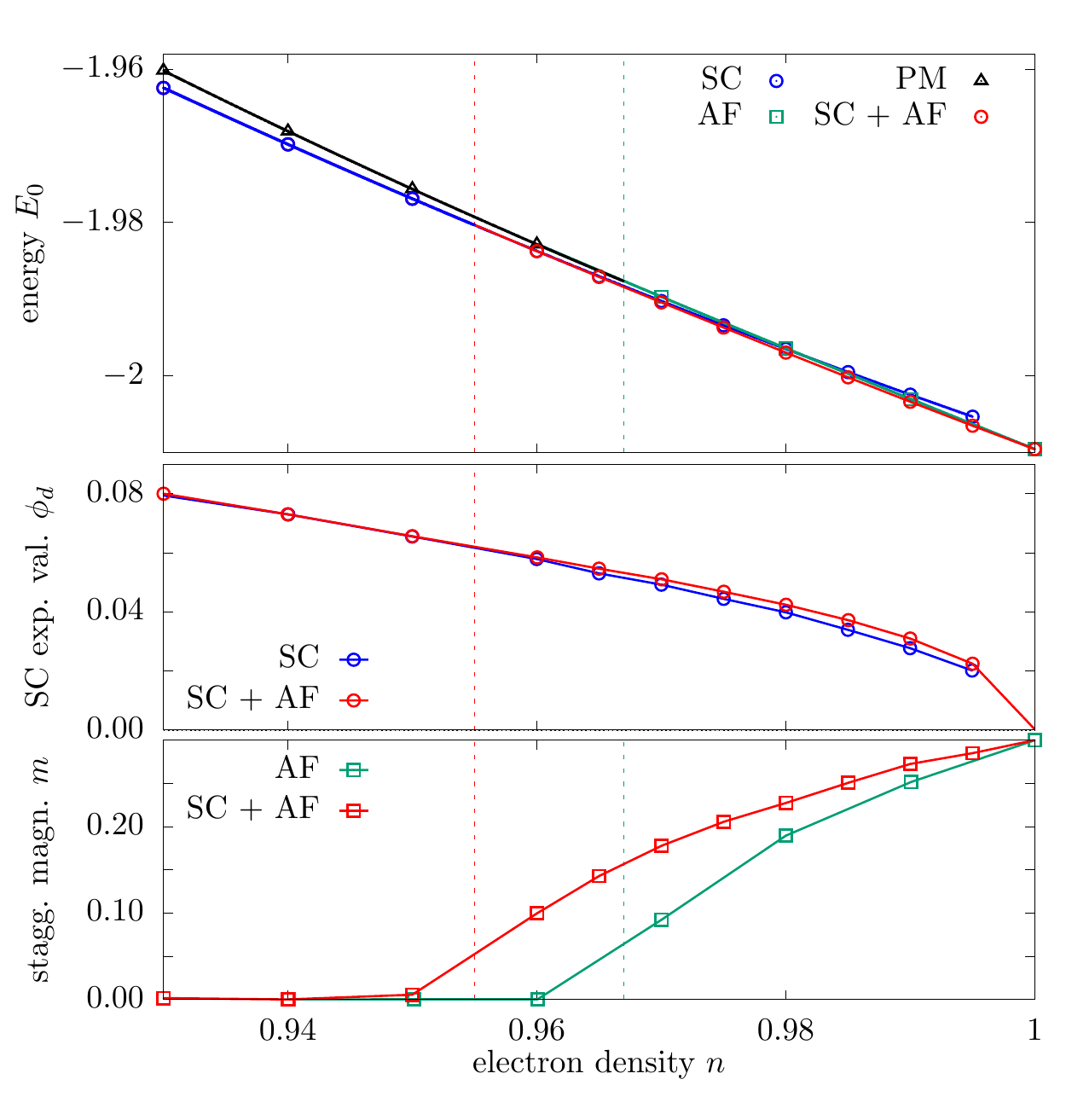}
		\caption{Results when considering d-wave SC and AF away from half-filling obtained using a $3\times2$ cluster at $J/t=1.8$. The top panel shows the ground state energy as a function of filling for the various cases indicated. The vertical lines indicate the transition points at which AF order vanishes when considering AF only or AF+SC, respectively. The center and bottom panels show the anomalous expectation value and the staggered magnetization as a function of electron density $n$ for the superconducting, the antiferromagnetic and the coexisting SC+AF solutions as indicated.}
		\label{fig_SC+AF_J18}
	\end{center}
\end{figure}

\begin{figure}[t!]
	\begin{center}
		\includegraphics[width=0.495\textwidth]{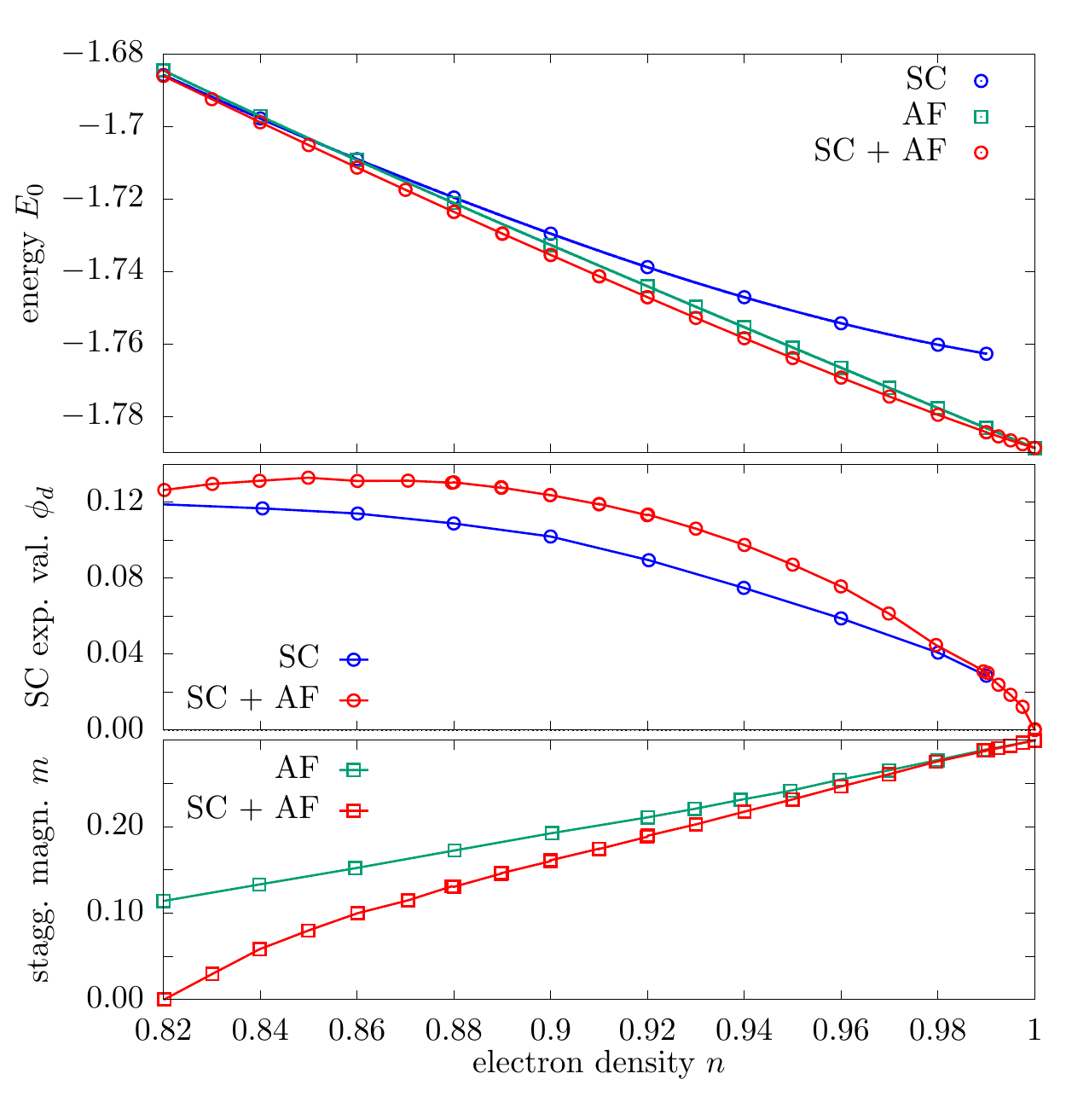}
		\caption{The same as in Fig.~\ref{fig_SC+AF_J18} showing the energy (top panel) and the order parameters (center and bottom panels) as a function of filling $n$, this time for $J/t=1.2$.}
		\label{fig_SC+AF_J12}
	\end{center}
\end{figure}

In case of the $3\times2$ cluster the critical coupling strength is $J_{c,AF}\approx 1.95$.
Here, we have a closer look at the behavior in the two AF phases identified in Sec.~\ref{KLM_AF_OffHF} by considering $J/t = 1.8$ in Fig.~\ref{fig_SC+AF_J18} and $J/t = 1.2$ in Fig.~\ref{fig_SC+AF_J12}, respectively. 
By keeping $J/t$ at these values, we avoid crossing the transition line between the two AF phases.

Figure \ref{fig_SC+AF_J18} shows both the pure antiferromagnetic and the pure d-wave superconducting solutions as well as a solution with coexistence of superconductivity and antiferromagnetism.
Starting with the 'pure' solutions at $J/t=1.8$, for small doping the antiferromagnetic solution has a lower energy than the superconducting one, but their energies cross at $n\approx 0.98$.
From considering these two phases only, the system would be antiferromagnetic for $0.98\leq n\leq1$ and d-wave superconducting for fillings smaller than $0.98$.
However, compared to the 'pure' phases, the solution with coexistence of AF and SC has the lowest energy and should therefore be realized in the system.

In the coexistence region the anomalous expectation value does not change much compared to the solely superconducting solution with $M_c=0$.
At the same time the staggered magnetization is enhanced compared to the AF solution.
The coexistence region therefore enlarges the superconducting region to a value close to half-filling and extends the antiferromagnetic region down to an electron density of $n\approx0.955$.
In this sense, the interplay of antiferromagnetism and d-wave superconductivity at $J/t=1.8$ can be considered to be cooperative.
Outside of the coexistence region, no antiferromagnetic order is present and the solution coincides with the one shown in Fig.~\ref{fig_SC_d_PD_3x2} where d-wave superconductivity without antiferromagnetism was considered.

When reducing the coupling strength, the interplay between antiferromagnetism and superconductivity changes.
Figure \ref{fig_SC+AF_J12} shows the same quantities for a coupling of $J/t=1.2$.
Still, there exists a coexistence region which has lowest energy and which is therefore preferred as compared to the pure AF and SC solutions.
Close to half-filling ($n\gtrsim0.98$), the coexistent solution has comparable order parameters as the pure solutions.
For smaller electron density, the staggered magnetization of this solution is reduced compared to the pure antiferromagnetic phase, but the anomalous expectation value is perceptibly larger than the one of the pure d-wave solution.
When extrapolating the staggered magnetization, it becomes clear that the antiferromagnetic region will be reduced compared to the antiferromagnetic phase diagram shown in Fig.~\ref{fig_3x2_AF_OffHF_PD}. 
At the electron density $n_c$ where the antiferromagnetic order breaks down, the pure superconducting solution should be recovered.
Until then (i.e. for $n_c<n<0.98$), the anomalous expectation value is larger than in the pure d-wave solution.

The interplay of both symmetry-breaking mechanisms is therefore characterized by a competition between superconductivity and antiferromagnetism at $J/t=1.2$. 
\begin{figure}[b!]
	\begin{center}
		\includegraphics[width=0.495\textwidth]{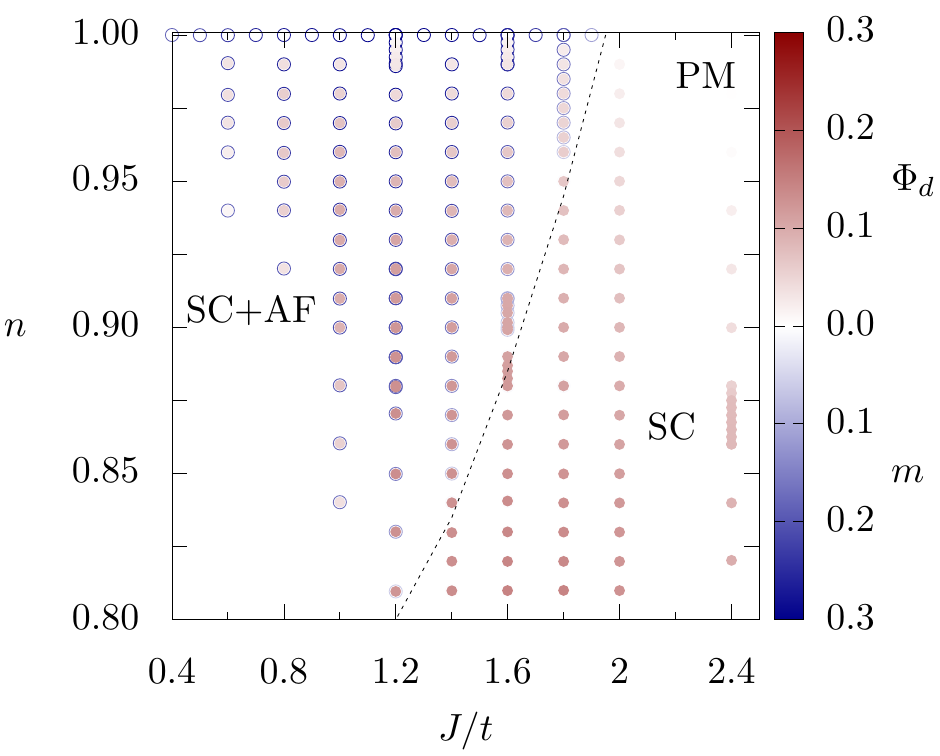}
		\caption{Values of the order parameters as a function of $J/t$ and $n$ when simultaneously considering d-wave SC and AF. The circles are data points. The values for the anomalous expectation value (intensity color coded in red) and for the staggered magnetization (color coded in blue) as obtained using a $3\times2$ cluster are displayed. The dashed line indicates the transition line at which the AF continuously disappears when increasing $J/t$.}
		\label{fig_SC+AF_PD}
	\end{center}
\end{figure}

The results of the investigation of the interplay of antiferromagnetism and d-wave superconductivity are summarized in Fig.~\ref{fig_SC+AF_PD}.
It shows the coexistence region of antiferromagnet and superconductor.
This region is limited by an electron density of $n=1$, where superconductivity breaks down, and by a critical coupling strength $J_c(n)$ (indicated by a dashed line in Fig.~\ref{fig_SC+AF_PD}), which marks the breakdown of the antiferromagnet.

Note that the transition between two AF phases with different Fermi surface topology discussed in Sec.~\ref{KLM_AF_OffHF} seems to be absent when considering AF and SC simultaneously.
Instead, the staggered magnetization $m$ changes smoothly around $J/t\sim1.24$.
This can be explained by the increase of  $m$ in the 'cooperative' region $J/t\gtrsim 1.24$ and the reduction of $m$ in the 'competing' region $J/t\lesssim 1.24$ compared to the 'pure' AF solution.
Hence, we do not find supporting evidence for the existence of two distinct SC+AF phases, while a more careful investigation might still identify subtle behavior related to this effect.

\section{Summary and Outlook}  \label{Summary}
Exploiting the strength of VCA that phases with broken symmetry can either be probed or actively avoided by choosing a suitable variational space, the different phases of the KLM were analyzed separately, as well as their interplay.
{This leads to the main result of this work, which} is the phase diagram displayed in Fig.~\ref{fig:PD_sketch}.
As the half-filled KLM has special features in the phase diagram, it has been analyzed separately from the doped model.

{In the paramagnetic phase, d}ue to the absence of Weiss fields, the set of variational parameters is comparatively small so that it was possible to investigate cluster size effects and to identify a necessary minimal set of variational parameters.
Differences between the $2\times2$, the $3\times2$, and the $4\times2$ clusters were found at small couplings, however, the influence of multiple (anisotropic) hopping strengths for asymmetric clusters turned out not to change the results qualitatively.
At half-filling and strong coupling, all clusters lead to a Kondo insulator with 
{quasiparticle gap, which is essentially independent from the cluster size}.
However, the asymmetric $3\times2$ and $4\times2$ clusters showed an unexpected transition to a paramagnetic metal at $J/t\approx 0.8$.
As confirmed by adding an AF Weiss-field, in this region a gapped long-range antiferromagnetic order emerges due to effective RKKY interaction, so that the metallic phase at half-filling has not been investigated in further detail. 
Doping the system away from half-filling at strong couplings resulted in a metallic phase with a large Fermi surface.
The Fermi surface area in this region amounts to the sum of electron density and f-spin density, which indicates the participation of f-spins in the charge-transfer process via mobile Kondo singlets.

The addition of an AF Weiss field as variational parameter allows for the investigation of an emerging antiferromagnetic phase below a critical coupling strength $J_c/t$.
At half-filling,
{finite-size extrapolation is possible}.
Despite the smallness of the clusters, the extrapolation reveals an infinite-cluster-size value of $J_c/t^{\mathrm{VCA}}\approx1.48\pm0.28$, which agrees within error bars with the one obtained by numerically exact QMC methods \cite{Ass99} at half-filling, $J_c/t^{\mathrm{QMC}} \approx 1.45 \pm 0.05$.
Although only three cluster sizes for ladder-shaped clusters could be used, it is impressive to see that the extrapolated value nicely fits to these exact results, giving us confidence for the further results obtained by VCA. 
{Unfortunately, only very limited cluster sizes can be treated for this model, so that a finite-size extrapolation in most cases is not possible. 
However, due to the excellent results at half-filling, we expect that such an extrapolation would lead to results similar to those obtained} with numerically exact approaches such as tensor-network methods \cite{VC04,VMC08,COBV10}.
{I}t would be desirable to develop improved cluster solvers 
{for larger clusters, so that in future work a finite-size extrapolation within the VCA for arbitrary parameters would become possible}.

{At half-filling, the transition from PM to AF insulator is seen in the spin-spin correlator $\langle \vec{S} \cdot \vec{s} \rangle$, the local susceptibility $\chi_f$, and the DOS.}
Off half-filling ($1 > n  \gtrsim 0.8$), the AF solution becomes metallic and the staggered magnetization 
{decreases} when reducing the electron filling.
At some critical filling $n_c(J)$, the magnetization vanishes smoothly and the system continuously goes over to a PM metal. 

{The AF} metal possesses two different regions, one at weak coupling with a small Fermi surface and the other one at larger coupling showing closed pocket structures.
This needs further investigations, and we believe that using larger cluster sizes would be very helpful to better understand these features of the Fermi surface. 
Our results for the spin-spin correlator, the local susceptibility, and the DOS indicate the existence of Kondo singlets in the antiferromagnetic phase over a wide region of parameters, and they seem to persist down to weak coupling strengths.
Such an existence of Kondo singlets in the AF phase has been reported before, e.g., by variational Monte Carlo (VMC) \cite{WO07,ABF13}, real-space DMFT (rDMFT) \cite{PK15}, and dynamical cluster approximation (DCA) \cite{MA08}, and contradicts the Kondo breakdown scenario of mean-field theory \cite{ABF13}. 
However, when considering only AF Weiss fields in the VCA, we also identify a discontinuous transition within the AF phase at lower values of the filling, which turns continuous for $n \gtrsim 0.97$. 
This is similar to what has been reported in Refs.~\onlinecite{WO07,ABF13}, while rDMFT \cite{PK15} and DCA \cite{MA08} studies identify a continuous transition, albeit at finite temperature.
 
Note that as the VCA in its current formulation is limited to the electronic degrees of freedom and does not include excitations of the $f$-spins in the Green function, it is difficult to address the question of Kondo screening in a direct approach.
Including spin excitations in an extension of standard VCA might allow for a detailed future investigation of Kondo screening.

In addition to the AF properties, we investigated the phase diagram for possible s-wave SC phases.
This was recently reported by a DMFT+NRG approach \cite{BZV+13}, but, however, has not been found with DMFT or DMFT-like techniques using different impurity solvers since.
Already at half-filling using the $2\times2$ cluster a seemingly superconducting solution could be traced back to a solution of atomic mean-field nature.
Extending the variational space, additional stationary points in the self-energy functional were investigated. 
{Most of them can be discarded, since they either correspond to artificial mean-field solutions of isolated SC sites, or because they violate thermodynamic stability.}
To exclude artificial solutions, a parameter regime for the Weiss field strength was identified and investigated for the most promising stationary points using the $3\times2$ cluster.
In this parameter regime of small Weiss fields, the only stable solutions correspond to a PM metal and not to s-wave SC.
Also allowing for a possible coexistence with AF did not lead to stable s-wave SC solutions.
Interestingly, EOM for the s-wave pairing susceptibility show no $J$ dependence, which might be a further indication that s-wave SC is suppressed in the KLM on a square lattice.

In contrast, d-wave SC is stabilized over a wide range of the parameters treated here. 
However, also here care needs to be taken: including only the d-wave SC Weiss field at half-filling leads to a stable SC solution for weak couplings, which disappears when also including AF in the treatment.
Off half-filling, also in the presence of AF, d-wave SC is stabilized, and a coexistence region for couplings $J< J_c$ is found, which persists down to a critical density $n_c(J)$.
Inside this coexistence phase two regions were found, in which within the VCA treatment at small couplings AF and d-wave SC seem to be in competition, while close to $J_c$ both appear to act cooperatively.

In order to establish a connection to experiments, the model should be extended
{, in particular also concerning the existence of a quantum critical point.}
{Possibilities are to include additional terms suppressing Kondo singlets, which might lead to a tunable Kondo breakdown. 
An example are long-range hopping terms, which lead to frustration and have been used in the past \cite{CN10,MBA10, AFB14,WT15}.}
Another 
{possibility is to extend} the model to the Heisenberg-Kondo lattice model \cite{PPN07,XD08,HK13,YGL14}.
In both cases the modifications of the Kondo lattice model can also lead to changes in the superconducting channel \cite{AFB14, XD08, YGL14}. 
{The additional Heisenberg interaction leads to a} $d$-wave SC condensate, which consists of magnetic pairs of $f$ electrons on neighboring sites, as well as composite pairs containing two conduction electrons \cite{FC10}.
Recently, it was shown that singlet pairing correlations are enhanced in the vicinity of a Kondo-destruction quantum critical point \cite{PDIS15}, which motivates the study of $d$- and $s$-wave SC for the Heisenberg-Kondo lattice model.
However, {in order to be able to study these scenarios with the VCA, one needs to include the possibility to treat non-local interactions}.

In order to better connect with results obtained by techniques such as DMFT, dual fermions or DCA, the VCA study could also be extended by using clusters with additional bath sites.
This would allow to consider in addition to the spatial fluctuations also dynamical fluctuations between cluster and bath sites, an aspect which might help to better understand the discrepancy between DMFT+NRG and VCA results with respect to s-wave superconductivity.
Including bath sites in VCA also offers a route to investigate the transition from d- to p-wave superconductivity found in Ref. \onlinecite{Ots15} with a complementary cluster technique. 

\acknowledgments
We are grateful to the late T. Pruschke who initiated this project. 
We acknowledge helpful discussions with O. Bodensiek as well as computer support by the GWDG and the GoeGrid project. B.L. and S.R.M. are grateful for financial support by the Deutsche Forschungsgemeinschaft (DFG) through research unit FOR 1807, Project No. P7.


\appendix
\section{Detailed analysis of the paramagnet at half-filling}
\label{App:Details_PM_HF}
Within VCA, the spin-spin correlator between $f$-spins and $c$-electrons can be easily obtained as a functional derivative of the self-energy functional at its stationary point: $\langle \vec{S_i}\cdot\vec{s_i}\rangle=\partial\Omega/\partial J$.
For perfect N\'eel order, the f-spins and conduction electrons order and a spin-spin correlator of $-1/4$ per site would be expected as a staggered Weiss field breaks the local spin symmetry.

For this reason, the behavior of the local spin-spin correlator can be used to shed some light onto the possible breakdown of Kondo singlet formation: 
Throughout the paper, we assumed for $0<J<J_c$ that the ground state can be constructed by considering three different contributions, corresponding to a state without correlation between electron and f-spin, an AF ordered N\'eel state, and a Kondo singlet state, respectively.

In this picture, at least in the region in which $\langle \vec{S}\cdot\vec{s}\rangle<-0.25$, a considerable admixture of Kondo singlets should be present in the ground state in order to obtain an expectation value, which is lower than the one of the N\'eel state. 
Even for small coupling strengths where the spin-spin correlator is larger, the absence of a kink for $J<J_c$ casts the occurrence of a sudden Kondo breakdown into doubt.
Hence, considering only the local spin-spin correlator in principle still renders two scenarios possible:
either Kondo singlets survive down to small coupling strengths within the whole antiferromagnetic phase or they smoothly decay when reducing the coupling strength.

\begin{figure}[t!]
	\begin{center}
		\includegraphics[width=0.45\textwidth]{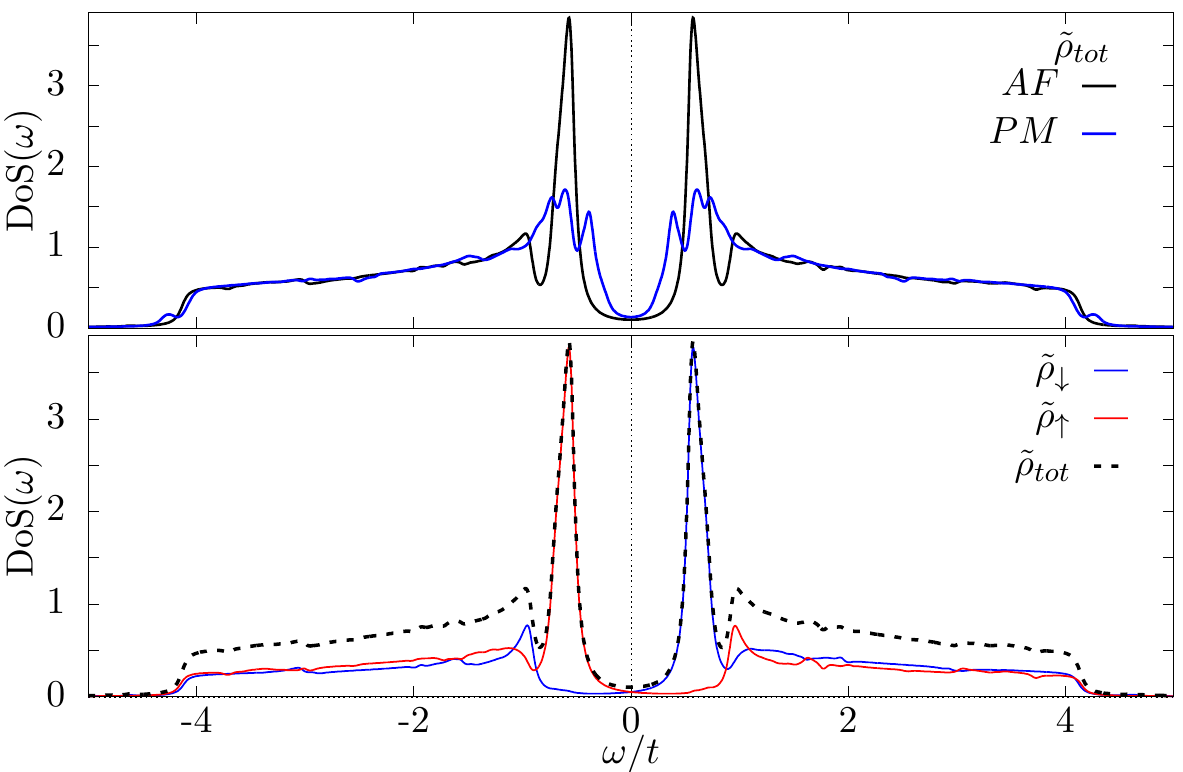}				
		\caption{Top: density of states (DOS) at $J/t=1.0$ and half-filling comparing the AF and PM insulator. Bottom: composition of the AF total DOS (dashed line) by staggered up- and down-spin densities $\tilde{\rho}_{\uparrow}$ (red) and $\tilde{\rho}_{\downarrow}$ (blue), which are explained in the text. The DOS has been obtained on a $k$-grid of $100\times100$ points by using the $3\times2$ cluster and an artificial broadening of $\eta=0.05$.}
		\label{fig_3x2_AF_HF_DoS_J1}
	\end{center}
\end{figure}
In the following, we briefly compare the density of states (DOS) of the PM and AF solutions.
Figure~\ref{fig_3x2_AF_HF_DoS_J1} shows the DOS at $J/t=1.0$ for both solutions.
The paramagnetic DOS consists in equal parts of the density of states of electrons with up- and down-spin.
In contrast, the antiferromagnetic DOS has different contributions for up- and down-electrons when focusing on one of the sublattices A and B. 
Due to the particle-hole symmetry of the DOS at half-filling, we limit the discussion to the electronic part only and focus on the region $\omega\in (-1.5,1.5)$.
Note that although Fig.~\ref{fig_3x2_AF_HF_DoS_J1} suggests finite spectral weight at $\omega=0$, this is only due to the large broadening ($\eta=0.05$). 
Both solutions indeed show a quasiparticle gap when reducing the broadening.
In the AF solution, the DOS is characterized by a sharp resonance right at the border of the quasiparticle gap, followed by a side resonance at larger frequency.
The paramagnet has a smaller quasiparticle gap and three distinct peaks can be identified.
Compared to the antiferromagnet, where the main resonance is separated from the side resonance by a dip, the three peaks of the paramagnet are quite close to each other.

\section{Variation of the cluster hopping strength}
\label{App:Hopping}

It is shown in Appendix~\ref{swaveHF} that it is crucial to include the intra-cluster hopping $t^{\prime}$ into the set of variational parameters when investigating superconductivity, in order to exclude unphysical solutions.
Here, we show the influence of $t^{\prime}$ on the normal and antiferromagnetic solution.
\begin{figure}[t!]
	\begin{center}
		\includegraphics[width=0.495\textwidth]{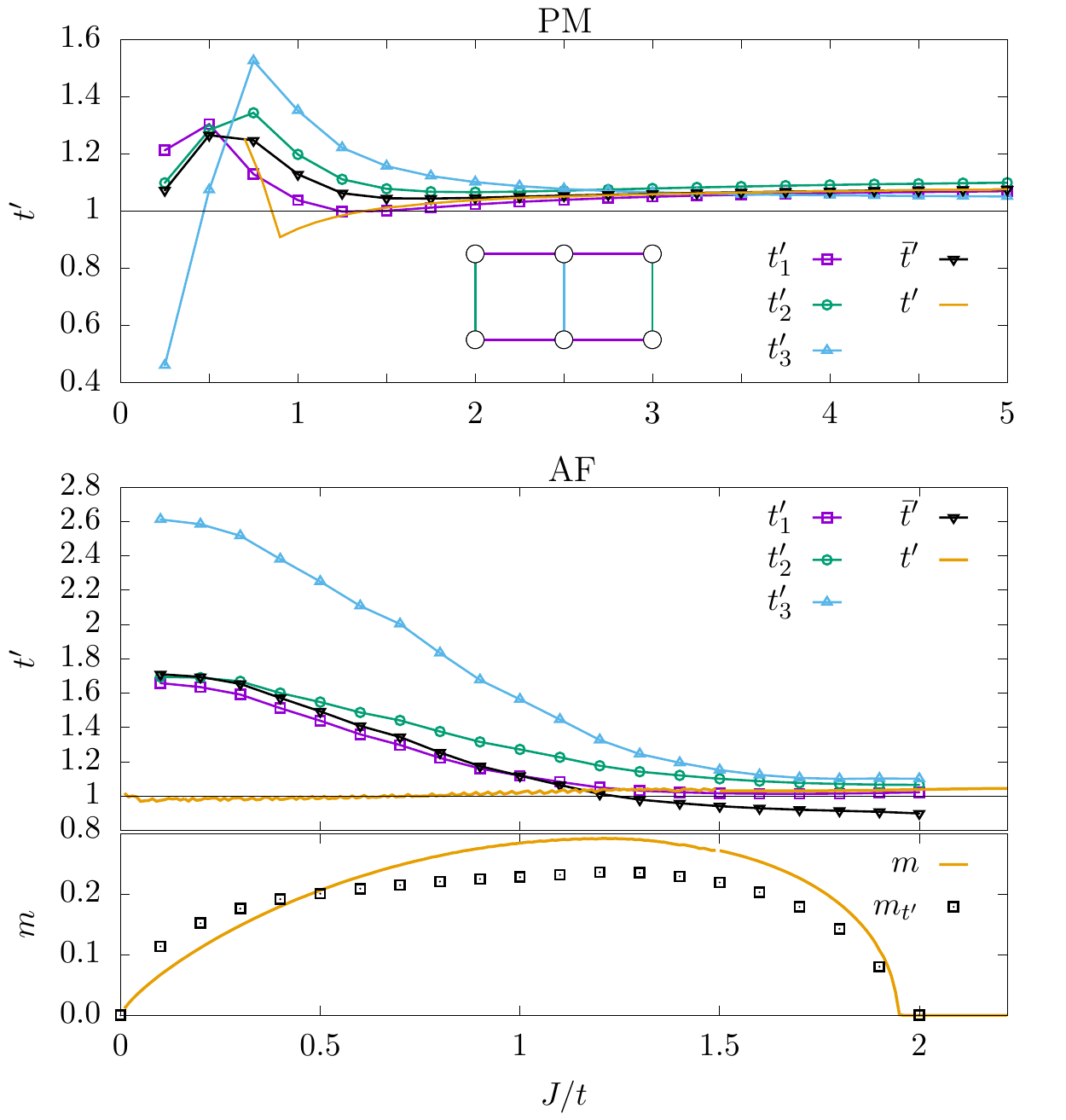}
		\caption{Variation of the cluster hopping parameters inside the $3\times2$ cluster at half-filling. Using three different cluster hopping parameters $t_1^{\prime},t_2^{\prime},t_3^{\prime}$ as indicated by different colors in the inset instead of only one parameter $t^{\prime}$ leads to a reduction in the free energy, but not to qualitative differences for the paramagnetic (top panel) and antiferromagnetic case (lower panel). For the latter, especially the value $J_c/t$ for the onset of antiferromagnetism and the value of the staggered magnetization do not change much when using anisotropic cluster hopping (black circles) in comparison to the result when varying only $t^\prime$ (continuous yellow line). 
		}
		\label{fig:hops_HF}
	\end{center}
\end{figure}

In the top panel of Fig.~\ref{fig:hops_HF}, the values of the cluster hopping parameters are shown at the stationary point in the paramagnetic case at half-filling as a function of coupling strength.
When considering isotropic hopping with strength $t^{\prime}$ on the cluster, $t^{\prime}$ shows a cusp anomaly at $J/t\sim0.9$.
This coincides with the point, where the paramagnetic solution has zero quasiparticle gap and hence changes from a paramagnetic insulator at large coupling to a paramagnetic metal.
Even when considering anisotropic hopping on the cluster with hopping strengths $t^{\prime}_1,t^{\prime}_2,t^{\prime}_3$ as shown in the sketch in Fig.~\ref{fig:hops_HF}, the anomaly shifts to smaller coupling strength $J/t$, but persists.
The transition can also be seen in the arithmetic mean of the three cluster hopping terms $\bar{t^{\prime}}=(4t_1^{\prime}+2t_2^{\prime}+t_3^{\prime})/7$.

When allowing for antiferromagnetism by adding a staggered Weiss field, at half-filling the system is insulating for all coupling strengths and the cusp singularity in $t^{\prime}$ is absent (see lower panel of Fig.~\ref{fig:hops_HF}).
This is still the case when choosing anisotropic hopping parameters on the cluster.
Furthermore anisotropic cluster hopping only leads to minor changes of the critical coupling strength $J_c/t$, where AF sets in, and of the value of the staggered magnetization $m$. 

\begin{figure}[t!]
	\begin{center}
		\includegraphics[width=0.495\textwidth]{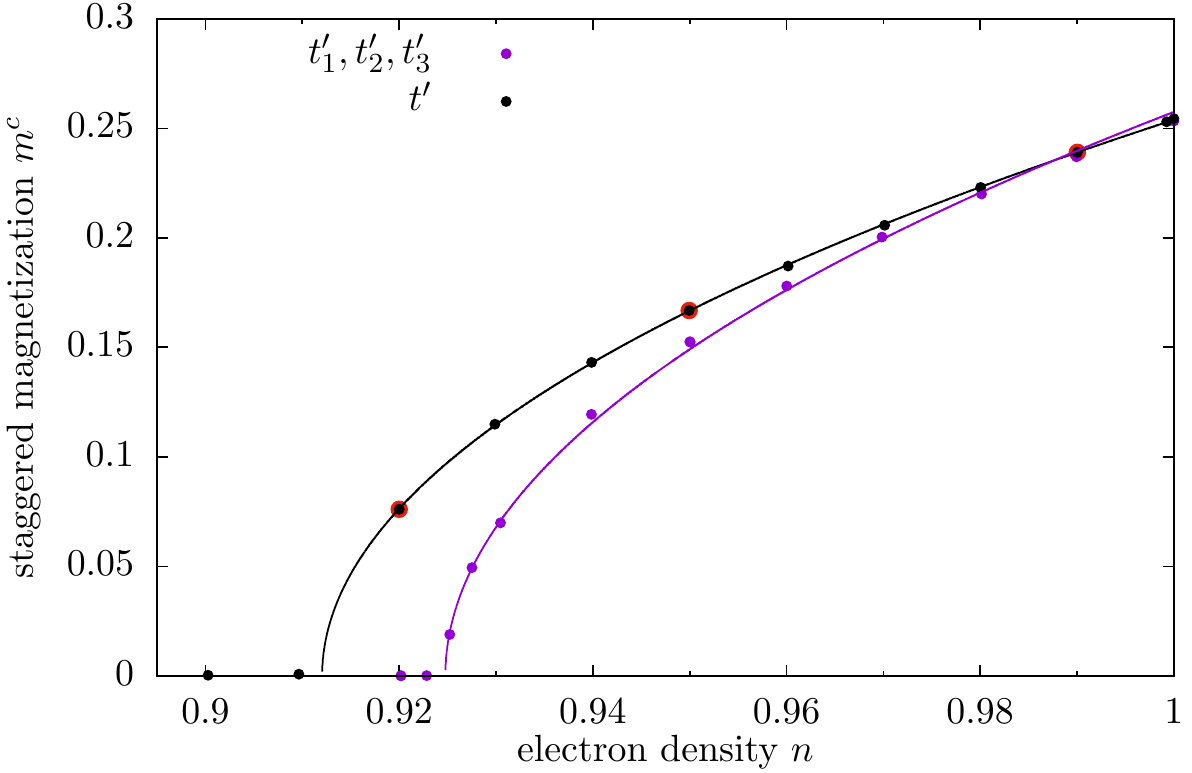}
		\includegraphics[width=0.495\textwidth]{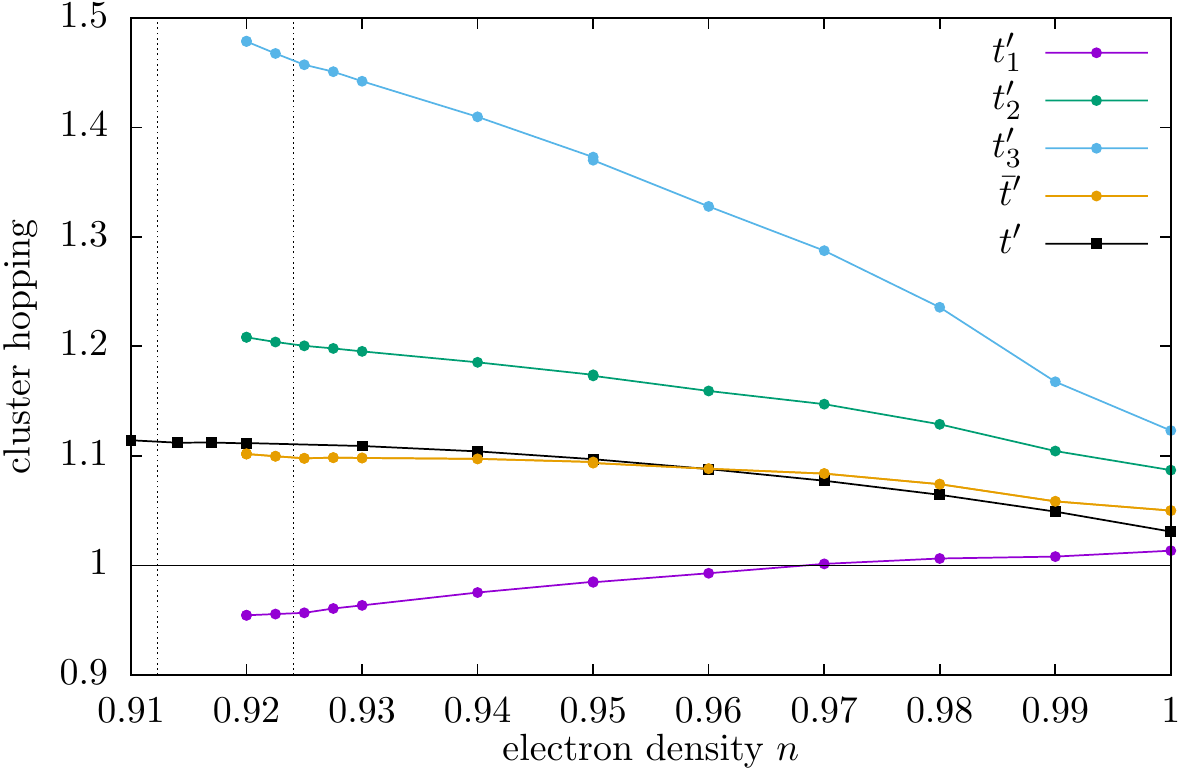}
		\caption{Variation of the cluster hopping parameters inside the $3\times2$ cluster at $J/t=1.6$ as a function of electron filling. Variation of three cluster hopping parameters $t_1^{\prime},t_2^{\prime},t_3^{\prime}$ instead of one parameter $t^{\prime}$ leads to a slightly higher critical filling $n_c$ (top panel), where AF sets in, but not to qualitative differences. Lines indicate square root fits. Especially the spreading of the different values of the hopping parameters increases when approaching the transition (lower panel), although their mean value is still comparable to $t^{\prime}$.}
		\label{fig:hops_offHF}
	\end{center}
\end{figure}

Away from half-filling, allowing for anisotropic cluster hopping leads to an increase of the critical electron filling $n_c$, where AF sets in, see top panel of Fig.~\ref{fig:hops_offHF}.
The AF region is thereby reduced, but the shape of the staggered magnetization curve as a function of electron density $n$ stays qualitatively the same.
When considering the values of the cluster hopping terms in the lower panel, it can be seen that the anisotropy in the cluster increases when reducing the electron density, although the arithmetic mean value $\hat{t^{\prime}}$ is still comparable to $t^{\prime}$.

\section{Luttinger's sum rule and Kondo breakdown}
\label{App:LSR}
{In this appendix, we discuss in more detail the relation of the FS to the spectral function $A(\mathbf{k},\omega)$ shown in Sec.~\ref{KLM_KB}. 
A consistency check for the validity of these results is to consider Luttinger's sum rule.
To do this, in the KLM one needs to take care due to the strongly correlated character of the system.}
In the following, we approach this within the VCA.\newline

For Fermi liquids at zero temperature, Luttinger's theorem states that the volume of the electrons' Fermi surface amounts to the number of electrons \cite{Lut60}.

A non-perturbative proof of Luttinger's theorem for the Kondo lattice model was given by \textit{Oshikawa} \cite{Osh00}, but since it assumes Fermi liquid behavior, it cannot be directly transferred to the AF metal for intermediate values of $J/t$. 
Nevertheless, the theorem can be extended to the case of non-Fermi liquids (see, e.g., Ref.~\onlinecite{Dzy03}).

In addition, it was shown by \textit{Ortloff et al.} \cite{OBP07} that the Luttinger theorem can be violated by conserving approximations based on the self-energy functional theory of \textit{Potthoff} such as VCA.
For finite systems, they showed that the (extended) Luttinger sum rule should rather be formulated as
\begin{equation}
2\sum_k\sum_m\alpha_m(\mbf{k})\Theta(\omega_m(\mbf{k})) = 2\sum_{\mbf{k}}(P_{\mbf{k}} - Z_{\mbf{k}}),
\label{Eq:LSR_1}
\end{equation}
where $P_{\mbf{k}}=\sum_m \Theta(\omega_m(\mbf{k}))$ denotes the sum of all positive poles $\omega_m({\mbf{k}})$ and $Z_{\mbf{k}}=\sum_n\Theta(\zeta_n(\mbf{k}))$ the sum of all positive zeros $\zeta_n(\mbf{k})$ of the Green function 
$$G_{\mbf{k}}(\omega) = \sum_m\frac{\alpha_m(\mbf{k})}{\omega-\omega_m(\mbf{k})}.$$
Right at the Fermi energy $\omega=0$ the sum of all poles constitutes the Fermi surface, whereas the sum of all zeros constitutes the Luttinger surface \cite{Dzy96,Ros07}.
The spectral function is $A(k,\omega+i\eta) = -\frac{1}{\pi}\mathrm{Im}~G_k(\omega+i\eta) = -\sum_m\frac{\alpha_m(k)\eta}{(\omega-\omega_m(k))^2+\eta^2}$, which means that the Fermi surface can be read off from $\lim_{\eta\rightarrow0^+}A(k,0+i\eta)$. 
However, poles with small weight $\alpha_m$ can be easily missed when reconstructing the Fermi surface from the spectral function, so that care needs to be taken.

The left hand side of Eq.~\eqref{Eq:LSR_1} amounts to the summation of the weight of the spectral function up to the Fermi energy.
Within our VCA approach the Green function only includes the propagation of conduction band electrons, which means that 
\begin{equation}
n_e^c=\frac{2}{N_k}\sum_{\mbf{k}}\int_{-\infty}^0 d\omega\,A(\mbf{k},\omega+i0^+)
\label{Eq:LSR_2}
\end{equation}
is always fulfilled due to thermodynamical consistency of the approach ($N_k$ denotes the number of $\mbf{k}$-points in the Brillouin zone).
Since between two poles $\omega_{m}$ and $\omega_{m+1}$ there exists exactly one zero $\zeta_{\tilde{m}}$, the right-hand side of Eq.~\eqref{Eq:LSR_1} amounts to sum all zeros $\zeta^{*}$ between the largest negative and the smallest positive pole if $\zeta^{*}>0$ \cite{ESO11}.
In other words, the electron density is obtained by \cite{Ros07}
\begin{equation}
n_e = \frac{2}{N_k}\sum_{\mbf{k}}\Theta\left(G(\omega=0,\mbf{k})\right).
\label{Eq:LSR_3}
\end{equation}

\begin{figure}[t!]
	\begin{center}
		\includegraphics[width=0.495\textwidth]{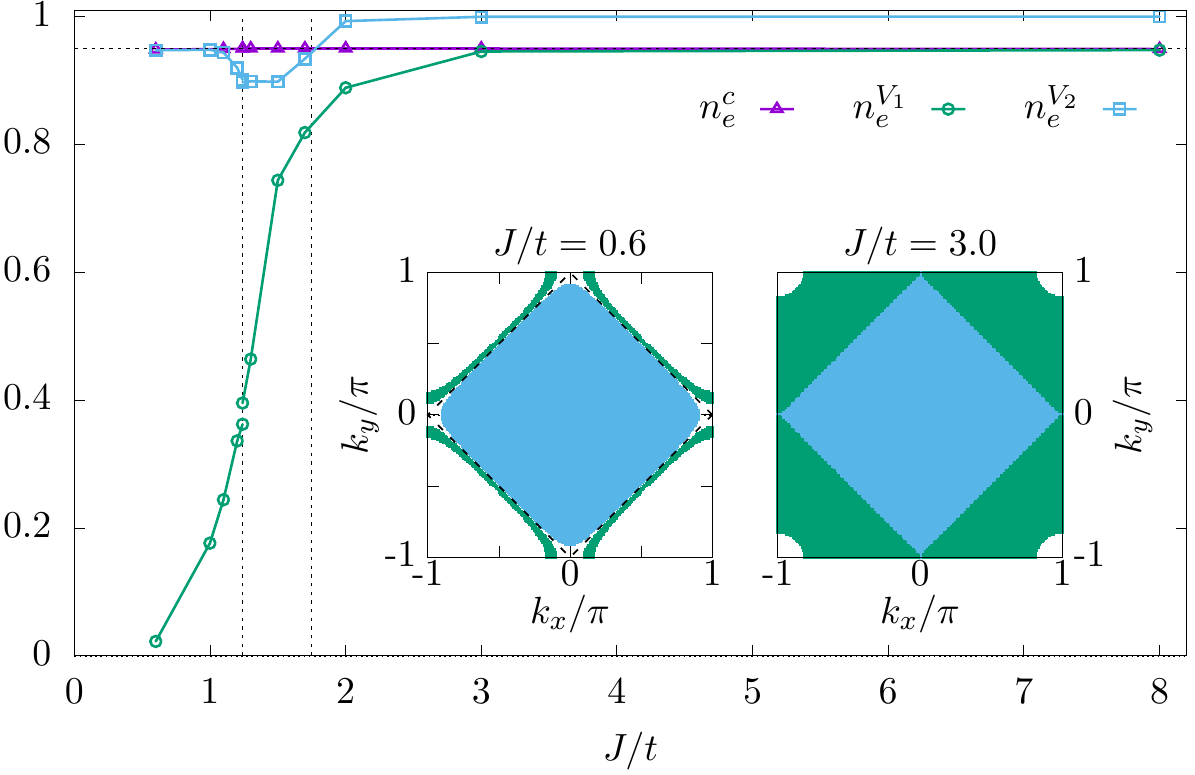}
		\caption{Electron densities $n_e^{V1/V2}$ calculated according to Eq.~\eqref{Eq:LSR_3} and $n_e^c$ according to Eq.~\eqref{Eq:LSR_2} as a function of coupling strength $J/t$ for the $3\times2$ cluster. 
		$n_e$ is split into the contribution of the ``outer'' part of the Brillouin zone ($n_e^{V_1}$) and the ``inner'' part ($n_e^{V_2}$).
		The solid horizontal line marks the predefined electron filling $n_e=0.95$, the dashed vertical lines indicate the phase transitions from AF${}_1$ to AF${}_2$ and from AF${}_2$ to PM.
		In the insets, $\Theta(G(\omega=0,\mbf{k}))$ is shown inside the Brillouin zone for small and strong coupling values. Contributions to $n_e^{V1(2)}$ are indicated by green (blue) color.}
		\label{fig:LSR_3x2}
	\end{center}
\end{figure}

The value of this sum is shown in Fig. \ref{fig:LSR_3x2} as a function of coupling strength $J/t$. 
An additional complication occurs once antiferromagnetic order is present since there, the Brillouin zone is halved compared to the paramagnetic case at large coupling.
When plotting the same region in $\mbf{k}$-space, the Brillouin zone is captured twice, indicated by the dashed line in the inset of Fig. \ref{fig:LSR_3x2} for $J/t=0.6$.
For this reason, we plot the sum over the ``inner'' ($n_e ^{V_2}$) and ``outer'' ($n_e ^{V_1}$) half of the BZ as shown in the inset.

For a large Fermi surface, where the $f$-spins participate in the charge transport via Kondo singlet formation, the electron density is $n_e=n_e^c + 1$.
As shown in Fig.~\ref{fig:LSR_3x2}, this is nicely fulfilled by our VCA results.
In case of a small Fermi surface, the $f$ spins are frozen out in an antiferromagnetic order and $n_e = n_e^c.$
As one can see from Fig.~\ref{fig:LSR_3x2}, $n_e^{V_2}=n_e^c$ is fulfilled in the AF${}_1$ region and the contribution of $n_e^{V_1}$ tends to zero for weak-coupling strength.
In between, a drastic reduction of $n_e^{V_1}$ takes place and it seems that the FS with many hole structures in the AF$_2$ region weakly violates Luttinger's sum rule.
For a detailed study of this effect, one 
{needs a finite-size extrapolation in the cluster size, which is not possible since only small clusters can be treated for this model.}

\begin{figure}[t!]
	\begin{center}
		\includegraphics[width=0.495\textwidth]{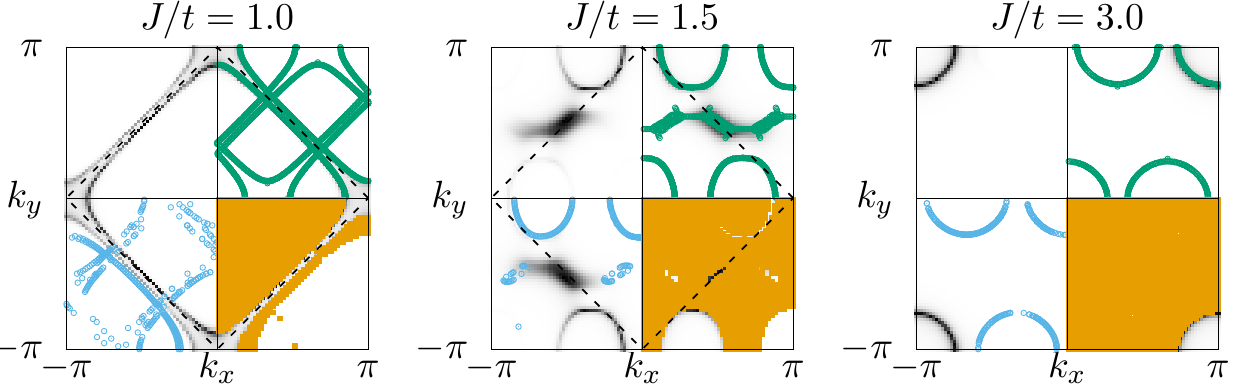}
		\caption{The spectral function at the Fermi energy with an artificial broadening of $\eta=0.01 t$, $A(\mathbf{k},\omega=0+i\eta)$ for three coupling strengths in the AF${}_1$,AF${}_2$ and PM regions (from left to right) at $n=0.90$.
		{We overlay to the data} the Fermi surface (top right), the Luttinger surface (bottom left), and the region with $\zeta^{*}>0$ (bottom right).	
		}
		\label{fig:LS_FS_3x2}
	\end{center}
\end{figure}

In Fig.~\ref{fig:LS_FS_3x2} we show how the FS, the Luttinger surface, and the spectral function relate to each other. 
We display both, the Fermi and the Luttinger surfaces, as well as $n_e$ according to Eq.~\eqref{Eq:LSR_3} for three characteristic values of the coupling strength.
The determination of $n_e$ is very stable within VCA since one does not need to resort to any kind of artificial broadening.
{For better comparison, these results are overlayed to the ones for the} spectral function at the Fermi energy [$A(\mathbf{k},\omega=0+i\eta),~ \eta=0.01t$], which is often used to extract the FS, especially within techniques, which do not have direct access to the poles of the propagator \cite{MBA10}.
Here, we use the so-called $\mathbf{Q}$-matrix formalism \cite{AAPH06a} and directly obtain all poles and weights of the (cluster) Green function.
This includes poles with negligibly small weight
{; note that it is these poles which lead to additional structures when comparing to} $A(\mathbf{k},0+i\eta)$, otherwise the behavior of the FS and of $\lim_{\eta\rightarrow0^+}A(\mathbf{k},0+i\eta)$ is identical, which justifies using the spectral function for the discussion in Sec.~\ref{KLM_KB}.
Note that the small but finite broadening of $\eta=0.01$ in Fig.~\ref{fig:LS_FS_3x2} makes it difficult to determine the Luttinger surface precisely.
Nevertheless, due to the small weight of some poles, zeros occur in their close vicinity and show up in the Luttinger surface (see, e.g., the plot for $J/t=3.0$ in Fig.~\ref{fig:LS_FS_3x2}).  
For this reason, poles with (very) small weight do not significantly add to the Luttinger sum Eq.~\eqref{Eq:LSR_1}.
In case of the PM, the poles that contribute finite weight to the spectral function contribute and the overall sum of Eq.~\eqref{Eq:LSR_1} corresponds to an electron density of $n_e=1.90$.

For the case of doped Mott insulators it was shown that the intricate interplay of zero and pole surfaces of the Green function are closely related to main features of the Mott physics \cite{SMI09,SMI09a}.
However, it is still debated whether the Luttinger theorem can be applied to strongly correlated metals \cite{SLGB96,GEH00,Ros07,SPC07,KP07,KP08,Far07,SSC+12}.
A further investigation of this issue for the Kondo lattice model would be interesting.
However, it would require the treatment of larger clusters and is left for future studies.

\section{Detailed analysis of s-wave superconductivity within VCA}
\label{Details_swave}
In the following, we show additional detailed calculations for putative s-wave SC in the KLM using VCA.
First of all, an s-wave solution at half-filling is identified and discussed in Appendix~\ref{swaveHF}.
To complement the discussion of putative s-wave SC off half-filling in Sec.~\ref{swaveOffHF}, we analyze possible additional solutions when treating AF and s-wave SC on equal footing in Appendix~\ref{swave_AF_off}.

\subsection{s-wave superconducting solutions at half-filling}
\label{swaveHF}
For small coupling strengths ($J/W\lesssim0.15$) \textit{Bodensiek et al.} report weak indications for superconductivity even at half-filling \cite{BZV+13}.
However, it should be noted that the anomalous expectation value is very small in this region ($\Phi_s\approx 0.002$) and DMFT suffered from convergence problems which made it difficult to stabilize the solution \cite{Bod16}.
Furthermore, this is the region where other DMFT studies without superconducting baths found an antiferromagnetic insulator \cite{PP07,OKK09}, so that the interplay between superconductivity and antiferromagnetism should be investigated in detail.

Indeed, when using only an s-wave SC Weiss field within VCA, the SEF possesses a minimum with respect to the superconducting Weiss field strength $D_s$ for large values of $J/t$, corresponding to a superconducting solution. 
This is unexpected since in the strong-coupling limit Kondo singlets should form, leading to an insulator instead. 
At weak couplings, the SEF possesses an additional maximum for larger $D_s$ and yet another minimum with larger free energy. 
This means, that -- according to Potthoff's rules for the selection of stationary points \cite{Pot12} -- still the minimum with lower free energy should be considered as the correct stationary point.

\begin{figure}[t!]
	\begin{center}
		\includegraphics[width=0.495\textwidth]{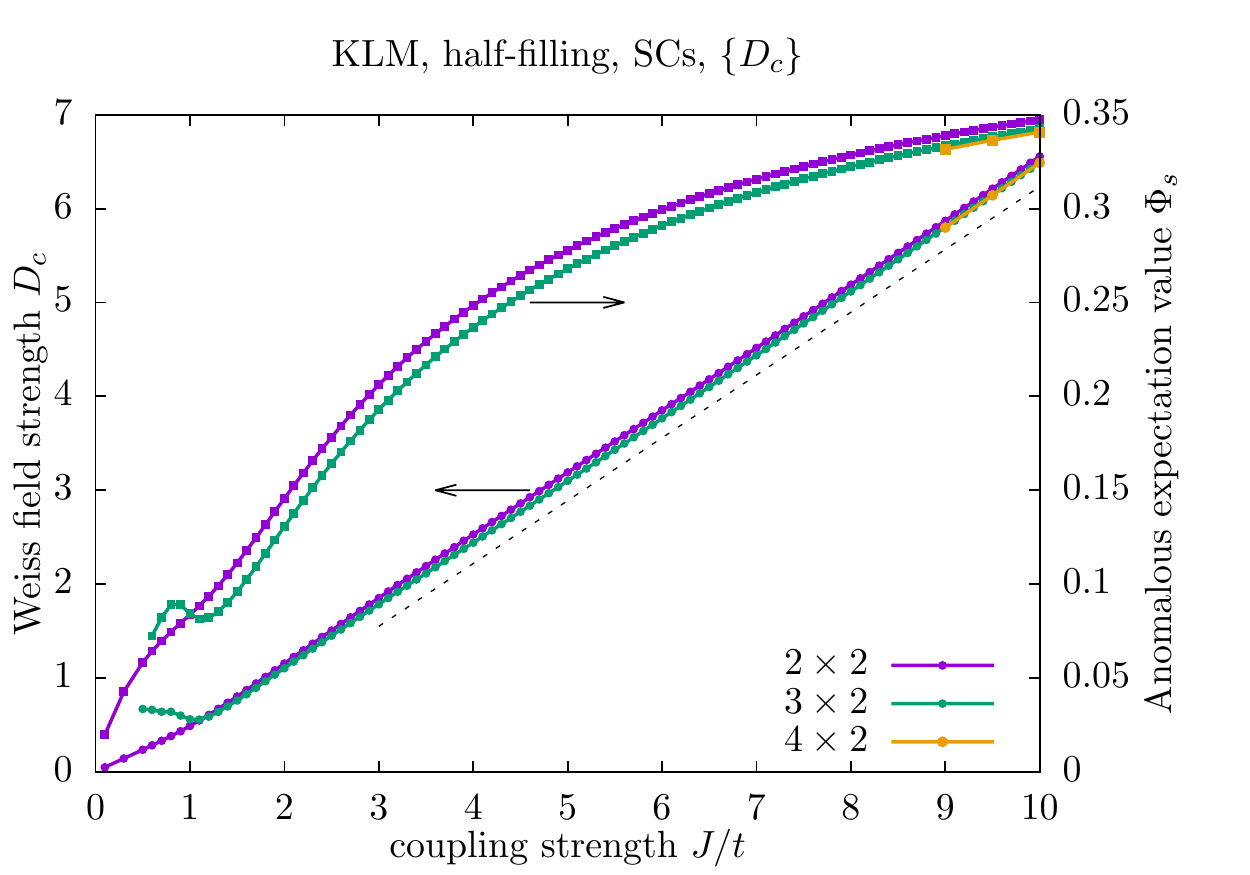}
		\includegraphics[width=0.495\textwidth]{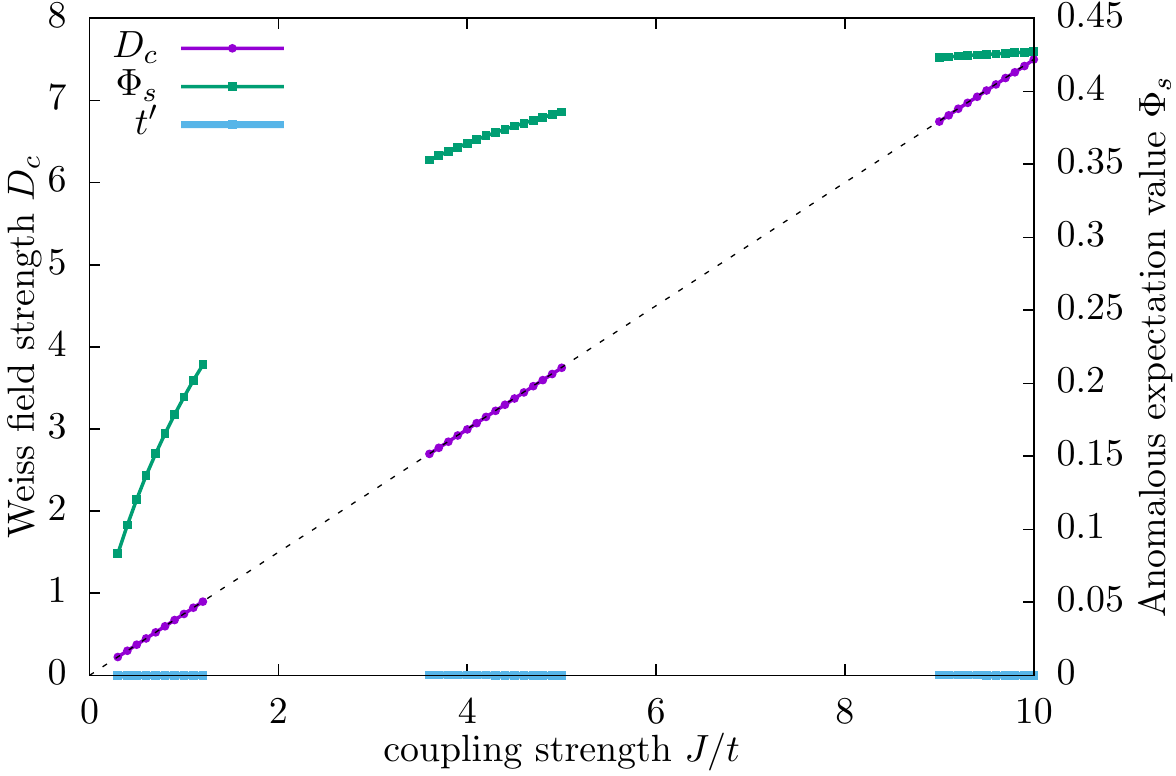}
		\caption{Results for the SC Weiss field $D_c$ and for the anomalous expectation value $\Phi_s$ at half-filling when considering s-wave SC. The set of variational parameters includes $\{ D, \mu^{\prime}\}$ (top panel) and $\{ D, \mu^{\prime}, t^{\prime}\}$ (bottom panel). In the bottom panel only a $3\times2$ cluster is considered whereas in the top panel the three indicated cluster sizes are shown. Note the hopping strength within the cluster is found to be zero within error bars and the Weiss field strength corresponds to the one of an isolated Kondo site, $D_c=3/4 J$.}
		\label{fig_3x2_SC_D_tc}
	\end{center}
\end{figure}

In contrast to the superconducting phase that was found in Ref.~\onlinecite{BZV+13} the solution at this stationary point exists for all coupling strengths (see the top panel of Fig.~\ref{fig_3x2_SC_D_tc}).
The strength of the Weiss field at the stationary point is quite large and for couplings larger than the hopping strength $J/t\gtrsim1$, it is even proportional to $J$. 
Furthermore, the Weiss field strength is for strong coupling independent of the cluster size as seen when comparing the results for the three clusters used in Fig.~\ref{fig_3x2_SC_D_tc}.
This indicates a local character of the solution, where the physics is dominated by the (local) Weiss field.

One key aspect of the approximation within VCA is the choice of the space of variational self-energies, which is given by the cluster geometry of the reference system and the choice of a set of one-body parameters which are used as variational parameters.
Only in few VCA studies the variation of the cluster hopping terms $t^{\prime}$ leads to significant improvements (see, e.g., Ref.~\onlinecite{RLRT15}). 
However, in the following it will be shown that including the hopping on the cluster as variational parameter in the case of the KLM is crucial for investigating s-wave superconductivity, as it leads to qualitatively different behavior.

When varying $D, \mu^{\prime}$, and $t^{\prime}$ for large $J/t$ (e.g. $J/t=10$ in Fig. \ref{fig_3x2_SC_D_tc}), and starting with the solution found previously with a large value of the pairing field, one arrives at a stationary point where the hopping on the cluster is zero. 
Nevertheless, the superconducting pairing $D$ on the cluster is finite and leads to a finite anomalous expectation value at the stationary point. 
This would correspond to a cluster ground state with local 'singlet-like' states of empty and doubly occupied sites, where single electrons are localized as they cannot move within the cluster. 
For all coupling strengths, the value of the Weiss field is given by $D_c=3/4 J$, which is exactly the Kondo singlet binding energy. 

At half-filling the only s-wave superconducting solution is therefore the somewhat artificial superconductor that consists of local Cooper pairs without any electron hopping between the sites.
Hence, no indication for a superconducting solution caused by correlation effects that would be comparable to the one proposed in Ref.~\onlinecite{BZV+13} is obtained.

\subsection{Allowing for s-wave superconductivity and antiferromagnetism off half-filling}
\label{swave_AF_off}
As no superconducting solution has been found so far, a possible next step would be to take additionally antiferromagnetism into consideration.
Naturally, one might think of antiferromagnetism and superconductivity as being two competing phases.
Treating antiferromagnetism and superconductivity on the same footing would then not change the results obtained until now.

\begin{figure}[t!]
	\begin{center}
	\includegraphics[width=0.45\textwidth]{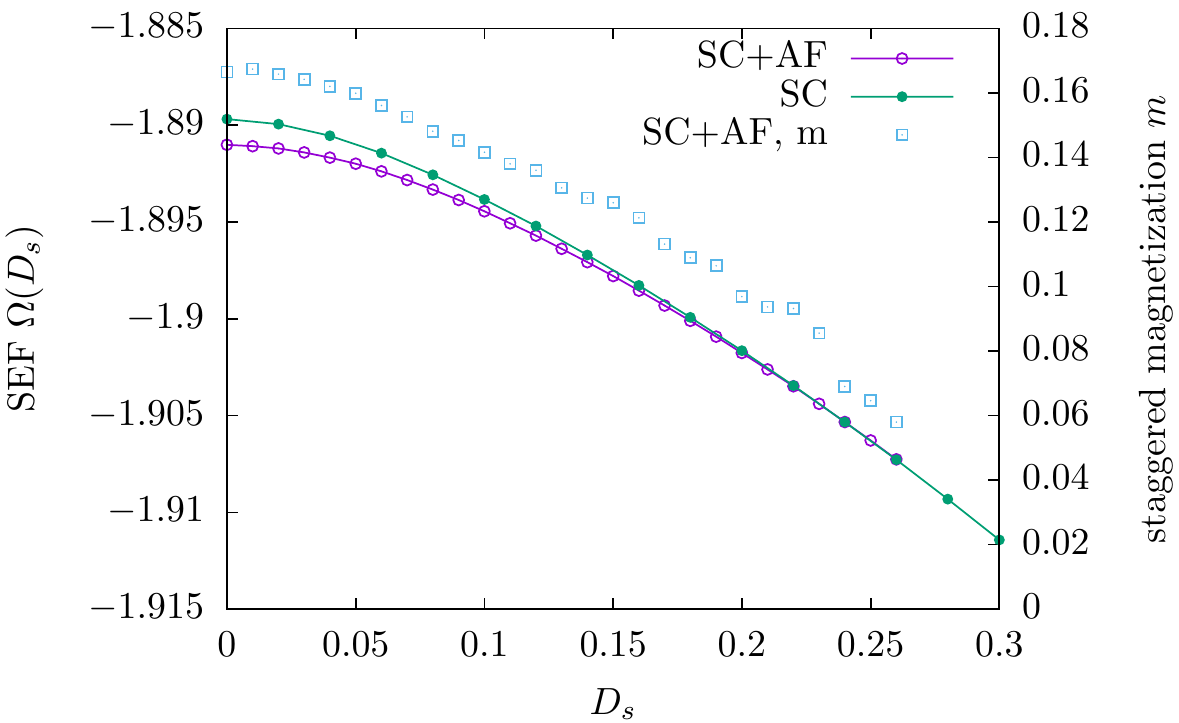}
	\caption{Results for the SEF when considering s-wave SC and AF order. Comparison of the results as a function of the strength of the superconducting Weiss field with ('SC+AF') and without ('SC') including $M_c$ into the set of variational parameters. The staggered magnetization of the 'SC+AF' solution diminishes for intermediate values of $D_s$, hence the self-energy functional approaches the 'SC' result.} 
	\label{fig_J16_SCsAF}
	\end{center}
\end{figure}

However, in both mechanisms for s-wave SC put forward by the authors of Ref.~\onlinecite{BZV+13}, spin fluctuations play an important role.
A first indication of the effect that the addition of an AF to the SC Weiss field might have, can be gained from considering the self-energy functional as a function of $D_s$ in both cases (see Fig.~\ref{fig_J16_SCsAF}).
Here, the solutions for the sets of variational parameters $\{\mu,\mu^{\prime},t^{\prime}\}$ and $\{\mu,\mu^{\prime},t^{\prime},M_c\}$ are compared as a function of the strength of a local SC Weiss field.
Due to the outcome of the discussion in the previous paragraph, the considered Weiss field strengths are in the region of  'sufficiently small' $D_s$, such that $t^{\prime}\neq 0$. 

The self-energy functional shows in both cases only one stationary point, namely, at $D_s=0$, which are the paramagnetic and antiferromagnetic solutions that are already known from Sec.~\ref{KLM_AF}; the antiferromagnetic solution has a smaller free energy and is therefore realized in the system.
When increasing the strength of the superconducting Weiss field $D_s$ the value of the antiferromagnetic Weiss field $M_c$ decreases and finally goes to zero.
The question as to whether a superconducting solution can be realized can hence be reformulated to asking whether the change from the AF to the PM solution can be found in some parameter regime to lead to an additional stationary point.

Investigating the self-energy functional as a function of $D_s$ for $J=1.6t\hat{=}0.2W$ and $2.4t\hat{=}0.3W$ for electron fillings down to $n=0.85$ leads to results similar to those shown in Fig.~\ref{fig_J24_SCs_SEF_t}.
At least for these coupling strengths and close to half-filling no AF is found for intermediate and strong superconducting Weiss fields, which means that the search for stationary points amounts to the situation without additional antiferromagnetic Weiss field.
This means that even in the region where in Ref.~\onlinecite{BZV+13} the largest anomalous expectation value was reported, no s-wave superconducting solutions are identified.

Due to the large variational space and comparatively high computational cost of the $3\times2$ cluster, the search for s-wave superconductivity has been restricted to the region of the phase diagram which seemed to be the most promising.
Based on this, the existence of s-wave superconductivity in the KLM within VCA is not excluded, but in the investigated region the calculations do not show evidence for physical s-wave SC solutions.

\section{Equations of motion for the pairing susceptibility}
\label{EOM:appendix}

As seen in Sec.~\ref{SCs} and Appendix~\ref{Details_swave}, there is no evidence for local s-wave SC in the VCA treatment of the KLM.
In addition, the time derivative in the EOM for the pairing susceptibility
\begin{equation} 
\chi_{jklm}(\omega) := 
\int_0^{\infty} dt \, e^{-i \omega t} 
 \left \langle \Delta_{jk}^{\dagger}(t) \Delta_{lm}^{\phantom{\dag}}(0) \right \rangle 
\label{eq:pairingsusc}
\end{equation}
  in the s-wave channel in Sec.~\ref{EOM} does not depend on the interaction $J$, further indicating the absence of s-wave SC. 
Here, we {complement the discussion of Sec.~\ref{EOM} by some computational details.}

For the derivation of the EOM, we need to take into account the interaction part of the KLM [Eq.~\eqref{KLM}], and in addition the kinetic part, which we denote by
\[
H_{\rm kin} = \sum_{\langle i,j \rangle, \alpha} c^{\dagger}_{i,\alpha} c^{\phantom \dagger}_{j,\alpha} \, 
\]
(we have set $t_{ij} \equiv 1$).

We now consider the EOM of the pairing susceptibility Eq.~\eqref{eq:pairingsusc}, which builds on the Green function
\begin{equation}
C(t) = \langle \Delta_{ij}^{\dagger}(t)\Delta^{\phantom \dagger}_{i^{\prime}j^{\prime}}(0)\rangle 
\label{eq:Green_appendix}
\end{equation}
for the pairing operators
$$\Delta_{ij}^{\dagger} = F(i,j)c^{\dagger}_{i\uparrow}c^{\dagger}_{j\downarrow}.$$ 
The function $F(i,j)$ takes into account the geometry of the pairing order parameter and is given by Eq.~\eqref{Delta}.

The first step to calculate the expression of the EOM is to differentiate Eq.~\eqref{eq:Green_appendix} with respect to time \cite{Pru14}. 
The time derivative of $\Delta^{\dag}_{ij}(t)$  is then obtained by means of the Heisenberg equation
\begin{equation}\label{EOM_Gap:appendix}
\frac{d}{dt} \Delta^{\dag}_{ij}(t)  =-i  \big[c^{\dag}_{i,\sigma}(t)c^{\dag}_{j,\bar \sigma}(t) , H_{I / kin} \big] \, , 
\end{equation}
where the EOM takes into account the dynamics driven by the interaction part or the kinetic part by considering $H_{I}$ or $H_{\rm kin}$, respectively.
The kinetic part, in general, will not vanish.
However, since it is the tight-binding part of the model describing non-interacting electrons, it does not contain any information about the coupling $J$, and we do not expect it to induce superconductivity.
For example, for the s-wave channel, we obtain
\[
\begin{split}
  -i  \big[c^{\dag}_{i,\sigma}(t)c^{\dag}_{j,\bar \sigma}(t)& , H_{\rm kin} \big] =\\
  &i \sum_m \left( c^\dagger_{i,\sigma}(t) c^\dagger_{m, \bar{\sigma}}(t) + c^\dagger_{m,\sigma}(t) c^\dagger_{i, \bar{\sigma} }(t) \right). 
\end{split}
  \]
This contributes to the dynamical properties of the pairing susceptibility, but it will be more interesting to consider the effect of the interaction term on the EOM.
To do so, we insert Eq.~\eqref{KLM} into Eq.~\eqref{EOM_Gap:appendix} and make use of the additivity of the commutator in order to split the calculation into three parts:
\begin{eqnarray}
\label{Kterms}
R1 &:=& i F(i,j) \frac{J_z}{4} \sum\limits_{l,s} \left[\Delta_{ij}(t), n_{l,s} N_{l,s} \right], \nonumber \\
R2 &:=&  i F(i,j) \frac{J_z}{4}\sum\limits_{l,s} \left[\Delta_{ij}(t), -n_{l,\bar s} N_{l,s} \right],\\
R3 &:=& i F(i,j) \frac{J_{\perp}}{2}\sum\limits_{l,s} \left[\Delta_{ij}(t), c^{\dag}_{l,s} c^{\dag}_{l,\bar s} 
f^{\dag}_{l,\bar s} f_{l,s} \right] , \nonumber
\end{eqnarray}
which produces 
\newline
\begin{eqnarray*}
R1 &=& i F(i,j) \frac{J_z}{4}  \biggl( c^{\dag}_{j,\bar\sigma}(t) c^{\dag}_{i,\sigma}(t)  N_{j,\bar \sigma}(t) \\
&&- c^{\dag}_{i,\sigma}(t) c^{\dag}_{j,\bar\sigma}(t) N_{i,\sigma}(t) \biggl)\\
&=& -i F(i,j)\frac{J_z}{4} c^{\dag}_{i,\sigma}(t)  c^{\dag}_{j,\bar\sigma}(t)\biggl(N_{j\bar\sigma}(t)+N_{i\sigma}(t)\biggl)      , 
\end{eqnarray*}
\begin{eqnarray*}
R2 &=& -i F(i,j) \frac{J_z}{4}  \biggl( c^{\dag}_{j,\bar\sigma}(t) c^{\dag}_{i,\sigma}(t)  N_{j, \sigma}(t) \\
&&- c^{\dag}_{i,\sigma}(t) c^{\dag}_{j,\bar\sigma}(t) N_{i,\bar \sigma}(t) \biggl)\\
&=& i F(i,j)\frac{J_z}{4} c^{\dag}_{i,\sigma}(t)  c^{\dag}_{j,\bar\sigma}(t) \biggl( N_{j\sigma}(t)+N_{i\bar \sigma}(t)\biggl)      ,  
\end{eqnarray*}
\begin{eqnarray}
  \label{Kterms1}
R3 &=& -i F(i,j) 
 \biggl(  c^{\dag}_{i,\sigma}(t) c^{\dag}_{j,\sigma}(t) f^{\dag}_{j,\bar \sigma}(t) f_{j,\sigma}(t)\nonumber\\
 &&+ c^{\dag}_{i,\bar \sigma}(t) c^{\dag}_{j,\bar \sigma}(t) f^{\dag}_{i,\sigma}(t) f_{i,\bar \sigma}(t)\biggr). \nonumber\\
\end{eqnarray}
 
In order to obtain the analytical expression of Eq.~\eqref{EOM_Gap:appendix}, we sum the three contributions of Eq.~\eqref{Kterms1} to obtain
\begin{eqnarray} \label{EOM_Allg}
 \frac{d}{dt} \Delta^{\dag}_{ij}(t) &=& i F(i,j)\sum\limits\limits_{\sigma} \biggl[ \frac{J_z}{2} c^{\dag}_{i,\sigma}(t) c^{\dag}_{j,\bar \sigma}(t)\nonumber \\
  && \biggl( (-1)^{\bar \sigma_j}S_j^z(t) - (-1)^{\bar \sigma_i}S_i^z(t) \biggr) - \nonumber\\
  && \frac{J_{\perp}}{2} \biggl(
      c^{\dag}_{i,\sigma}(t) c^{\dag}_{j,\sigma}(t) S^{\beta}_{j}(t) +
   c^{\dag}_{i,\bar \sigma}(t) c^{\dag}_{j, \bar \sigma}(t) S^{\alpha}_{i}(t) \biggr)
    \biggr] .\nonumber \\
    &&
\end{eqnarray}
The indices $\alpha, \beta$ stand for $S^{\alpha}=S^+ (S^-)$ when $\sigma= \uparrow (\downarrow)$ and $S^{\beta}= S^+ (S^-)$ when $\bar \sigma= \uparrow (\downarrow)$.
Note that until now we have not further specified the various cases introduced in Eq.~\eqref{Delta}, so that the result is general. 

Let us now consider the s-wave case. 
By substituting Eq.~\eqref{Delta} into Eq.~\eqref{Kterms1} one sees by a direct calculation that both the sum $R1+R2$ and $R3$ vanish
\begin{eqnarray}
\label{Kterms3}
(R1 + R2) &\propto& c^{\dag}_{i, \sigma}(t) c^{\dag}_{i,\bar \sigma}(t) \big( N_{i,\bar\sigma}(t)-N_{i,\sigma}(t) \big) \\ \nonumber
&&  -c^{\dag}_{i, \sigma}(t) c^{\dag}_{i,\bar \sigma}(t) \big( N_{i,\bar\sigma}(t)-N_{i,\sigma}(t) \big)\\ \nonumber
&=&0  ,\\
R3 &\propto& c^{\dag}_{i, \sigma}(t) c^{\dag}_{i,\sigma}(t) f_i^\dag(t) f_i(t) + c^{\dag}_{i, \bar\sigma}(t) c^{\dag}_{i,\bar\sigma}(t) f_i^\dag(t) f_i(t) \nonumber \\ 
&=&0
. 
\end{eqnarray}
R3 is equal to zero because of Pauli's principle, since we have two $c$ operators with the same quantum numbers on the same site, leading to  
\[
\frac{d}{dt} \Delta^{\dag}_{ij}(t)  =0 \qquad \mbox{for s-wave pairing,}
\]
as discussed in Sec.~\ref{EOM}.
Hence, the result for Eq.~\eqref{eq:Green_appendix} will not depend on the interaction $J$, and the susceptibility Eq.~\eqref{eq:pairingsusc} will have in the s-wave channel the same $\omega$-dependence as a system of free electrons.

This result, however, does not hold for the extended s-wave and d-wave channels because in the corresponding expression of $F(i,j)$ there is no Kronecker delta. 
For example, the result regarding $d_{x^2-y^2}$-wave ordering is
\begin{widetext}  
\begin{equation}\label{Delta_dwawe}
\frac{d}{dt} \Delta^{\dag}_{ij}(t) = \\ 
\left
\{
\begin{array}{cl}
i\frac{J_z}{2} \sum\limits_{\sigma}  c^{\dag}_{i,\sigma}(t) c^{\dag}_{j,\bar \sigma}(t) \left( (-1)^{\bar \sigma_j}S_j^z(t) - (-1)^{\bar \sigma_i}S_i^z(t) \right)  \\
- i\frac{J_{\perp}}{2} \sum\limits_{\sigma}
 \left( c^{\dag}_{i,\sigma}(t) c^{\dag}_{j,\sigma}(t) S^{\beta}_{j}(t) +
 c^{\dag}_{i,\bar \sigma}(t) c^{\dag}_{j, \bar \sigma}(t) S^{\alpha}_{i}(t)\right) & \text{x-direction} \\
 &\\
i \frac{J_z}{2} \sum\limits_{\sigma}  c^{\dag}_{i,\sigma}(t) c^{\dag}_{j,\bar \sigma}(t) \left(     (-1)^{\bar \sigma_i}S_i^z(t) - (-1)^{\bar \sigma_j}S_j^z(t)  \right)  \\
+ i \frac{J_{\perp}}{2} \sum\limits_{\sigma}
\left( c^{\dag}_{i,\sigma}(t) c^{\dag}_{j,\sigma}(t) S^{\beta}_{j}(t) +
c^{\dag}_{i,\bar \sigma}(t) c^{\dag}_{j, \bar \sigma}(t) S^{\alpha}_{i}(t)\right). & \text{y-direction}
\end{array}
\right.
\end{equation} 
\end{widetext}
The expression for extended s-wave ordering can be easily derived from this equation as its symmetry is closely related to the one of $d_{x^2-y^2}$-wave [see Eq.~\eqref{Delta}].
As expected for d-wave order, we see from Eq.~\eqref{Delta_dwawe} that the expressions for the $x$ direction and for the $y$ direction only differ by a factor of $-1$. 
Furthermore, the $d_{xy}$-wave channel can be accessed likewise by a similar calculation.


\bibliography{Literature}

\begin{thebibliography}{107}%
\makeatletter
\providecommand \@ifxundefined [1]{%
 \@ifx{#1\undefined}
}%
\providecommand \@ifnum [1]{%
 \ifnum #1\expandafter \@firstoftwo
 \else \expandafter \@secondoftwo
 \fi
}%
\providecommand \@ifx [1]{%
 \ifx #1\expandafter \@firstoftwo
 \else \expandafter \@secondoftwo
 \fi
}%
\providecommand \natexlab [1]{#1}%
\providecommand \enquote  [1]{``#1''}%
\providecommand \bibnamefont  [1]{#1}%
\providecommand \bibfnamefont [1]{#1}%
\providecommand \citenamefont [1]{#1}%
\providecommand \href@noop [0]{\@secondoftwo}%
\providecommand \href [0]{\begingroup \@sanitize@url \@href}%
\providecommand \@href[1]{\@@startlink{#1}\@@href}%
\providecommand \@@href[1]{\endgroup#1\@@endlink}%
\providecommand \@sanitize@url [0]{\catcode `\\12\catcode `\$12\catcode
  `\&12\catcode `\#12\catcode `\^12\catcode `\_12\catcode `\%12\relax}%
\providecommand \@@startlink[1]{}%
\providecommand \@@endlink[0]{}%
\providecommand \url  [0]{\begingroup\@sanitize@url \@url }%
\providecommand \@url [1]{\endgroup\@href {#1}{\urlprefix }}%
\providecommand \urlprefix  [0]{URL }%
\providecommand \Eprint [0]{\href }%
\@ifxundefined \urlstyle {%
  \providecommand \doi  [0]{\begingroup \@sanitize@url \@doi}%
  \providecommand \@doi [1]{\endgroup \@@startlink {\doibase
  #1}doi:\discretionary {}{}{}#1\@@endlink }%
}{%
  \providecommand \doi  [0]{doi:\discretionary{}{}{}\begingroup
  \urlstyle{rm}\Url }%
}%
\providecommand \doibase [0]{http://dx.doi.org/}%
\providecommand \Doi [0]{\begingroup \@sanitize@url \@Doi }%
\providecommand \@Doi  [1]{\endgroup\@@startlink{\doibase#1}\@@Doi}%
\providecommand \@@Doi [1]{#1\@@endlink}%
\providecommand \selectlanguage [0]{\@gobble}%
\providecommand \bibinfo  [0]{\@secondoftwo}%
\providecommand \bibfield  [0]{\@secondoftwo}%
\providecommand \translation [1]{[#1]}%
\providecommand \BibitemOpen [0]{}%
\providecommand \bibitemStop [0]{}%
\providecommand \bibitemNoStop [0]{.\EOS\space}%
\providecommand \EOS [0]{\spacefactor3000\relax}%
\providecommand \BibitemShut  [1]{\csname bibitem#1\endcsname}%
\bibitem [{\citenamefont {Schroder}\ \emph {et~al.}(2000)\citenamefont
  {Schroder}, \citenamefont {Aeppli}, \citenamefont {Coldea}, \citenamefont
  {Adams}, \citenamefont {Stockert}, \citenamefont {Lohneysen}, \citenamefont
  {Bucher}, \citenamefont {Ramazashvili},\ and\ \citenamefont
  {Coleman}}]{SAC+00}%
  \BibitemOpen
  \bibfield  {author} {\bibinfo {author} {\bibfnamefont {A.}~\bibnamefont
  {Schroder}}, \bibinfo {author} {\bibfnamefont {G.}~\bibnamefont {Aeppli}},
  \bibinfo {author} {\bibfnamefont {R.}~\bibnamefont {Coldea}}, \bibinfo
  {author} {\bibfnamefont {M.}~\bibnamefont {Adams}}, \bibinfo {author}
  {\bibfnamefont {O.}~\bibnamefont {Stockert}}, \bibinfo {author}
  {\bibfnamefont {H.}~\bibnamefont {Lohneysen}}, \bibinfo {author}
  {\bibfnamefont {E.}~\bibnamefont {Bucher}}, \bibinfo {author} {\bibfnamefont
  {R.}~\bibnamefont {Ramazashvili}}, \ and\ \bibinfo {author} {\bibfnamefont
  {P.}~\bibnamefont {Coleman}},\ }\Doi {10.1038/35030039} {\bibfield  {journal}
  {\bibinfo  {journal} {Nature},\ }\textbf {\bibinfo {volume} {407}},\ \bibinfo
  {pages} {351} (\bibinfo {year} {2000})},\ ISSN \bibinfo {issn}
  {0028-0836}\BibitemShut {NoStop}%
\bibitem [{\citenamefont {Si}\ \emph {et~al.}(2001)\citenamefont {Si},
  \citenamefont {Rabello}, \citenamefont {Ingersent},\ and\ \citenamefont
  {Smith}}]{SRIS01}%
  \BibitemOpen
  \bibfield  {author} {\bibinfo {author} {\bibfnamefont {Q.}~\bibnamefont
  {Si}}, \bibinfo {author} {\bibfnamefont {S.}~\bibnamefont {Rabello}},
  \bibinfo {author} {\bibfnamefont {K.}~\bibnamefont {Ingersent}}, \ and\
  \bibinfo {author} {\bibfnamefont {J.~L.}\ \bibnamefont {Smith}},\ }\Doi
  {10.1038/35101507} {\bibfield  {journal} {\bibinfo  {journal} {Nature},\
  }\textbf {\bibinfo {volume} {413}},\ \bibinfo {pages} {804} (\bibinfo {year}
  {2001})},\ ISSN \bibinfo {issn} {0028-0836}\BibitemShut {NoStop}%
\bibitem [{\citenamefont {Grempel}\ and\ \citenamefont {Si}(2003)}]{GS03}%
  \BibitemOpen
  \bibfield  {author} {\bibinfo {author} {\bibfnamefont {D.~R.}\ \bibnamefont
  {Grempel}}\ and\ \bibinfo {author} {\bibfnamefont {Q.}~\bibnamefont {Si}},\
  }\Doi {10.1103/PhysRevLett.91.026401} {\bibfield  {journal} {\bibinfo
  {journal} {Phys. Rev. Lett.},\ }\textbf {\bibinfo {volume} {91}},\ \bibinfo
  {pages} {026401} (\bibinfo {year} {2003})}\BibitemShut {NoStop}%
\bibitem [{\citenamefont {Si}(2006)}]{Si06}%
  \BibitemOpen
  \bibfield  {author} {\bibinfo {author} {\bibfnamefont {Q.}~\bibnamefont
  {Si}},\ }\href
  {http://www.sciencedirect.com/science/article/pii/S092145260600007X}
  {\bibfield  {journal} {\bibinfo  {journal} {Physica B: Condensed Matter},\
  }\textbf {\bibinfo {volume} {378}},\ \bibinfo {pages} {23} (\bibinfo {year}
  {2006})},\ ISSN \bibinfo {issn} {0921-4526}\BibitemShut {NoStop}%
\bibitem [{\citenamefont {Gegenwart}\ \emph {et~al.}(2008)\citenamefont
  {Gegenwart}, \citenamefont {Si},\ and\ \citenamefont {Steglich}}]{GSS08}%
  \BibitemOpen
  \bibfield  {author} {\bibinfo {author} {\bibfnamefont {P.}~\bibnamefont
  {Gegenwart}}, \bibinfo {author} {\bibfnamefont {Q.}~\bibnamefont {Si}}, \
  and\ \bibinfo {author} {\bibfnamefont {F.}~\bibnamefont {Steglich}},\ }\Doi
  {10.1038/nphys892} {\bibfield  {journal} {\bibinfo  {journal} {Nat Phys},\
  }\textbf {\bibinfo {volume} {4}},\ \bibinfo {pages} {186} (\bibinfo {year}
  {2008})},\ ISSN \bibinfo {issn} {1745-2473}\BibitemShut {NoStop}%
\bibitem [{\citenamefont {Hartmann}\ \emph {et~al.}(2010)\citenamefont
  {Hartmann}, \citenamefont {Oeschler}, \citenamefont {Krellner}, \citenamefont
  {Geibel}, \citenamefont {Paschen},\ and\ \citenamefont {Steglich}}]{HOK+10}%
  \BibitemOpen
  \bibfield  {author} {\bibinfo {author} {\bibfnamefont {S.}~\bibnamefont
  {Hartmann}}, \bibinfo {author} {\bibfnamefont {N.}~\bibnamefont {Oeschler}},
  \bibinfo {author} {\bibfnamefont {C.}~\bibnamefont {Krellner}}, \bibinfo
  {author} {\bibfnamefont {C.}~\bibnamefont {Geibel}}, \bibinfo {author}
  {\bibfnamefont {S.}~\bibnamefont {Paschen}}, \ and\ \bibinfo {author}
  {\bibfnamefont {F.}~\bibnamefont {Steglich}},\ }\Doi
  {10.1103/PhysRevLett.104.096401} {\bibfield  {journal} {\bibinfo  {journal}
  {Phys. Rev. Lett.},\ }\textbf {\bibinfo {volume} {104}},\ \bibinfo {pages}
  {096401} (\bibinfo {year} {2010})}\BibitemShut {NoStop}%
\bibitem [{\citenamefont {Si}(2010)}]{Si10}%
  \BibitemOpen
  \bibfield  {author} {\bibinfo {author} {\bibfnamefont {Q.}~\bibnamefont
  {Si}},\ }\Doi {10.1002/pssb.200983082} {\bibfield  {journal} {\bibinfo
  {journal} {physica status solidi (b)},\ }\textbf {\bibinfo {volume} {247}},\
  \bibinfo {pages} {476} (\bibinfo {year} {2010})},\ ISSN \bibinfo {issn}
  {1521-3951}\BibitemShut {NoStop}%
\bibitem [{\citenamefont {Si}\ and\ \citenamefont {Steglich}(2010)}]{SS10}%
  \BibitemOpen
  \bibfield  {author} {\bibinfo {author} {\bibfnamefont {Q.}~\bibnamefont
  {Si}}\ and\ \bibinfo {author} {\bibfnamefont {F.}~\bibnamefont {Steglich}},\
  }\Doi {10.1126/science.1191195} {\bibfield  {journal} {\bibinfo  {journal}
  {Science},\ }\textbf {\bibinfo {volume} {329}},\ \bibinfo {pages} {1161}
  (\bibinfo {year} {2010})}\BibitemShut {NoStop}%
\bibitem [{\citenamefont {Gegenwart}\ \emph {et~al.}(2015)\citenamefont
  {Gegenwart}, \citenamefont {Steglich}, \citenamefont {Geibel},\ and\
  \citenamefont {Brando}}]{GSGB15}%
  \BibitemOpen
  \bibfield  {author} {\bibinfo {author} {\bibfnamefont {P.}~\bibnamefont
  {Gegenwart}}, \bibinfo {author} {\bibfnamefont {F.}~\bibnamefont {Steglich}},
  \bibinfo {author} {\bibfnamefont {C.}~\bibnamefont {Geibel}}, \ and\ \bibinfo
  {author} {\bibfnamefont {M.}~\bibnamefont {Brando}},\ }\Doi
  {10.1140/epjst/e2015-02442-7} {\bibfield  {journal} {\bibinfo  {journal} {The
  European Physical Journal Special Topics},\ }\textbf {\bibinfo {volume}
  {224}},\ \bibinfo {pages} {975} (\bibinfo {year} {2015})},\ ISSN \bibinfo
  {issn} {1951-6401}\BibitemShut {NoStop}%
\bibitem [{\citenamefont {Aeppli}\ \emph {et~al.}(1989)\citenamefont {Aeppli},
  \citenamefont {Bishop}, \citenamefont {Broholm}, \citenamefont {Bucher},
  \citenamefont {Siemensmeyer}, \citenamefont {Steiner},\ and\ \citenamefont
  {St\"usser}}]{ABB+89}%
  \BibitemOpen
  \bibfield  {author} {\bibinfo {author} {\bibfnamefont {G.}~\bibnamefont
  {Aeppli}}, \bibinfo {author} {\bibfnamefont {D.}~\bibnamefont {Bishop}},
  \bibinfo {author} {\bibfnamefont {C.}~\bibnamefont {Broholm}}, \bibinfo
  {author} {\bibfnamefont {E.}~\bibnamefont {Bucher}}, \bibinfo {author}
  {\bibfnamefont {K.}~\bibnamefont {Siemensmeyer}}, \bibinfo {author}
  {\bibfnamefont {M.}~\bibnamefont {Steiner}}, \ and\ \bibinfo {author}
  {\bibfnamefont {N.}~\bibnamefont {St\"usser}},\ }\Doi
  {10.1103/PhysRevLett.63.676} {\bibfield  {journal} {\bibinfo  {journal}
  {Phys. Rev. Lett.},\ }\textbf {\bibinfo {volume} {63}},\ \bibinfo {pages}
  {676} (\bibinfo {year} {1989})}\BibitemShut {NoStop}%
\bibitem [{\citenamefont {Tou}\ \emph {et~al.}(1998)\citenamefont {Tou},
  \citenamefont {Kitaoka}, \citenamefont {Ishida}, \citenamefont {Asayama},
  \citenamefont {Kimura}, \citenamefont {Onuki}, \citenamefont {Yamamoto},
  \citenamefont {Haga},\ and\ \citenamefont {Maezawa}}]{TKI+98}%
  \BibitemOpen
  \bibfield  {author} {\bibinfo {author} {\bibfnamefont {H.}~\bibnamefont
  {Tou}}, \bibinfo {author} {\bibfnamefont {Y.}~\bibnamefont {Kitaoka}},
  \bibinfo {author} {\bibfnamefont {K.}~\bibnamefont {Ishida}}, \bibinfo
  {author} {\bibfnamefont {K.}~\bibnamefont {Asayama}}, \bibinfo {author}
  {\bibfnamefont {N.}~\bibnamefont {Kimura}}, \bibinfo {author} {\bibfnamefont
  {Y.}~\bibnamefont {Onuki}}, \bibinfo {author} {\bibfnamefont
  {E.}~\bibnamefont {Yamamoto}}, \bibinfo {author} {\bibfnamefont
  {Y.}~\bibnamefont {Haga}}, \ and\ \bibinfo {author} {\bibfnamefont
  {K.}~\bibnamefont {Maezawa}},\ }\Doi {10.1103/PhysRevLett.80.3129} {\bibfield
   {journal} {\bibinfo  {journal} {Phys. Rev. Lett.},\ }\textbf {\bibinfo
  {volume} {80}},\ \bibinfo {pages} {3129} (\bibinfo {year}
  {1998})}\BibitemShut {NoStop}%
\bibitem [{\citenamefont {Sato}\ \emph {et~al.}(2001)\citenamefont {Sato},
  \citenamefont {Aso}, \citenamefont {Miyake}, \citenamefont {Shiina},
  \citenamefont {Thalmeier}, \citenamefont {Varelogiannis}, \citenamefont
  {Geibel}, \citenamefont {Steglich}, \citenamefont {Fulde},\ and\
  \citenamefont {Komatsubara}}]{SAM+01}%
  \BibitemOpen
  \bibfield  {author} {\bibinfo {author} {\bibfnamefont {N.~K.}\ \bibnamefont
  {Sato}}, \bibinfo {author} {\bibfnamefont {N.}~\bibnamefont {Aso}}, \bibinfo
  {author} {\bibfnamefont {K.}~\bibnamefont {Miyake}}, \bibinfo {author}
  {\bibfnamefont {R.}~\bibnamefont {Shiina}}, \bibinfo {author} {\bibfnamefont
  {P.}~\bibnamefont {Thalmeier}}, \bibinfo {author} {\bibfnamefont
  {G.}~\bibnamefont {Varelogiannis}}, \bibinfo {author} {\bibfnamefont
  {C.}~\bibnamefont {Geibel}}, \bibinfo {author} {\bibfnamefont
  {F.}~\bibnamefont {Steglich}}, \bibinfo {author} {\bibfnamefont
  {P.}~\bibnamefont {Fulde}}, \ and\ \bibinfo {author} {\bibfnamefont
  {T.}~\bibnamefont {Komatsubara}},\ }\href
  {http://dx.doi.org/10.1038/35066519} {\bibfield  {journal} {\bibinfo
  {journal} {Nature},\ }\textbf {\bibinfo {volume} {410}},\ \bibinfo {pages}
  {340} (\bibinfo {year} {2001})},\ ISSN \bibinfo {issn}
  {0028-0836}\BibitemShut {NoStop}%
\bibitem [{\citenamefont {Bauer}\ \emph {et~al.}(2002)\citenamefont {Bauer},
  \citenamefont {Frederick}, \citenamefont {Ho}, \citenamefont {Zapf},\ and\
  \citenamefont {Maple}}]{BFH+02}%
  \BibitemOpen
  \bibfield  {author} {\bibinfo {author} {\bibfnamefont {E.~D.}\ \bibnamefont
  {Bauer}}, \bibinfo {author} {\bibfnamefont {N.~A.}\ \bibnamefont
  {Frederick}}, \bibinfo {author} {\bibfnamefont {P.-C.}\ \bibnamefont {Ho}},
  \bibinfo {author} {\bibfnamefont {V.~S.}\ \bibnamefont {Zapf}}, \ and\
  \bibinfo {author} {\bibfnamefont {M.~B.}\ \bibnamefont {Maple}},\ }\Doi
  {10.1103/PhysRevB.65.100506} {\bibfield  {journal} {\bibinfo  {journal}
  {Phys. Rev. B},\ }\textbf {\bibinfo {volume} {65}},\ \bibinfo {pages}
  {100506} (\bibinfo {year} {2002})}\BibitemShut {NoStop}%
\bibitem [{\citenamefont {Yuan}\ \emph {et~al.}(2003)\citenamefont {Yuan},
  \citenamefont {Grosche}, \citenamefont {Deppe}, \citenamefont {Geibel},
  \citenamefont {Sparn},\ and\ \citenamefont {Steglich}}]{YGD+03}%
  \BibitemOpen
  \bibfield  {author} {\bibinfo {author} {\bibfnamefont {H.~Q.}\ \bibnamefont
  {Yuan}}, \bibinfo {author} {\bibfnamefont {F.~M.}\ \bibnamefont {Grosche}},
  \bibinfo {author} {\bibfnamefont {M.}~\bibnamefont {Deppe}}, \bibinfo
  {author} {\bibfnamefont {C.}~\bibnamefont {Geibel}}, \bibinfo {author}
  {\bibfnamefont {G.}~\bibnamefont {Sparn}}, \ and\ \bibinfo {author}
  {\bibfnamefont {F.}~\bibnamefont {Steglich}},\ }\Doi
  {10.1126/science.1091648} {\bibfield  {journal} {\bibinfo  {journal}
  {Science},\ }\textbf {\bibinfo {volume} {302}},\ \bibinfo {pages} {2104}
  (\bibinfo {year} {2003})}\BibitemShut {NoStop}%
\bibitem [{\citenamefont {Huy}\ \emph {et~al.}(2007)\citenamefont {Huy},
  \citenamefont {Gasparini}, \citenamefont {de~Nijs}, \citenamefont {Huang},
  \citenamefont {Klaasse}, \citenamefont {Gortenmulder}, \citenamefont
  {de~Visser}, \citenamefont {Hamann}, \citenamefont {G\"orlach},\ and\
  \citenamefont {L\"ohneysen}}]{HGdN+07}%
  \BibitemOpen
  \bibfield  {author} {\bibinfo {author} {\bibfnamefont {N.~T.}\ \bibnamefont
  {Huy}}, \bibinfo {author} {\bibfnamefont {A.}~\bibnamefont {Gasparini}},
  \bibinfo {author} {\bibfnamefont {D.~E.}\ \bibnamefont {de~Nijs}}, \bibinfo
  {author} {\bibfnamefont {Y.}~\bibnamefont {Huang}}, \bibinfo {author}
  {\bibfnamefont {J.~C.~P.}\ \bibnamefont {Klaasse}}, \bibinfo {author}
  {\bibfnamefont {T.}~\bibnamefont {Gortenmulder}}, \bibinfo {author}
  {\bibfnamefont {A.}~\bibnamefont {de~Visser}}, \bibinfo {author}
  {\bibfnamefont {A.}~\bibnamefont {Hamann}}, \bibinfo {author} {\bibfnamefont
  {T.}~\bibnamefont {G\"orlach}}, \ and\ \bibinfo {author} {\bibfnamefont
  {H.~v.}\ \bibnamefont {L\"ohneysen}},\ }\Doi {10.1103/PhysRevLett.99.067006}
  {\bibfield  {journal} {\bibinfo  {journal} {Phys. Rev. Lett.},\ }\textbf
  {\bibinfo {volume} {99}},\ \bibinfo {pages} {067006} (\bibinfo {year}
  {2007})}\BibitemShut {NoStop}%
\bibitem [{\citenamefont {Nair}\ \emph {et~al.}(2010)\citenamefont {Nair},
  \citenamefont {Stockert}, \citenamefont {Witte}, \citenamefont {Nicklas},
  \citenamefont {Schedler}, \citenamefont {Kiefer}, \citenamefont {Thompson},
  \citenamefont {Bianchi}, \citenamefont {Fisk}, \citenamefont {Wirth},\ and\
  \citenamefont {Steglich}}]{NSW+10}%
  \BibitemOpen
  \bibfield  {author} {\bibinfo {author} {\bibfnamefont {S.}~\bibnamefont
  {Nair}}, \bibinfo {author} {\bibfnamefont {O.}~\bibnamefont {Stockert}},
  \bibinfo {author} {\bibfnamefont {U.}~\bibnamefont {Witte}}, \bibinfo
  {author} {\bibfnamefont {M.}~\bibnamefont {Nicklas}}, \bibinfo {author}
  {\bibfnamefont {R.}~\bibnamefont {Schedler}}, \bibinfo {author}
  {\bibfnamefont {K.}~\bibnamefont {Kiefer}}, \bibinfo {author} {\bibfnamefont
  {J.~D.}\ \bibnamefont {Thompson}}, \bibinfo {author} {\bibfnamefont {A.~D.}\
  \bibnamefont {Bianchi}}, \bibinfo {author} {\bibfnamefont {Z.}~\bibnamefont
  {Fisk}}, \bibinfo {author} {\bibfnamefont {S.}~\bibnamefont {Wirth}}, \ and\
  \bibinfo {author} {\bibfnamefont {F.}~\bibnamefont {Steglich}},\ }\Doi
  {10.1073/pnas.1004958107} {\bibfield  {journal} {\bibinfo  {journal} {PNAS},\
  }\textbf {\bibinfo {volume} {107}},\ \bibinfo {pages} {9537} (\bibinfo {year}
  {2010})}\BibitemShut {NoStop}%
\bibitem [{\citenamefont {Stockert}\ \emph {et~al.}(2011)\citenamefont
  {Stockert}, \citenamefont {Arndt}, \citenamefont {Faulhaber}, \citenamefont
  {Geibel}, \citenamefont {Jeevan}, \citenamefont {Kirchner}, \citenamefont
  {Loewenhaupt}, \citenamefont {Schmalzl}, \citenamefont {Schmidt},
  \citenamefont {Si},\ and\ \citenamefont {Steglich}}]{SAF+11}%
  \BibitemOpen
  \bibfield  {author} {\bibinfo {author} {\bibfnamefont {O.}~\bibnamefont
  {Stockert}}, \bibinfo {author} {\bibfnamefont {J.}~\bibnamefont {Arndt}},
  \bibinfo {author} {\bibfnamefont {E.}~\bibnamefont {Faulhaber}}, \bibinfo
  {author} {\bibfnamefont {C.}~\bibnamefont {Geibel}}, \bibinfo {author}
  {\bibfnamefont {H.~S.}\ \bibnamefont {Jeevan}}, \bibinfo {author}
  {\bibfnamefont {S.}~\bibnamefont {Kirchner}}, \bibinfo {author}
  {\bibfnamefont {M.}~\bibnamefont {Loewenhaupt}}, \bibinfo {author}
  {\bibfnamefont {K.}~\bibnamefont {Schmalzl}}, \bibinfo {author}
  {\bibfnamefont {W.}~\bibnamefont {Schmidt}}, \bibinfo {author} {\bibfnamefont
  {Q.}~\bibnamefont {Si}}, \ and\ \bibinfo {author} {\bibfnamefont
  {F.}~\bibnamefont {Steglich}},\ }\Doi {10.1038/nphys1852} {\bibfield
  {journal} {\bibinfo  {journal} {Nat Phys},\ }\textbf {\bibinfo {volume}
  {7}},\ \bibinfo {pages} {119} (\bibinfo {year} {2011})},\ ISSN \bibinfo
  {issn} {1745-2473}\BibitemShut {NoStop}%
\bibitem [{\citenamefont {Steglich}\ \emph {et~al.}(2013)\citenamefont
  {Steglich}, \citenamefont {Stockert}, \citenamefont {Wirth}, \citenamefont
  {Geibel}, \citenamefont {Yuan}, \citenamefont {Kirchner},\ and\ \citenamefont
  {Si}}]{SSW+13}%
  \BibitemOpen
  \bibfield  {author} {\bibinfo {author} {\bibfnamefont {F.}~\bibnamefont
  {Steglich}}, \bibinfo {author} {\bibfnamefont {O.}~\bibnamefont {Stockert}},
  \bibinfo {author} {\bibfnamefont {S.}~\bibnamefont {Wirth}}, \bibinfo
  {author} {\bibfnamefont {C.}~\bibnamefont {Geibel}}, \bibinfo {author}
  {\bibfnamefont {H.~Q.}\ \bibnamefont {Yuan}}, \bibinfo {author}
  {\bibfnamefont {S.}~\bibnamefont {Kirchner}}, \ and\ \bibinfo {author}
  {\bibfnamefont {Q.}~\bibnamefont {Si}},\ }\href
  {http://stacks.iop.org/1742-6596/449/i=1/a=012028} {\bibfield  {journal}
  {\bibinfo  {journal} {Journal of Physics: Conference Series},\ }\textbf
  {\bibinfo {volume} {449}},\ \bibinfo {pages} {012028} (\bibinfo {year}
  {2013})}\BibitemShut {NoStop}%
\bibitem [{\citenamefont {Steglich}(2014)}]{Ste14}%
  \BibitemOpen
  \bibfield  {author} {\bibinfo {author} {\bibfnamefont {F.}~\bibnamefont
  {Steglich}},\ }\Doi {10.1080/14786435.2014.956835} {\bibfield  {journal}
  {\bibinfo  {journal} {Philosophical Magazine},\ }\textbf {\bibinfo {volume}
  {94}},\ \bibinfo {pages} {3259} (\bibinfo {year} {2014})}\BibitemShut
  {NoStop}%
\bibitem [{\citenamefont {Enayat}\ \emph {et~al.}(2016)\citenamefont {Enayat},
  \citenamefont {Sun}, \citenamefont {Maldonado}, \citenamefont {Suderow},
  \citenamefont {Seiro}, \citenamefont {Geibel}, \citenamefont {Wirth},
  \citenamefont {Steglich},\ and\ \citenamefont {Wahl}}]{ESM+16}%
  \BibitemOpen
  \bibfield  {author} {\bibinfo {author} {\bibfnamefont {M.}~\bibnamefont
  {Enayat}}, \bibinfo {author} {\bibfnamefont {Z.}~\bibnamefont {Sun}},
  \bibinfo {author} {\bibfnamefont {A.}~\bibnamefont {Maldonado}}, \bibinfo
  {author} {\bibfnamefont {H.}~\bibnamefont {Suderow}}, \bibinfo {author}
  {\bibfnamefont {S.}~\bibnamefont {Seiro}}, \bibinfo {author} {\bibfnamefont
  {C.}~\bibnamefont {Geibel}}, \bibinfo {author} {\bibfnamefont
  {S.}~\bibnamefont {Wirth}}, \bibinfo {author} {\bibfnamefont
  {F.}~\bibnamefont {Steglich}}, \ and\ \bibinfo {author} {\bibfnamefont
  {P.}~\bibnamefont {Wahl}},\ }\Doi {10.1103/PhysRevB.93.045123} {\bibfield
  {journal} {\bibinfo  {journal} {Phys. Rev. B},\ }\textbf {\bibinfo {volume}
  {93}},\ \bibinfo {pages} {045123} (\bibinfo {year} {2016})}\BibitemShut
  {NoStop}%
\bibitem [{\citenamefont {Steglich}\ and\ \citenamefont {Wirth}(2016)}]{SW16}%
  \BibitemOpen
  \bibfield  {author} {\bibinfo {author} {\bibfnamefont {F.}~\bibnamefont
  {Steglich}}\ and\ \bibinfo {author} {\bibfnamefont {S.}~\bibnamefont
  {Wirth}},\ }\href {http://stacks.iop.org/0034-4885/79/i=8/a=084502}
  {\bibfield  {journal} {\bibinfo  {journal} {Reports on Progress in Physics},\
  }\textbf {\bibinfo {volume} {79}},\ \bibinfo {pages} {084502} (\bibinfo
  {year} {2016})}\BibitemShut {NoStop}%
\bibitem [{\citenamefont {Doniach}(1977)}]{Don77}%
  \BibitemOpen
  \bibfield  {author} {\bibinfo {author} {\bibfnamefont {S.}~\bibnamefont
  {Doniach}},\ }\href@noop {} {\bibfield  {journal} {\bibinfo  {journal}
  {Physica B+C},\ }\textbf {\bibinfo {volume} {91}},\ \bibinfo {pages} {231}
  (\bibinfo {year} {1977})}\BibitemShut {NoStop}%
\bibitem [{\citenamefont {Zhu}\ \emph {et~al.}(2003)\citenamefont {Zhu},
  \citenamefont {Grempel},\ and\ \citenamefont {Si}}]{ZGS03}%
  \BibitemOpen
  \bibfield  {author} {\bibinfo {author} {\bibfnamefont {J.-X.}\ \bibnamefont
  {Zhu}}, \bibinfo {author} {\bibfnamefont {D.~R.}\ \bibnamefont {Grempel}}, \
  and\ \bibinfo {author} {\bibfnamefont {Q.}~\bibnamefont {Si}},\ }\Doi
  {10.1103/PhysRevLett.91.156404} {\bibfield  {journal} {\bibinfo  {journal}
  {Phys. Rev. Lett.},\ }\textbf {\bibinfo {volume} {91}},\ \bibinfo {pages}
  {156404} (\bibinfo {year} {2003})}\BibitemShut {NoStop}%
\bibitem [{\citenamefont {Hertz}(1976)}]{Her76}%
  \BibitemOpen
  \bibfield  {author} {\bibinfo {author} {\bibfnamefont {J.~A.}\ \bibnamefont
  {Hertz}},\ }\Doi {10.1103/PhysRevB.14.1165} {\bibfield  {journal} {\bibinfo
  {journal} {Phys. Rev. B},\ }\textbf {\bibinfo {volume} {14}},\ \bibinfo
  {pages} {1165} (\bibinfo {year} {1976})}\BibitemShut {NoStop}%
\bibitem [{\citenamefont {Millis}(1993)}]{Mil93}%
  \BibitemOpen
  \bibfield  {author} {\bibinfo {author} {\bibfnamefont {A.~J.}\ \bibnamefont
  {Millis}},\ }\Doi {10.1103/PhysRevB.48.7183} {\bibfield  {journal} {\bibinfo
  {journal} {Phys. Rev. B},\ }\textbf {\bibinfo {volume} {48}},\ \bibinfo
  {pages} {7183} (\bibinfo {year} {1993})}\BibitemShut {NoStop}%
\bibitem [{\citenamefont {Moriya}\ and\ \citenamefont {Takimoto}(1995)}]{MT95}%
  \BibitemOpen
  \bibfield  {author} {\bibinfo {author} {\bibfnamefont {T.}~\bibnamefont
  {Moriya}}\ and\ \bibinfo {author} {\bibfnamefont {T.}~\bibnamefont
  {Takimoto}},\ }\Doi {10.1143/JPSJ.64.960} {\bibfield  {journal} {\bibinfo
  {journal} {Journal of the Physical Society of Japan},\ }\textbf {\bibinfo
  {volume} {64}},\ \bibinfo {pages} {960} (\bibinfo {year} {1995})}\BibitemShut
  {NoStop}%
\bibitem [{\citenamefont {Movshovich}\ \emph {et~al.}(1996)\citenamefont
  {Movshovich}, \citenamefont {Graf}, \citenamefont {Mandrus}, \citenamefont
  {Thompson}, \citenamefont {Smith},\ and\ \citenamefont {Fisk}}]{MGM+96}%
  \BibitemOpen
  \bibfield  {author} {\bibinfo {author} {\bibfnamefont {R.}~\bibnamefont
  {Movshovich}}, \bibinfo {author} {\bibfnamefont {T.}~\bibnamefont {Graf}},
  \bibinfo {author} {\bibfnamefont {D.}~\bibnamefont {Mandrus}}, \bibinfo
  {author} {\bibfnamefont {J.~D.}\ \bibnamefont {Thompson}}, \bibinfo {author}
  {\bibfnamefont {J.~L.}\ \bibnamefont {Smith}}, \ and\ \bibinfo {author}
  {\bibfnamefont {Z.}~\bibnamefont {Fisk}},\ }\Doi {10.1103/PhysRevB.53.8241}
  {\bibfield  {journal} {\bibinfo  {journal} {Phys. Rev. B},\ }\textbf
  {\bibinfo {volume} {53}},\ \bibinfo {pages} {8241} (\bibinfo {year}
  {1996})}\BibitemShut {NoStop}%
\bibitem [{\citenamefont {Hoshino}(2015)}]{HW15}%
  \BibitemOpen
  \bibfield  {author} {\bibinfo {author} {\bibfnamefont {S.}~\bibnamefont
  {Hoshino}} \ and\ \bibinfo {author}{\bibfnamefont {P.}~\bibnamefont{Werner}},\ }\Doi {10.1103/PhysRevLett.115.247001} {\bibfield  {journal} {\bibinfo  {journal}
  {Phys. Rev. Lett.},\ }\textbf {\bibinfo {volume} {115}},\ \bibinfo {pages}
  {247001} (\bibinfo {year} {2015})}\BibitemShut {NoStop}%
\bibitem [{\citenamefont {Yashima}\ \emph {et~al.}(2007)\citenamefont
  {Yashima}, \citenamefont {Kawasaki}, \citenamefont {Mukuda}, \citenamefont
  {Kitaoka}, \citenamefont {Shishido}, \citenamefont {Settai},\ and\
  \citenamefont {Onuki}}]{YKM+07}%
  \BibitemOpen
  \bibfield  {author} {\bibinfo {author} {\bibfnamefont {M.}~\bibnamefont
  {Yashima}}, \bibinfo {author} {\bibfnamefont {S.}~\bibnamefont {Kawasaki}},
  \bibinfo {author} {\bibfnamefont {H.}~\bibnamefont {Mukuda}}, \bibinfo
  {author} {\bibfnamefont {Y.}~\bibnamefont {Kitaoka}}, \bibinfo {author}
  {\bibfnamefont {H.}~\bibnamefont {Shishido}}, \bibinfo {author}
  {\bibfnamefont {R.}~\bibnamefont {Settai}}, \ and\ \bibinfo {author}
  {\bibfnamefont {Y.}~\bibnamefont {Onuki}},\
  }\Doi {10.1103/PhysRevB.76.020509} {\bibfield  {journal} {\bibinfo  {journal}
  {Phys. Rev. B},\ }\textbf {\bibinfo {volume} {76}},\ \bibinfo {pages}
  {020509} (\bibinfo {year} {2007})}\BibitemShut {NoStop}%
\bibitem [{\citenamefont {Asadzadeh}\ \emph {et~al.}(2013)\citenamefont
  {Asadzadeh}, \citenamefont {Becca},\ and\ \citenamefont {Fabrizio}}]{ABF13}%
  \BibitemOpen
  \bibfield  {author} {\bibinfo {author} {\bibfnamefont {M.~Z.}\ \bibnamefont
  {Asadzadeh}}, \bibinfo {author} {\bibfnamefont {F.}~\bibnamefont {Becca}}, \
  and\ \bibinfo {author} {\bibfnamefont {M.}~\bibnamefont {Fabrizio}},\ }\Doi
  {10.1103/PhysRevB.87.205144} {\bibfield  {journal} {\bibinfo  {journal}
  {Phys. Rev. B},\ }\textbf {\bibinfo {volume} {87}},\ \bibinfo {pages}
  {205144} (\bibinfo {year} {2013})}\BibitemShut {NoStop}%
\bibitem [{\citenamefont {Asadzadeh}\ \emph {et~al.}(2014)\citenamefont
  {Asadzadeh}, \citenamefont {Fabrizio},\ and\ \citenamefont {Becca}}]{AFB14}%
  \BibitemOpen
  \bibfield  {author} {\bibinfo {author} {\bibfnamefont {M.~Z.}\ \bibnamefont
  {Asadzadeh}}, \bibinfo {author} {\bibfnamefont {M.}~\bibnamefont {Fabrizio}},
  \ and\ \bibinfo {author} {\bibfnamefont {F.}~\bibnamefont {Becca}},\ }\Doi
  {10.1103/PhysRevB.90.205113} {\bibfield  {journal} {\bibinfo  {journal}
  {Phys. Rev. B},\ }\textbf {\bibinfo {volume} {90}},\ \bibinfo {pages}
  {205113} (\bibinfo {year} {2014})}\BibitemShut {NoStop}%
\bibitem [{\citenamefont {Wu}\ and\ \citenamefont {Tremblay}(2015)}]{WT15}%
  \BibitemOpen
  \bibfield  {author} {\bibinfo {author} {\bibfnamefont {W.}~\bibnamefont
  {Wu}}\ and\ \bibinfo {author} {\bibfnamefont {A.-M.-S.}\ \bibnamefont
  {Tremblay}},\ }\Doi {10.1103/PhysRevX.5.011019} {\bibfield  {journal}
  {\bibinfo  {journal} {Phys. Rev. X},\ }\textbf {\bibinfo {volume} {5}},\
  \bibinfo {pages} {011019} (\bibinfo {year} {2015})}\BibitemShut {NoStop}%
\bibitem [{\citenamefont {Otsuki}(2015)}]{Ots15}%
  \BibitemOpen
  \bibfield  {author} {\bibinfo {author} {\bibfnamefont {J.}~\bibnamefont
  {Otsuki}},\ }\Doi {10.1103/PhysRevLett.115.036404} {\bibfield  {journal}
  {\bibinfo  {journal} {Phys. Rev. Lett.},\ }\textbf {\bibinfo {volume}
  {115}},\ \bibinfo {pages} {036404} (\bibinfo {year} {2015})}\BibitemShut
  {NoStop}%
\bibitem [{\citenamefont {Bodensiek}\ \emph {et~al.}(2013)\citenamefont
  {Bodensiek}, \citenamefont {{{\v Z}itko}}, \citenamefont {Vojta},
  \citenamefont {Jarrell},\ and\ \citenamefont {Pruschke}}]{BZV+13}%
  \BibitemOpen
  \bibfield  {author} {\bibinfo {author} {\bibfnamefont {O.}~\bibnamefont
  {Bodensiek}}, \bibinfo {author} {\bibfnamefont {R.}~\bibnamefont {{{\v
  Z}itko}}}, \bibinfo {author} {\bibfnamefont {M.}~\bibnamefont {Vojta}},
  \bibinfo {author} {\bibfnamefont {M.}~\bibnamefont {Jarrell}}, \ and\
  \bibinfo {author} {\bibfnamefont {T.}~\bibnamefont {Pruschke}},\ }\Doi
  {10.1103/PhysRevLett.110.146406} {\bibfield  {journal} {\bibinfo  {journal}
  {Phys. Rev. Lett.},\ }\textbf {\bibinfo {volume} {110}},\ \bibinfo {pages}
  {146406} (\bibinfo {year} {2013})}\BibitemShut {NoStop}%
\bibitem [{\citenamefont {Martin}\ and\ \citenamefont {Assaad}(2008)}]{MA08}%
  \BibitemOpen
  \bibfield  {author} {\bibinfo {author} {\bibfnamefont {L.~C.}\ \bibnamefont
  {Martin}}\ and\ \bibinfo {author} {\bibfnamefont {F.~F.}\ \bibnamefont
  {Assaad}},\ }\Doi {10.1103/PhysRevLett.101.066404} {\bibfield  {journal}
  {\bibinfo  {journal} {Phys. Rev. Lett.},\ }\textbf {\bibinfo {volume}
  {101}},\ \bibinfo {pages} {066404} (\bibinfo {year} {2008})}\BibitemShut
  {NoStop}%
\bibitem [{\citenamefont {Martin}\ \emph {et~al.}(2010)\citenamefont {Martin},
  \citenamefont {Bercx},\ and\ \citenamefont {Assaad}}]{MBA10}%
  \BibitemOpen
  \bibfield  {author} {\bibinfo {author} {\bibfnamefont {L.~C.}\ \bibnamefont
  {Martin}}, \bibinfo {author} {\bibfnamefont {M.}~\bibnamefont {Bercx}}, \
  and\ \bibinfo {author} {\bibfnamefont {F.~F.}\ \bibnamefont {Assaad}},\ }\Doi
  {10.1103/PhysRevB.82.245105} {\bibfield  {journal} {\bibinfo  {journal}
  {Phys. Rev. B},\ }\textbf {\bibinfo {volume} {82}},\ \bibinfo {pages}
  {245105} (\bibinfo {year} {2010})}\BibitemShut {NoStop}%
\bibitem [{\citenamefont {Peters}\ and\ \citenamefont {Kawakami}(2015)}]{PK15}%
  \BibitemOpen
  \bibfield  {author} {\bibinfo {author} {\bibfnamefont {R.}~\bibnamefont
  {Peters}}\ and\ \bibinfo {author} {\bibfnamefont {N.}~\bibnamefont
  {Kawakami}},\ }\Doi {10.1103/PhysRevB.92.075103} {\bibfield  {journal}
  {\bibinfo  {journal} {Phys. Rev. B},\ }\textbf {\bibinfo {volume} {92}},\
  \bibinfo {pages} {075103} (\bibinfo {year} {2015})}\BibitemShut {NoStop}%
\bibitem [{\citenamefont {Assaad}(1999)}]{Ass99}%
  \BibitemOpen
  \bibfield  {author} {\bibinfo {author} {\bibfnamefont {F.~F.}\ \bibnamefont
  {Assaad}},\ }\Doi {10.1103/PhysRevLett.83.796} {\bibfield  {journal}
  {\bibinfo  {journal} {Phys. Rev. Lett.},\ }\textbf {\bibinfo {volume} {83}},\
  \bibinfo {pages} {796} (\bibinfo {year} {1999})}\BibitemShut {NoStop}%
\bibitem [{\citenamefont {Potthoff}\ \emph {et~al.}(2003)\citenamefont
  {Potthoff}, \citenamefont {Aichhorn},\ and\ \citenamefont {Dahnken}}]{PAD03}%
  \BibitemOpen
  \bibfield  {author} {\bibinfo {author} {\bibfnamefont {M.}~\bibnamefont
  {Potthoff}}, \bibinfo {author} {\bibfnamefont {M.}~\bibnamefont {Aichhorn}},
  \ and\ \bibinfo {author} {\bibfnamefont {C.}~\bibnamefont {Dahnken}},\
  }\href@noop {} {\bibfield  {journal} {\bibinfo  {journal} {Phys. Rev.
  Lett.},\ }\textbf {\bibinfo {volume} {91}},\ \bibinfo {pages} {206402}
  (\bibinfo {year} {2003})}\BibitemShut {NoStop}%
\bibitem [{\citenamefont {Dahnken}\ \emph {et~al.}(2004)\citenamefont
  {Dahnken}, \citenamefont {Aichhorn}, \citenamefont {Hanke}, \citenamefont
  {Arrigoni},\ and\ \citenamefont {Potthoff}}]{DAH+04}%
  \BibitemOpen
  \bibfield  {author} {\bibinfo {author} {\bibfnamefont {C.}~\bibnamefont
  {Dahnken}}, \bibinfo {author} {\bibfnamefont {M.}~\bibnamefont {Aichhorn}},
  \bibinfo {author} {\bibfnamefont {W.}~\bibnamefont {Hanke}}, \bibinfo
  {author} {\bibfnamefont {E.}~\bibnamefont {Arrigoni}}, \ and\ \bibinfo
  {author} {\bibfnamefont {M.}~\bibnamefont {Potthoff}},\ }\href@noop {}
  {\bibfield  {journal} {\bibinfo  {journal} {Phys. Rev. B},\ }\textbf
  {\bibinfo {volume} {70}},\ \bibinfo {pages} {245110} (\bibinfo {year}
  {2004})}\BibitemShut {NoStop}%
\bibitem [{\citenamefont {S{\'e}n\'echal}\ \emph {et~al.}(2005)\citenamefont
  {S{\'e}n\'echal}, \citenamefont {Lavertu}, \citenamefont {Marois},\ and\
  \citenamefont {Tremblay}}]{SLMT05}%
  \BibitemOpen
  \bibfield  {author} {\bibinfo {author} {\bibfnamefont {D.}~\bibnamefont
  {S{\'e}n\'echal}}, \bibinfo {author} {\bibfnamefont {P.-L.}\ \bibnamefont
  {Lavertu}}, \bibinfo {author} {\bibfnamefont {M.-A.}\ \bibnamefont {Marois}},
  \ and\ \bibinfo {author} {\bibfnamefont {A.-M.~S.}\ \bibnamefont
  {Tremblay}},\ }\Doi {10.1103/PhysRevLett.94.156404} {\bibfield  {journal}
  {\bibinfo  {journal} {Phys. Rev. Lett.},\ }\textbf {\bibinfo {volume} {94}},\
  \bibinfo {pages} {156404} (\bibinfo {year} {2005})}\BibitemShut {NoStop}%
\bibitem [{\citenamefont {Aichhorn}\ \emph
  {et~al.}(2006){\natexlab{a}}\citenamefont {Aichhorn}, \citenamefont
  {Arrigoni}, \citenamefont {Potthoff},\ and\ \citenamefont {Hanke}}]{AAPH06a}%
  \BibitemOpen
  \bibfield  {author} {\bibinfo {author} {\bibfnamefont {M.}~\bibnamefont
  {Aichhorn}}, \bibinfo {author} {\bibfnamefont {E.}~\bibnamefont {Arrigoni}},
  \bibinfo {author} {\bibfnamefont {M.}~\bibnamefont {Potthoff}}, \ and\
  \bibinfo {author} {\bibfnamefont {W.}~\bibnamefont {Hanke}},\ }\href@noop {}
  {\bibfield  {journal} {\bibinfo  {journal} {Phys. Rev. B},\ }\textbf
  {\bibinfo {volume} {74}},\ \bibinfo {pages} {235117} (\bibinfo {year}
  {2006}{\natexlab{a}})}\BibitemShut {NoStop}%
\bibitem [{\citenamefont {Nuss}\ \emph {et~al.}(2012)\citenamefont {Nuss},
  \citenamefont {Arrigoni}, \citenamefont {Aichhorn},\ and\ \citenamefont
  {von~der Linden}}]{NAAvdL12}%
  \BibitemOpen
  \bibfield  {author} {\bibinfo {author} {\bibfnamefont {M.}~\bibnamefont
  {Nuss}}, \bibinfo {author} {\bibfnamefont {E.}~\bibnamefont {Arrigoni}},
  \bibinfo {author} {\bibfnamefont {M.}~\bibnamefont {Aichhorn}}, \ and\
  \bibinfo {author} {\bibfnamefont {W.}~\bibnamefont {von~der Linden}},\ }\Doi
  {10.1103/PhysRevB.85.235107} {\bibfield  {journal} {\bibinfo  {journal}
  {Phys. Rev. B},\ }\textbf {\bibinfo {volume} {85}},\ \bibinfo {pages}
  {235107} (\bibinfo {year} {2012})}\BibitemShut {NoStop}%
\bibitem [{\citenamefont {Masuda}\ and\ \citenamefont {Yamamoto}(2015)}]{MY15}%
  \BibitemOpen
  \bibfield  {author} {\bibinfo {author} {\bibfnamefont {K.}~\bibnamefont
  {Masuda}}\ and\ \bibinfo {author} {\bibfnamefont {D.}~\bibnamefont
  {Yamamoto}},\ }\Doi {10.1103/PhysRevB.91.104508} {\bibfield  {journal}
  {\bibinfo  {journal} {Phys. Rev. B},\ }\textbf {\bibinfo {volume} {91}},\
  \bibinfo {pages} {104508} (\bibinfo {year} {2015})}\BibitemShut {NoStop}%
\bibitem [{\citenamefont {Potthoff}(2003){\natexlab{a}}}]{Pot03a}%
  \BibitemOpen
  \bibfield  {author} {\bibinfo {author} {\bibfnamefont {M.}~\bibnamefont
  {Potthoff}},\ }\Doi {10.1140/epjb/e2003-00121-8} {\bibfield  {journal}
  {\bibinfo  {journal} {The European Physical Journal B - Condensed Matter and
  Complex Systems},\ }\textbf {\bibinfo {volume} {32}},\ \bibinfo {pages} {429}
  (\bibinfo {year} {2003}{\natexlab{a}})},\ ISSN \bibinfo {issn}
  {1434-6028}\BibitemShut {NoStop}%
\bibitem [{\citenamefont {Luttinger}\ and\ \citenamefont {Ward}(1960)}]{LW60}%
  \BibitemOpen
  \bibfield  {author} {\bibinfo {author} {\bibfnamefont {J.~M.}\ \bibnamefont
  {Luttinger}}\ and\ \bibinfo {author} {\bibfnamefont {J.~C.}\ \bibnamefont
  {Ward}},\ }\href@noop {} {\bibfield  {journal} {\bibinfo  {journal} {Phys.
  Rev.},\ }\textbf {\bibinfo {volume} {118}},\ \bibinfo {pages} {1417}
  (\bibinfo {year} {1960})}\BibitemShut {NoStop}%
\bibitem [{\citenamefont {Potthoff}(2006)}]{Pot06a}%
  \BibitemOpen
  \bibfield  {author} {\bibinfo {author} {\bibfnamefont {M.}~\bibnamefont
  {Potthoff}},\ }\href@noop {} {\bibfield  {journal} {\bibinfo  {journal}
  {Cond. Mat. Phys.},\ }\textbf {\bibinfo {volume} {9}},\ \bibinfo {pages}
  {557} (\bibinfo {year} {2006})}\BibitemShut {NoStop}%
\bibitem [{\citenamefont {Potthoff}(2003){\natexlab{b}}}]{Pot03}%
  \BibitemOpen
  \bibfield  {author} {\bibinfo {author} {\bibfnamefont {M.}~\bibnamefont
  {Potthoff}},\ }\Doi {10.1140/epjb/e2003-00352-7} {\bibfield  {journal}
  {\bibinfo  {journal} {The European Physical Journal B - Condensed Matter and
  Complex Systems},\ }\textbf {\bibinfo {volume} {36}},\ \bibinfo {pages} {335}
  (\bibinfo {year} {2003}{\natexlab{b}})},\ ISSN \bibinfo {issn}
  {1434-6028}\BibitemShut {NoStop}%
\bibitem [{\citenamefont {Nevidomskyy}\ \emph {et~al.}(2008)\citenamefont
  {Nevidomskyy}, \citenamefont {S\'en\'echal},\ and\ \citenamefont
  {Tremblay}}]{NST08}%
  \BibitemOpen
  \bibfield  {author} {\bibinfo {author} {\bibfnamefont {A.~H.}\ \bibnamefont
  {Nevidomskyy}}, \bibinfo {author} {\bibfnamefont {D.}~\bibnamefont
  {S\'en\'echal}}, \ and\ \bibinfo {author} {\bibfnamefont {A.-M.~S.}\
  \bibnamefont {Tremblay}},\ }\Doi {10.1103/PhysRevB.77.075105} {\bibfield
  {journal} {\bibinfo  {journal} {Phys. Rev. B},\ }\textbf {\bibinfo {volume}
  {77}},\ \bibinfo {pages} {075105} (\bibinfo {year} {2008})}\BibitemShut
  {NoStop}%
\bibitem [{\citenamefont {Potthoff}(2012)}]{Pot12}%
  \BibitemOpen
  \bibfield  {author} {\bibinfo {author} {\bibfnamefont {M.}~\bibnamefont
  {Potthoff}},\ }in\ \Doi {10.1007/978-3-642-21831-6_10} {\emph {\bibinfo
  {booktitle} {Strongly Correlated Systems}}},\ \bibinfo {series} {Springer
  Series in Solid-State Sciences}, Vol.\ \bibinfo {volume} {171},\ \bibinfo
  {editor} {edited by\ \bibinfo {editor} {\bibfnamefont {A.}~\bibnamefont
  {Avella}}\ and\ \bibinfo {editor} {\bibfnamefont {F.}~\bibnamefont
  {Mancini}}}\ (\bibinfo  {publisher} {Springer Berlin Heidelberg},\ \bibinfo
  {year} {2012})\ pp.\ \bibinfo {pages} {303--339}\BibitemShut {NoStop}%
\bibitem [{\citenamefont {Filor}\ and\ \citenamefont {Pruschke}(2014)}]{FP14}%
  \BibitemOpen
  \bibfield  {author} {\bibinfo {author} {\bibfnamefont {S.}~\bibnamefont
  {Filor}}\ and\ \bibinfo {author} {\bibfnamefont {T.}~\bibnamefont
  {Pruschke}},\ }\href {http://stacks.iop.org/1367-2630/16/i=6/a=063059}
  {\bibfield  {journal} {\bibinfo  {journal} {New Journal of Physics},\
  }\textbf {\bibinfo {volume} {16}},\ \bibinfo {pages} {063059} (\bibinfo
  {year} {2014})}\BibitemShut {NoStop}%
\bibitem [{Note1()}]{Note1}%
  \BibitemOpen
  \bibinfo {note} {Instead, \protect \textit {Laubach et al.} showed in Ref.
  \protect \rev@citealpnum {LJR+16} that it is possible to treat the Hubbard
  model in the limit $U\rightarrow \infty $ in order to study the Heisenberg
  model within VCA. However, the enlarged local Hilbert space of the Hubbard
  model as compared to the Heisenberg model prevented the authors from studying
  larger clusters.}\BibitemShut {Stop}%
\bibitem [{\citenamefont {Sinjukow}\ and\ \citenamefont
  {Nolting}(2002)}]{SN02}%
  \BibitemOpen
  \bibfield  {author} {\bibinfo {author} {\bibfnamefont {P.}~\bibnamefont
  {Sinjukow}}\ and\ \bibinfo {author} {\bibfnamefont {W.}~\bibnamefont
  {Nolting}},\ }\Doi {10.1103/PhysRevB.65.212303} {\bibfield  {journal}
  {\bibinfo  {journal} {Phys. Rev. B},\ }\textbf {\bibinfo {volume} {65}},\
  \bibinfo {pages} {212303} (\bibinfo {year} {2002})}\BibitemShut {NoStop}%
\bibitem [{\citenamefont {Coleman}\ \emph {et~al.}(2005)\citenamefont
  {Coleman}, \citenamefont {Paul},\ and\ \citenamefont {Rech}}]{CPR05}%
  \BibitemOpen
  \bibfield  {author} {\bibinfo {author} {\bibfnamefont {P.}~\bibnamefont
  {Coleman}}, \bibinfo {author} {\bibfnamefont {I.}~\bibnamefont {Paul}}, \
  and\ \bibinfo {author} {\bibfnamefont {J.}~\bibnamefont {Rech}},\ }\Doi
  {10.1103/PhysRevB.72.094430} {\bibfield  {journal} {\bibinfo  {journal}
  {Phys. Rev. B},\ }\textbf {\bibinfo {volume} {72}},\ \bibinfo {pages}
  {094430} (\bibinfo {year} {2005})}\BibitemShut {NoStop}%
\bibitem [{\citenamefont {Balzer}\ and\ \citenamefont {Potthoff}(2010)}]{BP10}%
  \BibitemOpen
  \bibfield  {author} {\bibinfo {author} {\bibfnamefont {M.}~\bibnamefont
  {Balzer}}\ and\ \bibinfo {author} {\bibfnamefont {M.}~\bibnamefont
  {Potthoff}},\ }\Doi {10.1103/PhysRevB.82.174441} {\bibfield  {journal}
  {\bibinfo  {journal} {Phys. Rev. B},\ }\textbf {\bibinfo {volume} {82}},\
  \bibinfo {pages} {174441} (\bibinfo {year} {2010})}\BibitemShut {NoStop}%
\bibitem [{\citenamefont {Aichhorn}\ \emph
  {et~al.}(2006){\natexlab{b}}\citenamefont {Aichhorn}, \citenamefont
  {Arrigoni}, \citenamefont {Potthoff},\ and\ \citenamefont {Hanke}}]{AAPH06}%
  \BibitemOpen
  \bibfield  {author} {\bibinfo {author} {\bibfnamefont {M.}~\bibnamefont
  {Aichhorn}}, \bibinfo {author} {\bibfnamefont {E.}~\bibnamefont {Arrigoni}},
  \bibinfo {author} {\bibfnamefont {M.}~\bibnamefont {Potthoff}}, \ and\
  \bibinfo {author} {\bibfnamefont {W.}~\bibnamefont {Hanke}},\ }\Doi
  {10.1103/PhysRevB.74.024508} {\bibfield  {journal} {\bibinfo  {journal}
  {Phys. Rev. B},\ }\textbf {\bibinfo {volume} {74}},\ \bibinfo {pages}
  {024508} (\bibinfo {year} {2006}{\natexlab{b}})}\BibitemShut {NoStop}%
\bibitem [{Note2()}]{Note2}%
  \BibitemOpen
  \bibinfo {note} {Although it would be interesting to investigate spin density
  waves with incommensurate ordering wavevectors, within VCA one usually uses
  commensurate $Q$ vectors. The reason is that the reference system is made up
  of identical clusters, which treat short-ranged spatial correlations inside
  the clusters exactly. A way to treat longer-ranged, incommensurate ordering
  vectors might consist in using supercluster constructions.}\BibitemShut
  {Stop}%
\bibitem [{\citenamefont {Langer}\ and\ \citenamefont
  {Ambegaokar}(1961)}]{LA61}%
  \BibitemOpen
  \bibfield  {author} {\bibinfo {author} {\bibfnamefont {J.~S.}\ \bibnamefont
  {Langer}}\ and\ \bibinfo {author} {\bibfnamefont {V.}~\bibnamefont
  {Ambegaokar}},\ }\Doi {10.1103/PhysRev.121.1090} {\bibfield  {journal}
  {\bibinfo  {journal} {Phys. Rev.},\ }\textbf {\bibinfo {volume} {121}},\
  \bibinfo {pages} {1090} (\bibinfo {year} {1961})}\BibitemShut {NoStop}%
\bibitem [{\citenamefont {Langreth}(1966)}]{Lan66}%
  \BibitemOpen
  \bibfield  {author} {\bibinfo {author} {\bibfnamefont {D.~C.}\ \bibnamefont
  {Langreth}},\ }\Doi {10.1103/PhysRev.150.516} {\bibfield  {journal} {\bibinfo
   {journal} {Phys. Rev.},\ }\textbf {\bibinfo {volume} {150}},\ \bibinfo
  {pages} {516} (\bibinfo {year} {1966})}\BibitemShut {NoStop}%
\bibitem [{\citenamefont {Martin}(1982)}]{Mar82}%
  \BibitemOpen
  \bibfield  {author} {\bibinfo {author} {\bibfnamefont {R.~M.}\ \bibnamefont
  {Martin}},\ }\Doi {10.1103/PhysRevLett.48.362} {\bibfield  {journal}
  {\bibinfo  {journal} {Phys. Rev. Lett.},\ }\textbf {\bibinfo {volume} {48}},\
  \bibinfo {pages} {362} (\bibinfo {year} {1982})}\BibitemShut {NoStop}%
\bibitem [{\citenamefont {Coleman}(2015)}]{Col15a}%
  \BibitemOpen
  \bibfield  {author} {\bibinfo {author} {\bibfnamefont {P.}~\bibnamefont
  {Coleman}},\ }\enquote {\bibinfo {title} {Lecture notes of the autumn school
  on correlated electrons: Many-body physics: From kondo to hubbard},}\ \
  (\bibinfo  {publisher} {Forschungszentrum Juelich},\ \bibinfo {year} {2015})\
  Chap.\ \bibinfo {chapter} {Heavy Fermions and the Kondo Lattice}\BibitemShut
  {NoStop}%
\bibitem [{\citenamefont {Bodensiek}(2013)}]{Bod13}%
  \BibitemOpen
  \bibfield  {author} {\bibinfo {author} {\bibfnamefont {O.}~\bibnamefont
  {Bodensiek}},\ }\emph {}\href@noop {} {Ph.D.
  thesis},\ \bibinfo  {school} {Georg-August-Universit{\"a}t G{\"o}ttingen}
  (\bibinfo {year} {2013})\BibitemShut {NoStop}%
\bibitem [{\citenamefont {S{\'e}n{\'e}chal}(2008)}]{S08}%
  \BibitemOpen
  \bibfield  {author} {\bibinfo {author} {\bibfnamefont {D.}~\bibnamefont
  {S{\'e}n{\'e}chal}},\ }\href@noop {} {\bibfield  {journal} {\bibinfo
  {journal} {E-print arXiv:0806.2690v2}} (\bibinfo {year} {2008})}\BibitemShut
  {NoStop}%
\bibitem [{\citenamefont {Hoshino}\ and\ \citenamefont
  {Kuramoto}(2013)}]{HK13}%
  \BibitemOpen
  \bibfield  {author} {\bibinfo {author} {\bibfnamefont {S.}~\bibnamefont
  {Hoshino}}\ and\ \bibinfo {author} {\bibfnamefont {Y.}~\bibnamefont
  {Kuramoto}},\ }\Doi {10.1103/PhysRevLett.111.026401} {\bibfield  {journal}
  {\bibinfo  {journal} {Phys. Rev. Lett.},\ }\textbf {\bibinfo {volume}
  {111}},\ \bibinfo {pages} {026401} (\bibinfo {year} {2013})}\BibitemShut
  {NoStop}%
\bibitem [{Note3()}]{Note3}%
  \BibitemOpen
  \bibinfo {note} {Since we do not have access to the $f$-spins via the Green
  function, this is the only way to access this susceptibility. To calculate
  the total local magnetic susceptibility (including the contribution of the
  $c$ electrons) one needs to use spin-dependent variational parameters \cite
  {BP10}, which strongly increases the computational cost. As both the
  $f$-spins and the $c$-electrons are part of the Kondo singlets at strong
  coupling, it is sufficient to investigate the local susceptibility of one of
  its constituents.}\BibitemShut {Stop}%
\bibitem [{Note4()}]{Note4}%
  \BibitemOpen
  \bibinfo {note} {In practice, we determined $\chi $ from a polynomial fit of
  $\Omega (h_f)$ in the AF and from $\Omega (h_f)\sim \protect \qopname \relax
  o{ln}(\protect \qopname \relax o{cosh}(h_f))$ in the PM. The latter
  corresponds to the expected behavior of the magnetization in case of a
  paramagnet, $m\sim \protect \qopname \relax o{cosh}(h_f)$.}\BibitemShut
  {Stop}%
\bibitem [{\citenamefont {Otsuki}\ \emph {et~al.}(2009)\citenamefont {Otsuki},
  \citenamefont {Kusunose},\ and\ \citenamefont {Kuramoto}}]{OKK09}%
  \BibitemOpen
  \bibfield  {author} {\bibinfo {author} {\bibfnamefont {J.}~\bibnamefont
  {Otsuki}}, \bibinfo {author} {\bibfnamefont {H.}~\bibnamefont {Kusunose}}, \
  and\ \bibinfo {author} {\bibfnamefont {Y.}~\bibnamefont {Kuramoto}},\ }\Doi
  {10.1103/PhysRevLett.102.017202} {\bibfield  {journal} {\bibinfo  {journal}
  {Phys. Rev. Lett.},\ }\textbf {\bibinfo {volume} {102}},\ \bibinfo {pages}
  {017202} (\bibinfo {year} {2009})}\BibitemShut {NoStop}%
\bibitem [{\citenamefont {Watanabe}\ and\ \citenamefont {Ogata}(2007)}]{WO07}%
  \BibitemOpen
  \bibfield  {author} {\bibinfo {author} {\bibfnamefont {H.}~\bibnamefont
  {Watanabe}}\ and\ \bibinfo {author} {\bibfnamefont {M.}~\bibnamefont
  {Ogata}},\ }\Doi {10.1103/PhysRevLett.99.136401} {\bibfield  {journal}
  {\bibinfo  {journal} {Phys. Rev. Lett.},\ }\textbf {\bibinfo {volume} {99}},\
  \bibinfo {pages} {136401} (\bibinfo {year} {2007})}\BibitemShut {NoStop}%
\bibitem [{\citenamefont {Laubach}\ \emph {et~al.}(2015)\citenamefont
  {Laubach}, \citenamefont {Thomale}, \citenamefont {Platt}, \citenamefont
  {Hanke},\ and\ \citenamefont {Li}}]{LTP+15}%
  \BibitemOpen
  \bibfield  {author} {\bibinfo {author} {\bibfnamefont {M.}~\bibnamefont
  {Laubach}}, \bibinfo {author} {\bibfnamefont {R.}~\bibnamefont {Thomale}},
  \bibinfo {author} {\bibfnamefont {C.}~\bibnamefont {Platt}}, \bibinfo
  {author} {\bibfnamefont {W.}~\bibnamefont {Hanke}}, \ and\ \bibinfo {author}
  {\bibfnamefont {G.}~\bibnamefont {Li}},\ }\Doi {10.1103/PhysRevB.91.245125}
  {\bibfield  {journal} {\bibinfo  {journal} {Phys. Rev. B},\ }\textbf
  {\bibinfo {volume} {91}},\ \bibinfo {pages} {245125} (\bibinfo {year}
  {2015})}\BibitemShut {NoStop}%
\bibitem [{\citenamefont {{Howczak}}\ \emph {et~al.}(2012)\citenamefont
  {{Howczak}}, \citenamefont {{Kaczmarczyk}},\ and\ \citenamefont
  {{Spa{\l}ek}}}]{HKS12}%
  \BibitemOpen
  \bibfield  {author} {\bibinfo {author} {\bibfnamefont {O.}~\bibnamefont
  {{Howczak}}}, \bibinfo {author} {\bibfnamefont {J.}~\bibnamefont
  {{Kaczmarczyk}}}, \ and\ \bibinfo {author} {\bibfnamefont {J.}~\bibnamefont
  {{Spa{\l}ek}}},\ }\href@noop {} {\bibfield  {journal} {\bibinfo  {journal}
  {E-print arXiv:1209.0621}} (\bibinfo {year} {2012})},\ \Eprint
  {http://arxiv.org/abs/1209.0621} {arXiv:1209.0621 [cond-mat.str-el]}
  \BibitemShut {NoStop}%
\bibitem [{\citenamefont {Masuda}\ and\ \citenamefont {Yamamoto}(2013)}]{MY13}%
  \BibitemOpen
  \bibfield  {author} {\bibinfo {author} {\bibfnamefont {K.}~\bibnamefont
  {Masuda}}\ and\ \bibinfo {author} {\bibfnamefont {D.}~\bibnamefont
  {Yamamoto}},\ }\Doi {10.1103/PhysRevB.87.014516} {\bibfield  {journal}
  {\bibinfo  {journal} {Phys. Rev. B},\ }\textbf {\bibinfo {volume} {87}},\
  \bibinfo {pages} {014516} (\bibinfo {year} {2013})}\BibitemShut {NoStop}%
\bibitem [{\citenamefont {{Hickel}}\ \emph {et~al.}(2003)\citenamefont
  {{Hickel}}, \citenamefont {{Roeseler}},\ and\ \citenamefont
  {{Nolting}}}]{HRN03}%
  \BibitemOpen
  \bibfield  {author} {\bibinfo {author} {\bibfnamefont {T.}~\bibnamefont
  {{Hickel}}}, \bibinfo {author} {\bibfnamefont {J.}~\bibnamefont
  {{Roeseler}}}, \ and\ \bibinfo {author} {\bibfnamefont {W.}~\bibnamefont
  {{Nolting}}},\ }\href@noop {} {\bibfield  {journal} {\bibinfo  {journal}
  {Acta Physica Polonica B},\ }\textbf {\bibinfo {volume} {34}},\ \bibinfo
  {pages} {1291} (\bibinfo {year} {2003})}\BibitemShut {NoStop}%
\bibitem [{\citenamefont {Wang}\ and\ \citenamefont {Gu}(2000)}]{WG00}%
  \BibitemOpen
  \bibfield  {author} {\bibinfo {author} {\bibfnamefont {X.}~\bibnamefont
  {Wang}}\ and\ \bibinfo {author} {\bibfnamefont {B.}~\bibnamefont {Gu}},\
  }\href {http://stacks.iop.org/0253-6102/34/i=4/a=623} {\bibfield  {journal}
  {\bibinfo  {journal} {Communications in Theoretical Physics},\ }\textbf
  {\bibinfo {volume} {34}},\ \bibinfo {pages} {623} (\bibinfo {year}
  {2000})}\BibitemShut {NoStop}%
\bibitem [{\citenamefont {Coqblin}\ \emph {et~al.}(2003)\citenamefont
  {Coqblin}, \citenamefont {Lacroix}, \citenamefont {Gusm{\~a}o},\ and\
  \citenamefont {Iglesias}}]{CLGI03}%
  \BibitemOpen
  \bibfield  {author} {\bibinfo {author} {\bibfnamefont {B.}~\bibnamefont
  {Coqblin}}, \bibinfo {author} {\bibfnamefont {C.}~\bibnamefont {Lacroix}},
  \bibinfo {author} {\bibfnamefont {M.~S.}\ \bibnamefont {Gusm{\~a}o}}, \ and\
  \bibinfo {author} {\bibfnamefont {J.~R.}\ \bibnamefont {Iglesias}},\ }\Doi
  {10.1103/PhysRevB.67.064417} {\bibfield  {journal} {\bibinfo  {journal}
  {Phys. Rev. B},\ }\textbf {\bibinfo {volume} {67}},\ \bibinfo {pages}
  {064417} (\bibinfo {year} {2003})}\BibitemShut {NoStop}%
\bibitem [{\citenamefont {Miyake}\ \emph {et~al.}(1986)\citenamefont {Miyake},
  \citenamefont {Schmitt-Rink},\ and\ \citenamefont {Varma}}]{MSV86}%
  \BibitemOpen
  \bibfield  {author} {\bibinfo {author} {\bibfnamefont {K.}~\bibnamefont
  {Miyake}}, \bibinfo {author} {\bibfnamefont {S.}~\bibnamefont
  {Schmitt-Rink}}, \ and\ \bibinfo {author} {\bibfnamefont {C.~M.}\
  \bibnamefont {Varma}},\ }\Doi {10.1103/PhysRevB.34.6554} {\bibfield
  {journal} {\bibinfo  {journal} {Phys. Rev. B},\ }\textbf {\bibinfo {volume}
  {34}},\ \bibinfo {pages} {6554} (\bibinfo {year} {1986})}\BibitemShut
  {NoStop}%
\bibitem [{\citenamefont {Scalapino}\ \emph {et~al.}(1986)\citenamefont
  {Scalapino}, \citenamefont {Loh},\ and\ \citenamefont {Hirsch}}]{SLH86}%
  \BibitemOpen
  \bibfield  {author} {\bibinfo {author} {\bibfnamefont {D.~J.}\ \bibnamefont
  {Scalapino}}, \bibinfo {author} {\bibfnamefont {E.}~\bibnamefont {Loh}}, \
  and\ \bibinfo {author} {\bibfnamefont {J.~E.}\ \bibnamefont {Hirsch}},\ }\Doi
  {10.1103/PhysRevB.34.8190} {\bibfield  {journal} {\bibinfo  {journal} {Phys.
  Rev. B},\ }\textbf {\bibinfo {volume} {34}},\ \bibinfo {pages} {8190}
  (\bibinfo {year} {1986})}\BibitemShut {NoStop}%
\bibitem [{\citenamefont {B\'eal-Monod}\ \emph {et~al.}(1986)\citenamefont
  {B\'eal-Monod}, \citenamefont {Bourbonnais},\ and\ \citenamefont
  {Emery}}]{BBE86}%
  \BibitemOpen
  \bibfield  {author} {\bibinfo {author} {\bibfnamefont {M.~T.}\ \bibnamefont
  {B\'eal-Monod}}, \bibinfo {author} {\bibfnamefont {C.}~\bibnamefont
  {Bourbonnais}}, \ and\ \bibinfo {author} {\bibfnamefont {V.~J.}\ \bibnamefont
  {Emery}},\ }\Doi {10.1103/PhysRevB.34.7716} {\bibfield  {journal} {\bibinfo
  {journal} {Phys. Rev. B},\ }\textbf {\bibinfo {volume} {34}},\ \bibinfo
  {pages} {7716} (\bibinfo {year} {1986})}\BibitemShut {NoStop}%
\bibitem [{\citenamefont {{Verstraete}}\ and\ \citenamefont
  {{Cirac}}(2004)}]{VC04}%
  \BibitemOpen
  \bibfield  {author} {\bibinfo {author} {\bibfnamefont {F.}~\bibnamefont
  {{Verstraete}}}\ and\ \bibinfo {author} {\bibfnamefont {J.~I.}\ \bibnamefont
  {{Cirac}}},\ }\href@noop {} {\bibfield  {journal} {\bibinfo  {journal}
  {E-print arXiv:cond-mat/0407066}} (\bibinfo {year} {2004})},\ \Eprint
  {http://arxiv.org/abs/cond-mat/0407066} {cond-mat/0407066} \BibitemShut
  {NoStop}%
\bibitem [{\citenamefont {Verstraete}\ \emph {et~al.}(2008)\citenamefont
  {Verstraete}, \citenamefont {Murg},\ and\ \citenamefont {Cirac}}]{VMC08}%
  \BibitemOpen
  \bibfield  {author} {\bibinfo {author} {\bibfnamefont {F.}~\bibnamefont
  {Verstraete}}, \bibinfo {author} {\bibfnamefont {V.}~\bibnamefont {Murg}}, \
  and\ \bibinfo {author} {\bibfnamefont {J.}~\bibnamefont {Cirac}},\ }\Doi
  {10.1080/14789940801912366} {\bibfield  {journal} {\bibinfo  {journal}
  {Advances in Physics},\ }\textbf {\bibinfo {volume} {57}},\ \bibinfo {pages}
  {143} (\bibinfo {year} {2008})}\BibitemShut {NoStop}%
\bibitem [{\citenamefont {Corboz}\ \emph {et~al.}(2010)\citenamefont {Corboz},
  \citenamefont {Or\'us}, \citenamefont {Bauer},\ and\ \citenamefont
  {Vidal}}]{COBV10}%
  \BibitemOpen
  \bibfield  {author} {\bibinfo {author} {\bibfnamefont {P.}~\bibnamefont
  {Corboz}}, \bibinfo {author} {\bibfnamefont {R.}~\bibnamefont {Or\'us}},
  \bibinfo {author} {\bibfnamefont {B.}~\bibnamefont {Bauer}}, \ and\ \bibinfo
  {author} {\bibfnamefont {G.}~\bibnamefont {Vidal}},\ }\Doi
  {10.1103/PhysRevB.81.165104} {\bibfield  {journal} {\bibinfo  {journal}
  {Phys. Rev. B},\ }\textbf {\bibinfo {volume} {81}},\ \bibinfo {pages}
  {165104} (\bibinfo {year} {2010})}\BibitemShut {NoStop}%
\bibitem [{\citenamefont {Coleman}\ and\ \citenamefont
  {Nevidomskyy}(2010)}]{CN10}%
  \BibitemOpen
  \bibfield  {author} {\bibinfo {author} {\bibfnamefont {P.}~\bibnamefont
  {Coleman}}\ and\ \bibinfo {author} {\bibfnamefont {A.}~\bibnamefont
  {Nevidomskyy}},\ }\Doi {10.1007/s10909-010-0213-4} {\bibfield  {journal}
  {\bibinfo  {journal} {Journal of Low Temperature Physics},\ }\textbf
  {\bibinfo {volume} {161}},\ \bibinfo {pages} {182} (\bibinfo {year}
  {2010})},\ ISSN \bibinfo {issn} {0022-2291}\BibitemShut {NoStop}%
\bibitem [{\citenamefont {Paul}\ \emph {et~al.}(2007)\citenamefont {Paul},
  \citenamefont {P\'epin},\ and\ \citenamefont {Norman}}]{PPN07}%
  \BibitemOpen
  \bibfield  {author} {\bibinfo {author} {\bibfnamefont {I.}~\bibnamefont
  {Paul}}, \bibinfo {author} {\bibfnamefont {C.}~\bibnamefont {P\'epin}}, \
  and\ \bibinfo {author} {\bibfnamefont {M.~R.}\ \bibnamefont {Norman}},\ }\Doi
  {10.1103/PhysRevLett.98.026402} {\bibfield  {journal} {\bibinfo  {journal}
  {Phys. Rev. Lett.},\ }\textbf {\bibinfo {volume} {98}},\ \bibinfo {pages}
  {026402} (\bibinfo {year} {2007})}\BibitemShut {NoStop}%
\bibitem [{\citenamefont {Xavier}\ and\ \citenamefont {Dagotto}(2008)}]{XD08}%
  \BibitemOpen
  \bibfield  {author} {\bibinfo {author} {\bibfnamefont {J.~C.}\ \bibnamefont
  {Xavier}}\ and\ \bibinfo {author} {\bibfnamefont {E.}~\bibnamefont
  {Dagotto}},\ }\Doi {10.1103/PhysRevLett.100.146403} {\bibfield  {journal}
  {\bibinfo  {journal} {Phys. Rev. Lett.},\ }\textbf {\bibinfo {volume}
  {100}},\ \bibinfo {pages} {146403} (\bibinfo {year} {2008})}\BibitemShut
  {NoStop}%
\bibitem [{\citenamefont {Yu}\ \emph {et~al.}(2014)\citenamefont {Yu},
  \citenamefont {Guang-Ming},\ and\ \citenamefont {Lu}}]{YGL14}%
  \BibitemOpen
  \bibfield  {author} {\bibinfo {author} {\bibfnamefont {L.}~\bibnamefont
  {Yu}}, \bibinfo {author} {\bibfnamefont {Z.}~\bibnamefont {Guang-Ming}}, \
  and\ \bibinfo {author} {\bibfnamefont {Y.}~\bibnamefont {Lu}},\ }\href
  {http://stacks.iop.org/0256-307X/31/i=8/a=087102} {\bibfield  {journal}
  {\bibinfo  {journal} {Chinese Physics Letters},\ }\textbf {\bibinfo {volume}
  {31}},\ \bibinfo {pages} {087102} (\bibinfo {year} {2014})}\BibitemShut
  {NoStop}%
\bibitem [{\citenamefont {Flint}\ and\ \citenamefont {Coleman}(2010)}]{FC10}%
  \BibitemOpen
  \bibfield  {author} {\bibinfo {author} {\bibfnamefont {R.}~\bibnamefont
  {Flint}}\ and\ \bibinfo {author} {\bibfnamefont {P.}~\bibnamefont
  {Coleman}},\ }\Doi {10.1103/PhysRevLett.105.246404} {\bibfield  {journal}
  {\bibinfo  {journal} {Phys. Rev. Lett.},\ }\textbf {\bibinfo {volume}
  {105}},\ \bibinfo {pages} {246404} (\bibinfo {year} {2010})}\BibitemShut
  {NoStop}%
\bibitem [{\citenamefont {Pixley}\ \emph {et~al.}(2015)\citenamefont {Pixley},
  \citenamefont {Deng}, \citenamefont {Ingersent},\ and\ \citenamefont
  {Si}}]{PDIS15}%
  \BibitemOpen
  \bibfield  {author} {\bibinfo {author} {\bibfnamefont {J.~H.}\ \bibnamefont
  {Pixley}}, \bibinfo {author} {\bibfnamefont {L.}~\bibnamefont {Deng}},
  \bibinfo {author} {\bibfnamefont {K.}~\bibnamefont {Ingersent}}, \ and\
  \bibinfo {author} {\bibfnamefont {Q.}~\bibnamefont {Si}},\ }\Doi
  {10.1103/PhysRevB.91.201109} {\bibfield  {journal} {\bibinfo  {journal}
  {Phys. Rev. B},\ }\textbf {\bibinfo {volume} {91}},\ \bibinfo {pages}
  {201109} (\bibinfo {year} {2015})}\BibitemShut {NoStop}%
\bibitem [{\citenamefont {Luttinger}(1960)}]{Lut60}%
  \BibitemOpen
  \bibfield  {author} {\bibinfo {author} {\bibfnamefont {J.~M.}\ \bibnamefont
  {Luttinger}},\ }\Doi {10.1103/PhysRev.119.1153} {\bibfield  {journal}
  {\bibinfo  {journal} {Phys. Rev.},\ }\textbf {\bibinfo {volume} {119}},\
  \bibinfo {pages} {1153} (\bibinfo {year} {1960})}\BibitemShut {NoStop}%
\bibitem [{\citenamefont {Oshikawa}(2000)}]{Osh00}%
  \BibitemOpen
  \bibfield  {author} {\bibinfo {author} {\bibfnamefont {M.}~\bibnamefont
  {Oshikawa}},\ }\Doi {10.1103/PhysRevLett.84.3370} {\bibfield  {journal}
  {\bibinfo  {journal} {Phys. Rev. Lett.},\ }\textbf {\bibinfo {volume} {84}},\
  \bibinfo {pages} {3370} (\bibinfo {year} {2000})}\BibitemShut {NoStop}%
\bibitem [{\citenamefont {Dzyaloshinskii}(2003)}]{Dzy03}%
  \BibitemOpen
  \bibfield  {author} {\bibinfo {author} {\bibfnamefont {I.}~\bibnamefont
  {Dzyaloshinskii}},\ }\Doi {10.1103/PhysRevB.68.085113} {\bibfield  {journal}
  {\bibinfo  {journal} {Phys. Rev. B},\ }\textbf {\bibinfo {volume} {68}},\
  \bibinfo {pages} {085113} (\bibinfo {year} {2003})}\BibitemShut {NoStop}%
\bibitem [{\citenamefont {Ortloff}\ \emph {et~al.}(2007)\citenamefont
  {Ortloff}, \citenamefont {Balzer},\ and\ \citenamefont {Potthoff}}]{OBP07}%
  \BibitemOpen
  \bibfield  {author} {\bibinfo {author} {\bibfnamefont {J.}~\bibnamefont
  {Ortloff}}, \bibinfo {author} {\bibfnamefont {M.}~\bibnamefont {Balzer}}, \
  and\ \bibinfo {author} {\bibfnamefont {M.}~\bibnamefont {Potthoff}},\ }\Doi
  {10.1140/epjb/e2007-00203-7} {\bibfield  {journal} {\bibinfo  {journal} {The
  European Physical Journal B},\ }\textbf {\bibinfo {volume} {58}},\ \bibinfo
  {pages} {37} (\bibinfo {year} {2007})},\ ISSN \bibinfo {issn}
  {1434-6036}\BibitemShut {NoStop}%
\bibitem [{\citenamefont {Dzyaloshinskii}(1996)}]{Dzy96}%
  \BibitemOpen
  \bibfield  {author} {\bibinfo {author} {\bibfnamefont {I.}~\bibnamefont
  {Dzyaloshinskii}},\ }\Doi {10.1051/jp1:1996127} {\bibfield  {journal}
  {\bibinfo  {journal} {J. Phys. I France},\ }\textbf {\bibinfo {volume} {6}},\
  \bibinfo {pages} {119} (\bibinfo {year} {1996})}\BibitemShut {NoStop}%
\bibitem [{\citenamefont {Rosch}(2007)}]{Ros07}%
  \BibitemOpen
  \bibfield  {author} {\bibinfo {author} {\bibfnamefont {A.}~\bibnamefont
  {Rosch}},\ }\Doi {10.1140/epjb/e2007-00312-3} {\bibfield  {journal} {\bibinfo
   {journal} {The European Physical Journal B},\ }\textbf {\bibinfo {volume}
  {59}},\ \bibinfo {pages} {495} (\bibinfo {year} {2007})},\ ISSN \bibinfo
  {issn} {1434-6036}\BibitemShut {NoStop}%
\bibitem [{\citenamefont {Eder}\ \emph {et~al.}(2011)\citenamefont {Eder},
  \citenamefont {Seki},\ and\ \citenamefont {Ohta}}]{ESO11}%
  \BibitemOpen
  \bibfield  {author} {\bibinfo {author} {\bibfnamefont {R.}~\bibnamefont
  {Eder}}, \bibinfo {author} {\bibfnamefont {K.}~\bibnamefont {Seki}}, \ and\
  \bibinfo {author} {\bibfnamefont {Y.}~\bibnamefont {Ohta}},\ }\Doi
  {10.1103/PhysRevB.83.205137} {\bibfield  {journal} {\bibinfo  {journal}
  {Phys. Rev. B},\ }\textbf {\bibinfo {volume} {83}},\ \bibinfo {pages}
  {205137} (\bibinfo {year} {2011})}\BibitemShut {NoStop}%
\bibitem [{\citenamefont {Sakai}\ \emph
  {et~al.}(2009){\natexlab{a}}\citenamefont {Sakai}, \citenamefont {Motome},\
  and\ \citenamefont {Imada}}]{SMI09}%
  \BibitemOpen
  \bibfield  {author} {\bibinfo {author} {\bibfnamefont {S.}~\bibnamefont
  {Sakai}}, \bibinfo {author} {\bibfnamefont {Y.}~\bibnamefont {Motome}}, \
  and\ \bibinfo {author} {\bibfnamefont {M.}~\bibnamefont {Imada}},\ }\Doi
  {10.1103/PhysRevLett.102.056404} {\bibfield  {journal} {\bibinfo  {journal}
  {Phys. Rev. Lett.},\ }\textbf {\bibinfo {volume} {102}},\ \bibinfo {pages}
  {056404} (\bibinfo {year} {2009}{\natexlab{a}})}\BibitemShut {NoStop}%
\bibitem [{\citenamefont {Sakai}\ \emph
  {et~al.}(2009){\natexlab{b}}\citenamefont {Sakai}, \citenamefont {Motome},\
  and\ \citenamefont {Imada}}]{SMI09a}%
  \BibitemOpen
  \bibfield  {author} {\bibinfo {author} {\bibfnamefont {S.}~\bibnamefont
  {Sakai}}, \bibinfo {author} {\bibfnamefont {Y.}~\bibnamefont {Motome}}, \
  and\ \bibinfo {author} {\bibfnamefont {M.}~\bibnamefont {Imada}},\ }\Doi
  {https://doi.org/10.1016/j.physb.2009.07.050} {\bibfield  {journal} {\bibinfo
   {journal} {Physica B: Condensed Matter},\ }\textbf {\bibinfo {volume}
  {404}},\ \bibinfo {pages} {3183 } (\bibinfo {year}
  {2009}{\natexlab{b}})}\BibitemShut {NoStop}%
\bibitem [{\citenamefont {Schmalian}\ \emph {et~al.}(1996)\citenamefont
  {Schmalian}, \citenamefont {Langer}, \citenamefont {Grabowski},\ and\
  \citenamefont {Bennemann}}]{SLGB96}%
  \BibitemOpen
  \bibfield  {author} {\bibinfo {author} {\bibfnamefont {J.}~\bibnamefont
  {Schmalian}}, \bibinfo {author} {\bibfnamefont {M.}~\bibnamefont {Langer}},
  \bibinfo {author} {\bibfnamefont {S.}~\bibnamefont {Grabowski}}, \ and\
  \bibinfo {author} {\bibfnamefont {K.~H.}\ \bibnamefont {Bennemann}},\ }\Doi
  {10.1103/PhysRevB.54.4336} {\bibfield  {journal} {\bibinfo  {journal} {Phys.
  Rev. B},\ }\textbf {\bibinfo {volume} {54}},\ \bibinfo {pages} {4336}
  (\bibinfo {year} {1996})}\BibitemShut {NoStop}%
\bibitem [{\citenamefont {Gr\"ober}\ \emph {et~al.}(2000)\citenamefont
  {Gr\"ober}, \citenamefont {Eder},\ and\ \citenamefont {Hanke}}]{GEH00}%
  \BibitemOpen
  \bibfield  {author} {\bibinfo {author} {\bibfnamefont {C.}~\bibnamefont
  {Gr\"ober}}, \bibinfo {author} {\bibfnamefont {R.}~\bibnamefont {Eder}}, \
  and\ \bibinfo {author} {\bibfnamefont {W.}~\bibnamefont {Hanke}},\ }\Doi
  {10.1103/PhysRevB.62.4336} {\bibfield  {journal} {\bibinfo  {journal} {Phys.
  Rev. B},\ }\textbf {\bibinfo {volume} {62}},\ \bibinfo {pages} {4336}
  (\bibinfo {year} {2000})}\BibitemShut {NoStop}%
\bibitem [{\citenamefont {Stanescu}\ \emph {et~al.}(2007)\citenamefont
  {Stanescu}, \citenamefont {Phillips},\ and\ \citenamefont {Choy}}]{SPC07}%
  \BibitemOpen
  \bibfield  {author} {\bibinfo {author} {\bibfnamefont {T.~D.}\ \bibnamefont
  {Stanescu}}, \bibinfo {author} {\bibfnamefont {P.}~\bibnamefont {Phillips}},
  \ and\ \bibinfo {author} {\bibfnamefont {T.-P.}\ \bibnamefont {Choy}},\ }\Doi
  {10.1103/PhysRevB.75.104503} {\bibfield  {journal} {\bibinfo  {journal}
  {Phys. Rev. B},\ }\textbf {\bibinfo {volume} {75}},\ \bibinfo {pages}
  {104503} (\bibinfo {year} {2007})}\BibitemShut {NoStop}%
\bibitem [{\citenamefont {Kokalj}\ and\ \citenamefont
  {Prelov\ifmmode~\check{s}\else \v{s}\fi{}ek}(2007)}]{KP07}%
  \BibitemOpen
  \bibfield  {author} {\bibinfo {author} {\bibfnamefont {J.}~\bibnamefont
  {Kokalj}}\ and\ \bibinfo {author} {\bibfnamefont {P.}~\bibnamefont
  {Prelov\ifmmode~\check{s}\else \v{s}\fi{}ek}},\ }\Doi
  {10.1103/PhysRevB.75.045111} {\bibfield  {journal} {\bibinfo  {journal}
  {Phys. Rev. B},\ }\textbf {\bibinfo {volume} {75}},\ \bibinfo {pages}
  {045111} (\bibinfo {year} {2007})}\BibitemShut {NoStop}%
\bibitem [{\citenamefont {Kokalj}\ and\ \citenamefont
  {Prelov\ifmmode~\check{s}\else \v{s}\fi{}ek}(2008)}]{KP08}%
  \BibitemOpen
  \bibfield  {author} {\bibinfo {author} {\bibfnamefont {J.}~\bibnamefont
  {Kokalj}}\ and\ \bibinfo {author} {\bibfnamefont {P.}~\bibnamefont
  {Prelov\ifmmode~\check{s}\else \v{s}\fi{}ek}},\ }\Doi
  {10.1103/PhysRevB.78.153103} {\bibfield  {journal} {\bibinfo  {journal}
  {Phys. Rev. B},\ }\textbf {\bibinfo {volume} {78}},\ \bibinfo {pages}
  {153103} (\bibinfo {year} {2008})}\BibitemShut {NoStop}%
\bibitem [{\citenamefont {{Farid}}(2007)}]{Far07}%
  \BibitemOpen
  \bibfield  {author} {\bibinfo {author} {\bibfnamefont {B.}~\bibnamefont
  {{Farid}}},\ }\href@noop {} {\bibfield  {journal} {\bibinfo  {journal} {ArXiv
  e-prints}} (\bibinfo {year} {2007})},\ \Eprint
  {http://arxiv.org/abs/0711.0952} {arXiv:0711.0952 [cond-mat.str-el]}
  \BibitemShut {NoStop}%
\bibitem [{\citenamefont {Sakai}\ \emph {et~al.}(2012)\citenamefont {Sakai},
  \citenamefont {Sangiovanni}, \citenamefont {Civelli}, \citenamefont {Motome},
  \citenamefont {Held},\ and\ \citenamefont {Imada}}]{SSC+12}%
  \BibitemOpen
  \bibfield  {author} {\bibinfo {author} {\bibfnamefont {S.}~\bibnamefont
  {Sakai}}, \bibinfo {author} {\bibfnamefont {G.}~\bibnamefont {Sangiovanni}},
  \bibinfo {author} {\bibfnamefont {M.}~\bibnamefont {Civelli}}, \bibinfo
  {author} {\bibfnamefont {Y.}~\bibnamefont {Motome}}, \bibinfo {author}
  {\bibfnamefont {K.}~\bibnamefont {Held}}, \ and\ \bibinfo {author}
  {\bibfnamefont {M.}~\bibnamefont {Imada}},\ }\Doi
  {10.1103/PhysRevB.85.035102} {\bibfield  {journal} {\bibinfo  {journal}
  {Phys. Rev. B},\ }\textbf {\bibinfo {volume} {85}},\ \bibinfo {pages}
  {035102} (\bibinfo {year} {2012})}\BibitemShut {NoStop}%
\bibitem [{\citenamefont {Bodensiek}(2016)}]{Bod16}%
  \BibitemOpen
  \bibfield  {author} {\bibinfo {author} {\bibfnamefont {O.}~\bibnamefont
  {Bodensiek}},\ }\href@noop {} {\enquote {\bibinfo {title} {Private
  communication},}\ } (\bibinfo {year} {2016})\BibitemShut {NoStop}%
\bibitem [{\citenamefont {Peters}\ and\ \citenamefont {Pruschke}(2007)}]{PP07}%
  \BibitemOpen
  \bibfield  {author} {\bibinfo {author} {\bibfnamefont {R.}~\bibnamefont
  {Peters}}\ and\ \bibinfo {author} {\bibfnamefont {T.}~\bibnamefont
  {Pruschke}},\ }\Doi {10.1103/PhysRevB.76.245101} {\bibfield  {journal}
  {\bibinfo  {journal} {Phys. Rev. B},\ }\textbf {\bibinfo {volume} {76}},\
  \bibinfo {pages} {245101} (\bibinfo {year} {2007})}\BibitemShut {NoStop}%
\bibitem [{\citenamefont {Rachel}\ \emph {et~al.}(2015)\citenamefont {Rachel},
  \citenamefont {Laubach}, \citenamefont {Reuther},\ and\ \citenamefont
  {Thomale}}]{RLRT15}%
  \BibitemOpen
  \bibfield  {author} {\bibinfo {author} {\bibfnamefont {S.}~\bibnamefont
  {Rachel}}, \bibinfo {author} {\bibfnamefont {M.}~\bibnamefont {Laubach}},
  \bibinfo {author} {\bibfnamefont {J.}~\bibnamefont {Reuther}}, \ and\
  \bibinfo {author} {\bibfnamefont {R.}~\bibnamefont {Thomale}},\ }\Doi
  {10.1103/PhysRevLett.114.167201} {\bibfield  {journal} {\bibinfo  {journal}
  {Phys. Rev. Lett.},\ }\textbf {\bibinfo {volume} {114}},\ \bibinfo {pages}
  {167201} (\bibinfo {year} {2015})}\BibitemShut {NoStop}%
\bibitem [{\citenamefont {Pruschke}(2014)}]{Pru14}%
  \BibitemOpen
  \bibfield  {author} {\bibinfo {author} {\bibfnamefont {T.}~\bibnamefont
  {Pruschke}},\ }\Doi {10.1088/978-1-627-05328-0} {\emph {\bibinfo {title}
  {Advanced Solid State Theory}}},\ 2053-2571\ (\bibinfo  {publisher} {Morgan
  \& Claypool Publishers},\ \bibinfo {year} {2014})\ ISBN \bibinfo {isbn}
  {978-1-627-05328-0}\BibitemShut {NoStop}%
\bibitem [{\citenamefont {Laubach}\ \emph {et~al.}(2016)\citenamefont
  {Laubach}, \citenamefont {Joshi}, \citenamefont {Reuther}, \citenamefont
  {Thomale}, \citenamefont {Vojta},\ and\ \citenamefont {Rachel}}]{LJR+16}%
  \BibitemOpen
  \bibfield  {author} {\bibinfo {author} {\bibfnamefont {M.}~\bibnamefont
  {Laubach}}, \bibinfo {author} {\bibfnamefont {D.~G.}\ \bibnamefont {Joshi}},
  \bibinfo {author} {\bibfnamefont {J.}~\bibnamefont {Reuther}}, \bibinfo
  {author} {\bibfnamefont {R.}~\bibnamefont {Thomale}}, \bibinfo {author}
  {\bibfnamefont {M.}~\bibnamefont {Vojta}}, \ and\ \bibinfo {author}
  {\bibfnamefont {S.}~\bibnamefont {Rachel}},\ }\Doi
  {10.1103/PhysRevB.93.041106} {\bibfield  {journal} {\bibinfo  {journal}
  {Phys. Rev. B},\ }\textbf {\bibinfo {volume} {93}},\ \bibinfo {pages}
  {041106} (\bibinfo {year} {2016})}\BibitemShut {NoStop}%
\end{thebibliography}%

\end{document}